\documentclass[11pt, oneside]{Thesis} 

\graphicspath{{Pictures/}} 
\usepackage[square, numbers, comma, sort&compress]{natbib}

\hypersetup{urlcolor=blue, colorlinks=true} 
\title{\ttitle} 
\usepackage{relsize}
\usepackage{amsmath}
\usepackage{cleveref}
\usepackage{adjustbox}
\usepackage{enumitem}
\newlist{arrowlist}{itemize}{1}
\setlist[arrowlist]{label=$\Rightarrow$}

\usepackage{algorithm}

\usepackage{leftidx}
\usepackage{algpseudocode}
\usepackage{mathtools}
\usepackage{tikz}
\usetikzlibrary{shapes.geometric, arrows}
\usepackage{tensor}
\newcommand\Perms[2]{\tensor[^{#2}]C{_{#1}}}

\newcommand{\tikzcircle}[2][red,fill=red]{\tikz[baseline=-0.5ex]\draw[#1,radius=#2] (0,0) circle ;}%

\renewcommand{\bibname}{References}

\begin{document}

\frontmatter 

\setstretch{1.25} 

\fancyhead{} 
\rhead{\thepage} 
\lhead{} 

\pagestyle{fancy} 

\newcommand{\HRule}{\rule{\linewidth}{0.5mm}} 

\hypersetup{pdftitle={\ttitle}}
\hypersetup{pdfsubject=\subjectname}
\hypersetup{pdfauthor=\authornames}
\hypersetup{pdfkeywords=\keywordnames}


\begin{titlepage}
\begin{center}

\textsc{\LARGE \univname}\\[1.5cm] 

\HRule \\[0.4cm] 
{\Large \bfseries \ttitle}\\[0.4cm] 
\HRule \\[1.5cm] 



\begin{figure}[h]
   \centering
   \includegraphics[scale=0.08]{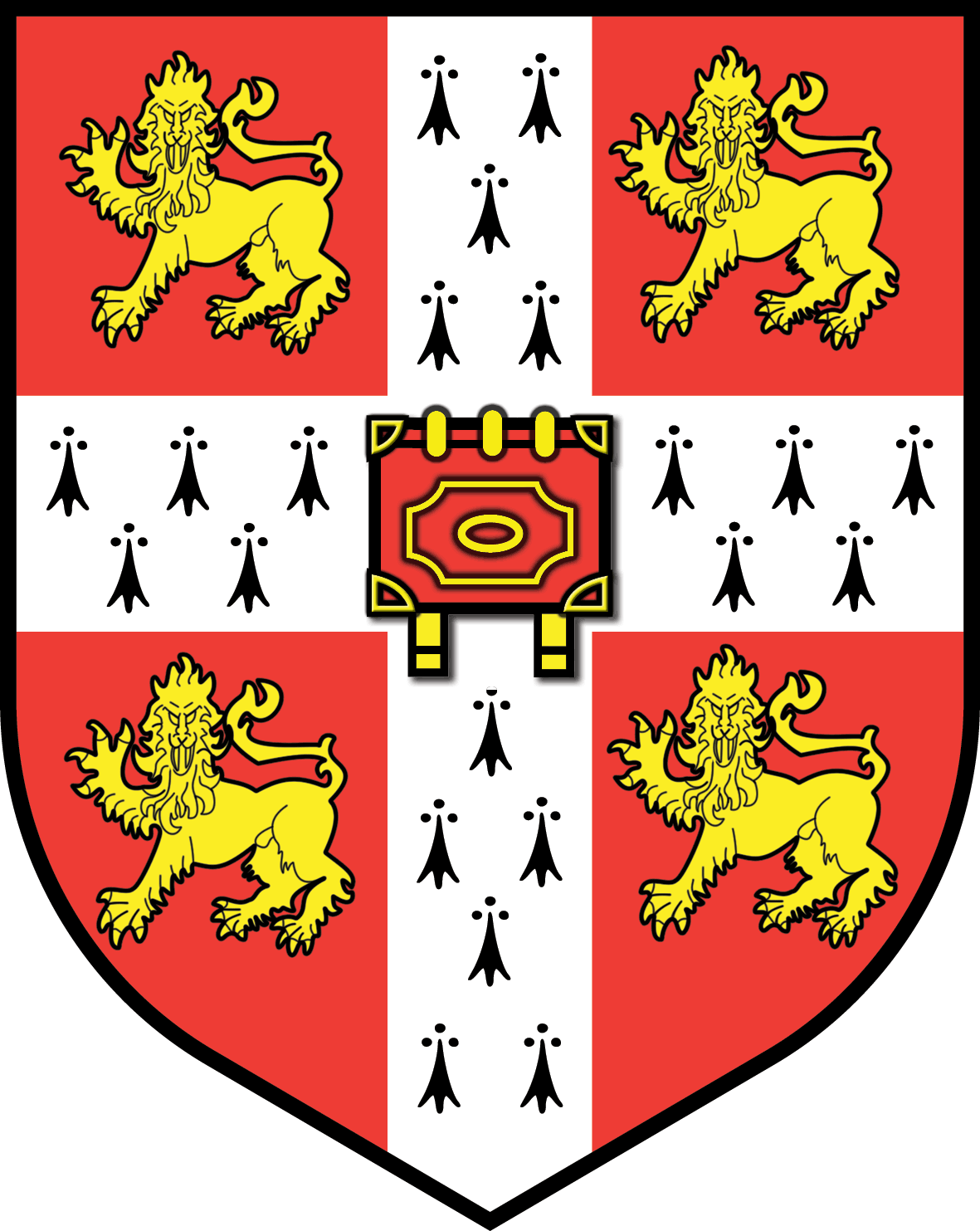}
\end{figure}

\vspace{1cm}

\begin{center}

{\LARGE \textbf{Nick Woods}} 
\vspace{0.5cm}

{\Large Clare College}

\vspace{0.5cm}

\begin{Large}
Theory of Condensed Matter, Cavendish Laboratory
\end{Large}
\end{center}






\large \textit{A thesis submitted in fulfilment of the requirements\\ for the degree of Master of Philosophy}\\[0.3cm] 
\textit{in the}\\[0.2cm]
Centre for Scientific Computing \\[0.8cm] 
Supervisors: Dr. Phil Hasnip, Prof. Mike Payne

\vspace{1cm}




\begin{minipage}{0.4\textwidth}
\begin{flushleft} \large
\end{flushleft}
\end{minipage}
\begin{minipage}{0.4\textwidth}
\begin{flushright} \large
\end{flushright}
\end{minipage}\\[1cm]

\end{center}

\vfill

\end{titlepage}

\newpage

\clearpage
\vspace*{\fill}
\begin{center}
\begin{minipage}{.7\textwidth}
\large This dissertation is substantially my own work and conforms to the University of Cambridge’s guidelines on plagiarism. Where reference has been made to other research this is acknowledged in the text and bibliography. 
\end{minipage}
\end{center}
\vfill 
\clearpage








\begin{abstract}

Density functional theory (DFT) in the Kohn-Sham (KS) framework presents a non-linear eigenvalue problem whereby, in order to solve the system, the particle density one calculates from the KS equations must be equal to the particle density one uses to construct the KS equations. Such a solution defines what it means to attain \textit{self-consistency}. This work studies the methods used to achieve and accelerate convergence toward a self-consistent solution of KS DFT. That is, the extent to which one can utilise prior information about an input (and the framework itself) to minimise the time spent in going from an initial guess solution, to a self-consistent solution. The ideas and techniques presented here are general and (mostly) implementation independent. However, this work will focus specifically on self-consistency within the \textsc{castep} software -- a planewave, pseudopotential implementation. Density (potential) mixing is the name formally given to the process one uses to combine iterative particle densities (potentials) in order to produce an optimal estimate for the subsequent particle density (potential) such that self-consistency is achieved. Density mixing is studied from two perspectives. First, from a mathematical perspective, where KS DFT is simply treated as a featureless non-linear system -- a `black-box'. Second, from a physical perspective, where the \textit{linear response} of KS DFT is studied in an attempt to augment the above-mentioned purely mathematical treatment; this defines preconditioning. The former perspective leads to the implementation of two recently proposed mathematical methods within \textsc{castep}: Marks $\&$ Luke (2011) and Banerjee \textit{et al}$.$ (2015). Both of these methods  are examined utilising a test suite of over fifteen representative input systems, and neither were found to provide a significant, systematic improvement over the default methods in \textsc{castep}. The latter perspective involves the implementation of an adaptable framework allowing one to posit a model for the inhomogeneous KS dielectric in either real or Fourier space, which is in turn used to precondition the self-consistent cycles. The scope for improvement with such a framework in place is discussed, and simple models utilising the framework are analysed with the aforementioned test suite.

\end{abstract}

\clearpage

\setstretch{1.3} 

\acknowledgements{\addtocontents{toc}{\vspace{1em}} 
I would first like to thank Phil Hasnip, my supervisor, for countless hours of invaluable help and guidance, despite the difficulties caused by remote supervising. Matt Smith, for many fun discussions in our (now 1\% completed) quest to understand \textsc{castep}, particularly in helping me understand \textsc{castep}'s parallelism, and for proof reading the thesis. Andrew Morris and his group, for kindly letting me sit in on (and occasionally contribute to) their group meetings, and providing me with an application project to run along side the development. Finally, Mike Payne and Michael Rutter, for help throughout the year with \textsc{castep}, computing, and logistical issues. 

}
\clearpage 

\clearpage

%

\pagestyle{fancy} 

\lhead{\emph{Contents}} 
\tableofcontents 


\mainmatter 

\pagestyle{fancy} 


\makeatletter
\def\@makechapterhead#1{%
  \vspace*{50\p@}%
  {\parindent \z@ \raggedright \normalfont
    \interlinepenalty\@M
    \Huge \bfseries #1\par\nobreak
    \vskip 40\p@
  }}
  \makeatother

\renewcommand{\chaptername}{}


\chapter{Introduction} 

\label{Chapter1} 

\lhead{1. \emph{Introduction}} 


\section{Overview}

The term \textit{ab initio}, meaning `from the beginning', is used in physics to refer to the current established level of physical theory -- that is, \textit{quantum mechanics}. Within the framework of quantum mechanics, the natural world is characterised completely by a quantum state obeying a certain mathematical structure. However, predicting all facets of the natural world from this mathematical structure is a near-impossible task, defining so-called \textit{emergence}. Analytic solutions for the quantum state are rare, and numerical solutions come up against the (in)famous exponential wall of computational complexity \citep{martin_electruc}. Finding a computationally tractable method for extracting exact quantum mechanical predictions has been the source of much work across many scientific disciplines, e.g$.$ quantum chemistry \citep{qchem}, materials science \citep{matsci}, nuclear physics \citep{nucphys}, and particilarly for the following work, solid state physics \citep{phil_dft}. Density functional theory (DFT) within the Kohn-Sham (KS) framework has emerged as the most widely used method for performing \textit{ab initio} calculations \citep{dft111,111122}. The success of KS DFT is, in part, due to its vast domain of applicability, particularly in predicting ground state properties of matter such as stable phases and cell parameters \citep{repdft}. Moreover, the construction and implementation of KS DFT allows these properties, and more, to be accessed in a feasible timeframe for systems containing over one thousand constituents \citep{onetep}. Efficient numerical implementation of KS DFT is therefore of paramount importance, and is also the focus of the work to follow.

The equations required to solve KS DFT define a \textit{non-linear eigenvalue problem}. That is, one first solves the so-called KS equations,
\begin{gather}
\hat{H}^{\textsc{ks}}[\rho(\textbf{r})] \phi_i(\textbf{r}) = \epsilon_i \phi_i(\textbf{r}), \label{ks111} \\
\int d\textbf{r} \text{ }  \phi^*_i(\textbf{r})  \phi_{j}(\textbf{r}) = \delta_{ij},
\end{gather}
taking the form of a \textit{linear} eigenvalue problem, where $\{ \phi_i(\textbf{r}) \}$ are the \textit{single particle wavefunctions}, and the spectrum of the KS Hamiltonian defines the corresponding single particle energies. For a system containing $N$ constituents, only the $N$ lowest energy eigenfunctions and eigenvalues of the KS Hamiltonian need be calculated, leading to a \textit{particle density} defined via
\begin{align}
\rho(\textbf{r}) = \sum_{i=1}^{N} |\phi_i(\textbf{r})|^2. \label{elecdens111}
\end{align}
This particle density, however, determines the form of the KS Hamiltonian in Eq$.$ (\ref{ks111}), the eigenfunctions of which are in turn used to define the particle density. A set of single particle wavefunctions is thus required that satisfies both Eq$.$ (\ref{ks111}) and Eq$.$ (\ref{elecdens111}) simultaneously, at which point the solution is said to be \textit{self-consistent}. This work will focus on the algorithms and strategies employed to determine this self-consistent solution, often referred to in literature as the \textit{self-consistent field} (SCF) procedure \citep{gonze}. Starting from an initial estimate of the particle density (or equivalently, the single particle wavefunctions), the SCF process defines an iterative sequence $\{\rho_j \text{ } | \text{ } j \in \mathbb{N}, \text{ } j \in [1,n] \}$ such that this sequence converges to the self-consistent solution, $\rho_n = \rho^*$. The optimal approach to generating this sequence as robustly and efficiently as possible for an arbitrary input to KS DFT is unknown. Contemporary strategies to achieve convergence do so from two perspectives. First, by utilising various methodologies from the theory of non-linear systems, where no explicit reference is made to the underlying mathematical structure of KS DFT. Second, by analysing the linear response of KS DFT to \textit{precondition} these methodologies. The purpose of this thesis is to provide a comprehensive review of the difficulties involved in attaining a self-consistent solution to KS DFT. Following this, various extensions to the current methods utilised in many KS DFT codes for achieving self-consistency  will be considered and tested in the \textsc{castep} software \citep{CASTEP}.

\section{Thesis Structure}

The majority of time spent on the work presented in this thesis was in understanding the mathematical and conceptual foundations of both the numerical analysis and linear response theory involved in obtaining a self-consistent solution to KS DFT. As such, the thesis is constructed to reflect this, where an outline of the contents is as follows.

Chapter \ref{sec_2} is a pedagogical introduction to the fundamental concepts required to follow the theory used in the remainder of the thesis. First, $\S$\ref{elec_struc_prob} will define the electronic structure problem, i.e$.$ the exponential increase in complexity inherent within the standard, first-quantisation formulation of quantum mechanics. Following this, DFT within the KS framework will be defined in detail; including, but not limited to, the Hohenberg-Kohn theorems, the KS system of `non-interacting'  particles, and the form of the KS Hamiltonian. A particular emphasis will be placed on the fundamental structure of the resultant equations as defined by the KS Hamiltonian. This will lead to a rigorous definition of self-consistency in the present context, thus formally characterising the overarching goals of the work presented in this thesis. A note on the numerical implementation of KS DFT in software will then be given in $\S$\ref{impsec}. In particular, the algorithms and approach employed by \textsc{castep} will be briefly reviewed in such a way that demonstrates the scope for improved methodology in achieving self-consistency. Then, $\S$\ref{DMsec} will outline the numerical analysis involved in solving the non-linear KS system from the point of view of \textit{density mixing}. Furthermore, the strategies used to precondition density mixing schemes utilised in state-of-the-art electronic structure codes will be presented. This involves an investigation into the \textit{linear response} of KS DFT, and how this linear response can be used to stabilise and accelerate the SCF iterations.

Chapter \ref{metho} will begin by describing the technical considerations on the implementation of density mixing within \textsc{castep}. Then, the conceptual and mathematical foundations behind three distinct attempts at improving density mixing will be detailed. First, a density mixing scheme based on the work Marks $\&$ Luke in Ref$.$ \citep{ML} will be derived. Crucially, this derivation deviates somewhat from the original work of Marks $\&$ Luke due to inherent differences between \textsc{castep} and the native KS DFT software of Marks $\&$ Luke, \textsc{wien2k} \citep{wein2k}. Second, the Periodic Pulay scheme presented by Banarjee \textit{et al}. in Ref$.$ \citep{PP} will be described. Finally, the linear response theory presented in $\S$\ref{precond} will be utilised by introducing a computationally efficient framework whereby one can propose an \textit{inhomogeneous} model of the dielectric response of a general KS system. This framework will take an inhomogeneous and \textit{local} model of the KS \textit{susceptibility} as input, and subsequently calculate the dielectric response in such a way to precondition the SCF cycles. Two preliminary, physically-motivated model forms of the KS susceptibility will then be presented in order showcase this framework.

Chapter \ref{results} will present the outcome of testing on the above-mentioned three proposed improvements to density mixing. First, the chapter will begin by defining what an exactly an `improvement' constitutes in the present context, and moreover the degree of improvement that can be realistically expected will be discussed. Following this, a test suite of representative KS DFT input systems will be presented and motivated. This test suite is designed such that definitive conclusions can be drawn regarding the effectiveness of a given proposed improvement. Furthermore, individual systems from the test suite will be isolated in order to study the convergence patterns of the presented methods in more detail. This individual analysis will highlight the key advantages and drawbacks of each method, and provide further context to the conclusions drawn from the full test suite analysis.

Finally, Chapter \ref{conc} will briefly summarise the pertinent conclusions of Chapter \ref{results}, and directions for future developments will be considered.


\chapter{Theory} 

\label{sec_2} 

\lhead{2. \emph{Theory}} 


The problem of achieving self-consistency in KS DFT is a multifaceted one, which spans subdisciplines such as numerical analysis of linear and non-linear systems, linear response theory, and electronic structure theory (not just limited to KS systems). This chapter will detail why these are all necessary ingredients in solving KS DFT. To do this, one must start with a statement of the overarching problem, which is calculating the electronic structure of a system using quantum theory. The necessity of a framework like KS DFT will then become apparent, and following this, the background of (KS) DFT will be reviewed. This review will end by isolating the specific subsection of KS DFT which the following work pertains to -- self-consistency and density mixing. Here, a clear and detailed statement of the problem to be solved will be provided. Finally, a discussion of the numerical methods employed within DFT software to achieve self-consistency, and how one can precondition these methods, will follow. 

\section{The Electronic Structure Problem}
\label{elec_struc_prob}

The principles of quantum mechanics, developed over the course of approximately two decades in the early twentieth century by some of history's greatest physicists, provides the physical theory underpinning the behaviour of fundamental particles. This includes systems containing the particles that make up everyday matter -- protons, neutrons, and electrons. Phenomena arising from the collective behaviour of these three constituents is dubbed condensed matter physics, and this emergent behaviour makes up much of the rich and varied natural world around us. Within quantum mechanics, the complete description of a condensed matter system is given by the quantum state $|\Psi\rangle$, obeying the dynamical equation of motion,
\begin{equation}
\hat{H} |\Psi\rangle = -i \partial_t |\Psi\rangle,
\end{equation} 
which applies universally, for an appropriately constructed Hamiltonian operator\footnote{This equation in fact \textit{does} apply relativistically, but the form of the Hamiltonian operator necessarily changes in accordance with relativity.}. Using eigenstates of the position operator as a basis leads to a Hamiltonian of the form
\begin{align}
\hat{H} = -\frac{1}{2m}\nabla^2 + v(\textbf{r}),\label{posbasisH}
\end{align}
now operating on a \textit{wavefunction} $\psi(\textbf{r})$, the position basis expansion of the quantum state $|\Psi\rangle$. Given the representation in Eq$.$ (\ref{posbasisH}), one can now begin to construct a Hamiltonian operator which describes a many-body interacting system of charged nuclei and electrons. The potential function will simply describe the Coulomb interactions between all constituents of the system. The Coulomb interaction decays proportional to inverse of distance, therefore a system of interacting nuclei (Greek indices) and electrons (Latin indices) with masses, charges and positions $\{ M_{\mu}, Z_{\mu}, \textbf{R}_{\mu} \}$ and $\{ m_e, e, \textbf{r}_i \}$ respectively is described by the Hamiltonian,
\begin{align}
\hat{H} = -\frac{1}{2} \sum_i \nabla_i^2 - \sum_{\mu} \frac{1}{2M_{\mu}}  \nabla_{\mu}^2 - \sum_{i,\mu} \frac{Z_{\mu}}{|\textbf{r}_i - \textbf{R}_{\mu}|} + \frac{1}{2} \sum_{\mu \neq \nu} \frac{Z_{\mu}Z_{\nu}}{|\textbf{R}_{\nu} - \textbf{R}_{\mu}|} + \frac{1}{2} \sum_{i \neq j} \frac{1}{|\textbf{r}_i - \textbf{r}_j|},\label{cmtham}
\end{align}
where now, and hereafter, atomic units are used, $m_e = e = \hbar = 4\pi \epsilon_0 = 1$. This expression includes (from left to right): the kinetic energy operators of the electrons and nuclei, the electron-nuclear, nuclear-nuclear, and electron-electron Coulomb interactions.

\begin{figure*}[htbp]
\centering
\includegraphics[width=5in]{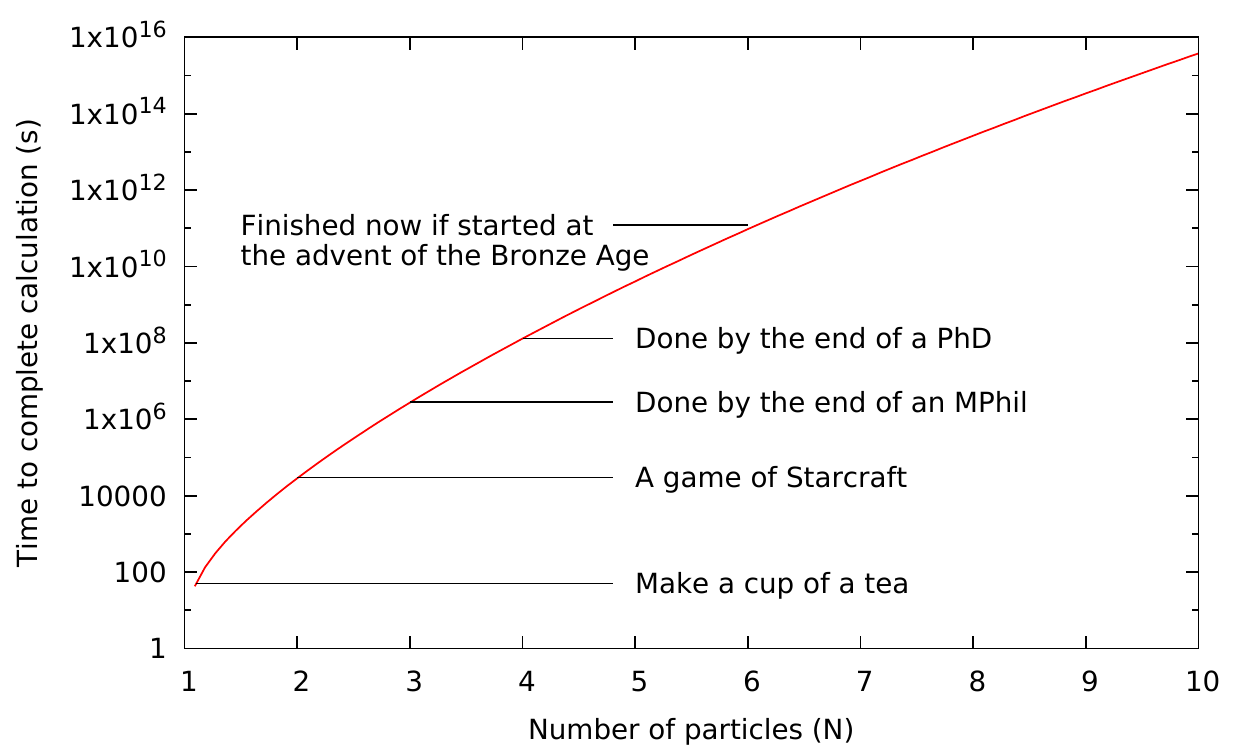}
\caption{Scaling of the many-body problem: the y-axis represents the combinatorial scaling $t \sim \mathcal{O}(^bC_N)$ with a prefactor for the one electron case of about 60 seconds.}
\label{fig:scaling}
\end{figure*}

 The first simplification in finding a solution to the Schr\"odinger equation is to notice that it is linear in time, meaning the solution is separable in the following way,
\begin{gather}
\Psi(\{\textbf{r}_i \}, \{ \textbf{R}_{\mu} \}; t) = \tilde{\Psi}(\{\textbf{r}_i \}, \{ \textbf{R}_{\mu} \}) \Theta(t),\\
 \Theta(t) = e^{-iEt}.
\end{gather}  
This allows time to be included explicitly, and one can focus on finding a \textit{stationary state} solution to the time-independent Schr\"odinger equation,
\begin{align}
\hat{H} |\Psi\rangle = E |\Psi\rangle.
\end{align}
One is left now with a quantum state, $|\Psi\rangle$, describing a composite system as an eigenstate of the Hamiltonian operator, living in a Hilbert space, $\mathcal{H}$, whose dimension is multiplicative with respect to the number of constituents. This is an important fact, and a quick example can show why the non-separable nature of the Hamiltonian via the Columb interaction vastly increases the complexity of the problem. Imagine two single particle states
\begin{align}
|\phi\rangle, |\psi\rangle \in \mathcal{H}
\end{align}
representing identical particles. Each can be expressed in an infinite (complete) basis $\{ |n\rangle \text{ } | \text{ }  n \in [1,\infty] \} $ that spans $\mathcal{H}$. Computationally, one can truncate the basis,  $\{ |n\rangle \text{ } | \text{ }  n \in [1,b] \}$, in such a way that it monotonically approaches the true Hilbert space with increasing $b$ (for example, using planewaves). Here, $b$ is the total number of basis states used in the finite representation, and thus parametrises the accuracy of the representation. In an interacting system of two identical particles,
\begin{align}
|\Psi\rangle \in \mathcal{H}^{\text{composite}} = \mathcal{H} \otimes \mathcal{H},
\end{align}
the dimension of the truncated composite Hilbert space will be $b^2$. This reflects the fact that all possible combinations of the individual single particle bases are now required to span the composite space. Hence, if a computist finds $b=10^2$ basis states provides a reasonable accuracy for resolving the individual single particle states, the solution of their composite system to the same accuracy will have an upper bound requirement of $10^4$ basis states. In reality, indistinguishability (i.e$.$ the same basis is used for each constituent) and the exclusion principle mean the complexity increases combinatorially in number of constituents as $^bC_N$. If the above $\{ N=1, b=10^2 \}$ calculation took, for example, 60 seconds, one might expect a $\{ N=20, b = \Perms{100}{20} \}$ calculation would take something of the order of $10^{22}$ seconds, which is a number perhaps best expressed in units of universe ages (see Fig$.$ (\ref{fig:scaling})). 

Clearly, one needs to find a way to suppress this exponential increase in complexity. The source of this complexity is the interaction terms in the Hamiltonian, which cannot be separated in an additive fashion with respect to the constituent variables. If they could, the exact solution would take the form of a product state,
\begin{align}
\Psi(\{ \textbf{r}_i \}) = \prod\limits_{i=1}^N \phi_i(\textbf{r}_i),
\end{align}
where each $\phi_i(\textbf{r}_i)$  solves its own separate Hamiltonian, requiring a basis resolution $b$. The total composite solution to a given single particle accuracy now only requires $Nb$ basis functions -- the effective dimension of the composite Hilbert space has become additive with respect to particle number. This defines a system of so-called `non-interacting' particles, which is a core concept behind KS DFT, and many other electronic structure methods.

The first simplifying approximation usually made to Eq$.$ (\ref{cmtham}) in an attempt to reduce the complexity is called the \textit{Born-Oppenheimer} approximation. This approximation principally functions by noticing that $1/M_{\mu}$ in the nuclear kinetic energy is small relative to $1/m_{e} = 1$ in the electronic kinetic energy. This leads to a weak coupling between the nuclear and electronic states. To elaborate, under certain conditions, the electronic state can be assumed to evolve \textit{adiabatically} to sufficiently small changes in the nuclear coordinates\footnote{Moreover, the nuclear coordinates can be, and often are, taken as the coordinates of \textit{classical} point charges. This is because the nuclear state is extremely localised, allowing the nuclei to be treated classically by placing a point charge at the centre of the localisation.}. Here, `sufficiently small changes' refers to the aforementioned observation that the kinetic energy of the nuclei is far lower than that of the electrons, where both the nuclei and electrons are subject to a Coulomb force of similar magnitude. Hence, changes in the nuclear state will be small relative to the evolution of the electronic state. That is, the electronic state will equilibrate to the electronic ground state for a given nuclear state before the nuclear state can mathematically couple to this equilibration. This defines adiabaticity, i.e$.$ to a good approximation the complete state of the system is determined as a weighted combination of \textit{continuously connected} electronic ground states over all sets of nuclear coordinates -- the \textit{potential energy surface}. As Ref$.$ \citep{martin_electruc} details, the Born-Oppenheimer approximation breaks down when these potential energy surfaces are not sufficiently separate. That is, the potential energy surface corresponding to the electronic ground state should be sufficiently separate to the potential energy surface of the first excited state, which should be sufficiently separate to the potential energy surface of the second excited state, and so on. This is often the case, and leads to a complete decoupling of the electronic and nuclear states as their contributions to the Hamiltonian under this approximation are entirely additive with respect to each other. Thus, the electron-nuclear interaction term is replaced by an electron moving in a background potential parametrised by the nuclear coordinates. The full wavefunction now takes the form,
\begin{align}
\Psi(\{ \textbf{r}_i \},\{ \textbf{R}_{\mu} \}) = \Psi^{\text{elec}}(\{ \textbf{r}_i \}) \Psi^{\text{nuc}}(\{ \textbf{R}_{\mu} \}),
\end{align}
where the electronic wavefunction now solves its own electronic Hamiltonian parametrised by the set of nuclear coordinates $\{ \textbf{R}_{\mu} \}$,
\begin{align}
\hat{H}^{\text{elec}} = -\frac{1}{2} \sum_i \nabla_i^2 + \frac{1}{2} \sum_{i \neq j} \frac{1}{|\textbf{r}_i - \textbf{r}_j|} + \sum_i v_{\text{ext}}(\textbf{r}_i),\label{elecham}
\end{align}
$v_{\text{ext}}$ being the parametrised term, the external potential. Solving the Schr\"odinger equation using this Hamiltonian defines what it means to determine the electronic structure of a system. In this sense, the Born-Oppenheimer approximation is responsible for the name of the subdiscipline electronic structure.

A `system' is now entirely specified by a given external potential and electron number. However, as it stands, one interaction term remains -- the electron-electron Coulomb interaction -- and dealing with this term has been the source of much work within the field of many-body physics. Some of the most popular methods to do this utilise the concept of a \textit{mean-field} in which independent (or non-interacting) particles respond. The mean-field potential is constructed such that it attempts to mimic the behaviour of the interaction term; KS DFT and Hartree-Fock (HF) theory are mean-field theories. This regime has also seen success as a post-DFT method for treating electron correlation with a technique called dynamical mean-field theory (DMFT) \citep{dmft}. Mean-field theories have historically been most successful in describing ground state structural properties of matter, e.g. cell parameters \citep{phil_dft}. For excited state (optical) properties however, many-body perturbation theory has been shown to provide a better treatment of the relevant physics \citep{martin_intelec}. For example, the GW approximation using Hedin's equations \citep{hedin}, and the Bethe-Salpeter equation (BSE) \citep{BSE}. To conclude, there are many methods available that attempt to treat the true interacting nature of a system within a computationally tractable framework, each finding success in its own domain of applicability \citep{configI,qmc,mppt}. However, due to its relative accuracy and transferability for the computational cost required to solve it, KS DFT has emerged as the most widely used method in electronic structure theory. It also provides the starting point for many of the post-DFT perturbative methods quoted, such as DFT+GW and DFT+GW+BSE \citep{martin_intelec}.



\section{Density Functional Theory}
\label{DFT}

The core concepts behind DFT are quite separate from the approximations one makes in order to implement it. In isolation from these approximations, DFT is an exact reformulation of quantum mechanics based on two astonishing theorems. These theorems state that, in principle, no information about a quantum system in the \textit{lowest energy state} of a Hamiltonian is lost by operating on the state such that one retrieves the electron density
\begin{align}
\rho(\textbf{r}) = N \int \prod_{i=2}^N d\textbf{r}_i |\Psi(\textbf{r}, \textbf{r}_2, \dots , \textbf{r}_N) |^2. \label{elecdens}
\end{align}
This quantity is a measure of the probability of \textit{any} electron being found at a position \textbf{r}, where here, and throughout the remainder of the thesis, spin will be ignored (although a note on spin in relation to the following work will be given in \S\ref{furtherwork}).
These theorems are called the Hohenberg-Kohn (HK) theorems \citep{HKthe}, and they result in all observable properties of a ground state system being determined by the three degrees of freedom in the real-valued electron density rather than the $3N$ degrees of freedom in the complex-valued wavefunction.

\textbf{HK Theorem I.} \textit{The ground state electron density for a system of interacting electrons is uniquely determined by the external potential.}

\textbf{HK Theorem II.} \textit{As a result of HK Theorem I, a universal (non-system-dependent) functional of the electron density can be defined such that this functional is minimised for the exact ground state electron density.}

These proofs are formulated under the requirement that the wavefunctions be non-degenerate, and the electron density is so-called $v$-representable. This is a problem, as it is not known if an arbitrary function $f(\textbf{r})$ can be represented as a ground state solution of the Schr\"odinger equation with an external potential $v_{\text{ext}}$. Hence, it may be possible that an (appropriately defined) energy functional can be minimised by a spurious non-$v$-representable electron density. Fortunately, the $v$-representability problem, and the non-degeneracy requirement, were solved by the alternate proof of Levy and Lieb \citep{levy}. Instead, a functional is defined such that a constrained search over all `$N$-representable' electron densities yields a unique minimum (for the ground state density). Unlike $v$-representibility, $N$-representibility is not a problem as it has been proven that an arbitrary (differentiable) function $f(\textbf{r})$ can be represented by some wavefunction using Eq$.$ (\ref{elecdens}) \citep{nrep} (so spurious minima in the energy functional are avoided). Neither proof will be presented here, but a vast literature discussing both is available \citep{birdseye}. 

As it stands, the HK theorems are not immediately beneficial in solving the electronic structure problem, as the form of the density functional is unknown. Thus, one still requires calculation of many-body body wavefunction in order to construct the correct ground state density. The Nobel prize winning framework able to utilise the concepts of DFT in practice is a theory proposed by Kohn and Sham in 1965 \citep{KSeq}. This theory remains formally exact, but shifts all unknowns into a single `exchange-correlation' term, requiring approximation. Over a wide range of systems this term has been demonstrated to contribute little to the overall energy of the system. Moreover, simple approximations of exchange and correlation are able to provide excellent agreement with formally exact methods, leading to KS DFT becoming one of the most popular techniques to determine electronic structure.

\subsection{Density Functional Theory in the Kohn-Sham Framework}

In order to develop a practical approach to DFT, one needs to define a total energy functional $E[\rho(\textbf{r})]$ such that the minimum of this functional provides the exact (or a good approximation to the exact) interacting ground state energy. The ansatz of Kohn and Sham was formulated by envisioning a non-interacting (and therefore computationally soluble) system of electrons in an effective potential $v_{\text{eff}}$. The question now is whether such a  $v_{\text{eff}}$ exists that can uniquely reproduce the exact interacting ground state energy and density. Kohn and Sham showed this to be the case, and the mapping between the non-interacting and interacting systems defines the famed auxiliary KS system of electrons.

An outline of the KS framework is as follows. Under the assumption that electrons are non-interacting, the many-body wavefunction takes the form of a product state: $\Psi(\{ \textbf{r}_i \}) = \prod\limits_{i=1}^N \phi_i(\textbf{r}_i)$ (or alternatively a Slater determinant). The wavefunctions $\{ \phi_i \}$ are often referred to as `single particle orbitals'. As a result of the exclusion principle allowing only one fermion per single particle eigenstate of the Hamiltonian, there exist a total of $N$ single particle orbitals (foregoing for now partial occupancies). The KS system is thus defined as
\begin{gather}
\Big( -\frac{1}{2} \nabla^2 + v_{\text{eff}}(\textbf{r}) \Big) \phi_i(\textbf{r}) = \epsilon_i \phi_i(\textbf{r}), \label{KS} \\
\rho^{\textsc{ks}}(\textbf{r}) = \sum_{i \in \text{occupied}} | \phi_i(\textbf{r}) |^2, \\
E^{\textsc{ks}} \sim \sum_{i \in \text{occupied}} \epsilon_i, \label{KSe}
\end{gather}
where all quantities are assumed to be ground state, and the first term in Eq$.$ (\ref{KS}) represents the kinetic energy of a non-interacting particle. This defines the most general non-interacting form of the time-independent Schr\"odinger equation. One now seeks to prove that a  $v_{\text{eff}}$  exists such that
\begin{gather}
\rho^{\textsc{ks}}(\textbf{r}) = \rho^{\text{int}}(\textbf{r}) = N \int \prod_{i=2}^N d\textbf{r}_i \text{ } |\Psi(\textbf{r}, \textbf{r}_2, \dots , \textbf{r}_N) |^2, \\
E^{\textsc{ks}} = E^{\text{int}} = \int  \prod_{i=1}^N d\textbf{r}_i  \text{ }  \Psi^*(\{ \textbf{r}_i \}) \hat{H}^{\text{int}} \Psi(\{ \textbf{r}_i \}).
\end{gather}
To do this, one defines the corresponding energy functional of the system in Eqs$.$ (\ref{KS}) - (\ref{KSe}) as 
\begin{align}
E^{\textsc{ks}}[\rho] = T_0[\rho] + \int d\textbf{r} \text{ } v_{\text{eff}}(\textbf{r}) \rho(\textbf{r}), \label{KS11}
\end{align}
where $T_0[\rho]$ is the kinetic energy functional of independent particles. The energy functional of the fully interacting system is
\begin{gather}
E^{\text{int}}[\rho] = F[\rho] +  \int d\textbf{r} \text{ } v_{\text{ext}}(\textbf{r}) \rho(\textbf{r}), \label{KS22} \\
F[\rho] = \langle \Psi_{\textsc{gs}} | T + v_{ee} |  \Psi_{\textsc{gs}} \rangle.
\end{gather}
Here, $T$ is the kinetic energy operating on the fully interacting system, and $v_{ee}$ is the Coulomb potential operator. To prove that such a $v_{\text{eff}}$ exists, one can check there exists a density which is the stationary point of both functionals, Eq$.$ (\ref{KS11}) and Eq$.$ (\ref{KS22}), simultaneously under some definition of $v_{\text{eff}}$, as will now be shown. Note that in determining the stationary points, a constrained variation is needed -- the ground state electron density is required to satisfy
\begin{align}
\int d\textbf{r} \text{ } \rho(\textbf{r}) = N.
\end{align}
Therefore, the variation is performed using Lagrange multipliers, which leads to
\begin{gather}
\frac{\delta E^{\textsc{ks}}[\rho]}{\delta \rho} = \frac{\delta T_0[\rho]}{ \delta \rho} + v_{\text{eff}} + \lambda_{\text{NI}} = 0,\label{eq122} \\
\frac{\delta E^{\text{int}}[\rho]}{\delta \rho} = \frac{\delta F[\rho]}{ \delta \rho} + v_{\text{ext}} + \lambda_{\text{I}} = 0,\label{eq222}
\end{gather} 
with $\lambda_{\text{I/NI}}$ being the interacting and non-interacting Lagrange multipliers. These can effectively be dropped here, as the behaviour of the variation can be defined up to a constant without loss of generality. Clearly, equating Eqs$.$ (\ref{eq122}) and (\ref{eq222}) yields an expression for the effective potential,
\begin{align}
v_{\text{eff}} = \frac{\delta F[\rho]}{ \delta \rho} +  v_{\text{ext}} -  \frac{\delta T_0[\rho]}{ \delta \rho}.\label{effKS}
\end{align}
The ansatz of Kohn and Sham has been proven: it is possible to define an auxiliary non-interacting system which exactly reproduces the behaviour of the interacting system\footnote{The problem of $v$-representability has resurfaced, however. It is not possible to say that any electron density resulting from a given external potential in the interacting system will be expressible as a density from some external potential in the non-interacting system (and vice versa). This is a problem yet to be solved, but is mostly benign \citep{vrepprob}.}. The universal functional $F[\rho]$ remains unknown, and to utilise this mapping in practice a computationally tractable approximate form is required. Kohn and Sham proposed an approximation analogous to Hartree theory. The electron-electron interaction is replaced with a mean-field `Hartree potential', and kinetic energy operations are performed on the non-interacting orbitals, rather than the many-body wavefunction\footnote{Pre-KS, the kinetic energy of the interacting system was often approximated using Thomas-Fermi theory, defining the first practical density functional \citep{tf1}.},
\begin{align}
F[\rho] = -  \frac{1}{2} \int d\textbf{r} \text{ }  \phi^*_i(\textbf{r}) \nabla^2_i \phi_i(\textbf{r}) + \frac{1}{2} \int d\textbf{r} d\textbf{r}' \text{ }  \frac{\rho(\textbf{r}) \rho(\textbf{r}')}{|\textbf{r} - \textbf{r}'|} + E_{\text{xc}}[\rho].\label{11111}
\end{align}
The final term here -- the exchange-correlation energy -- is defined as the difference between the ground state energy of the Hartree system and that of the exact interacting system. It is named as such because, so far, no explicit reference has been made to the fermionic nature of the system in relation to the statistics it obeys. That is, the many-body wavefunction changes sign upon exchange of any two electrons, and is therefore an \textit{antisymmetric} function with respect to its arguments. This has the important consequence of lowering the total energy by tending to separate electrons of like spin, and thereby lowering the total energy by an amount $E_{\text{x}}$, the \textit{exchange energy}. The exchange-correlation energy is simply the sum of the exchange energy and the `correlation energy'. Therefore, correlation (an accurate, computationally tractable treatment of which is the holy grail of electronic structure theory) is defined as the difference in energy between Hartree-Fock theory (Hartree theory with an exact treatment of exchange) and the fully interacting theory. Correlation then necessarily includes a contribution from the fact that the independent particle kinetic energy, obtained from the kinetic energy operator on the independent particle states, is not the same as the kinetic operator on the interacting wavefunction, which one no longer has access to. This difference is a significant contributor to correlation, as the Hartree potential, by virtue of being a mean-field, provides an exact description of the Coulomb potential felt by a test charge due to its surroundings\footnote{Although this is true for a \textit{test charge}, it is not true for constituents of the KS system. That is, the Hartree mean-field is constructed from the \textit{total electron density}, meaning KS constituents interact with themselves through this Hartree term.}.

Substituting Eq$.$ (\ref{11111}) into Eq$.$ (\ref{effKS}) finally provides a systematic method of approaching the electronic structure problem within a computationally feasible framework,
\begin{align}
v_{\text{eff}}(\textbf{r}) &= \int d\textbf{r} \text{ } \frac{\rho(\textbf{r})}{|\textbf{r} - \textbf{r}'|} + v_{\text{ext}}(\textbf{r}) + \frac{\delta E_{\text{xc}}[\rho(\textbf{r})]}{\delta \rho(\textbf{r})}, \nonumber \\
 &=  v_{\text{h}} +  v_{\text{ext}} +  v_{\text{xc}},
\end{align}
which is used to construct the KS Hamiltonian, solving the KS equations Eq$.$ (\ref{KS}). The exchange-correlation potential is the only term left requiring approximation, acting to parametrise the degree of ignorance about the correlated nature of electrons. It is worth noting here that there is a certain arbitrary nature with which the exchange-correlation potential was constructed. That is, as a functional \textit{explicitly} defined to satisfy a (non-physical\footnote{Non-physical, here, is intended to refer to the fact that a \textit{potential energy} is being constructed in order to account for a fundamental misrepresentation of the \textit{kinetic energy}.}) mathematical mapping between interacting and non-interacting systems. As such, one loses a degree of physical meaning within the non-interacting KS system. A direct consequence of this is that all but one of the individual energy eigenvalues of the KS system carry no formal physical meaning (other than their sum being the total energy, minus double counting). The eigenvalues have meaning within the framework itself (see, Janak's theorem \citep{janak}), but the only eigenvalue with true physical meaning (in finite systems) is the highest energy occupied one, equal to the ionisation energy \citep{highoc}. 


Despite these problems, it is an astonishing fact that for many materials the exchange-correlation term accounts for a quite a low
 percentage of the overall ground state energy. Consequently, ground state properties of matter, such as stable phases and cell parameters, are predicted by KS DFT to within 1$\%$ of experiment using only a rudimentary treatment of exchange and correlation \citep{111122}. For example, by using the exact exchange-correlation energy functional of the homogeneous electron gas -- the \textit{local density approximation} (LDA) \citep{lda1}. Increasing the accuracy of KS DFT therefore amounts to finding, in some sense, a `better' exchange-correlation functional to model the system at hand. It could be argued therefore that KS DFT is not an \textit{ab initio} theory, because it does not approach the exact solution in some well defined limit (as, for example, a perturbative method would). In this context, the problem of increasing accuracy is explained well in Ref$.$ \citep{jladder}, which conceptualises a `Jacob's ladder' of increasing accuracy, starting from no exchange-correlation to exact exchange-correlation --  passing through the LDA, gradient-based extensions of the LDA, exact (non-local) exchange, and more.

\subsection{Solving the Kohn-Sham Equations}
\label{selfconssec1}

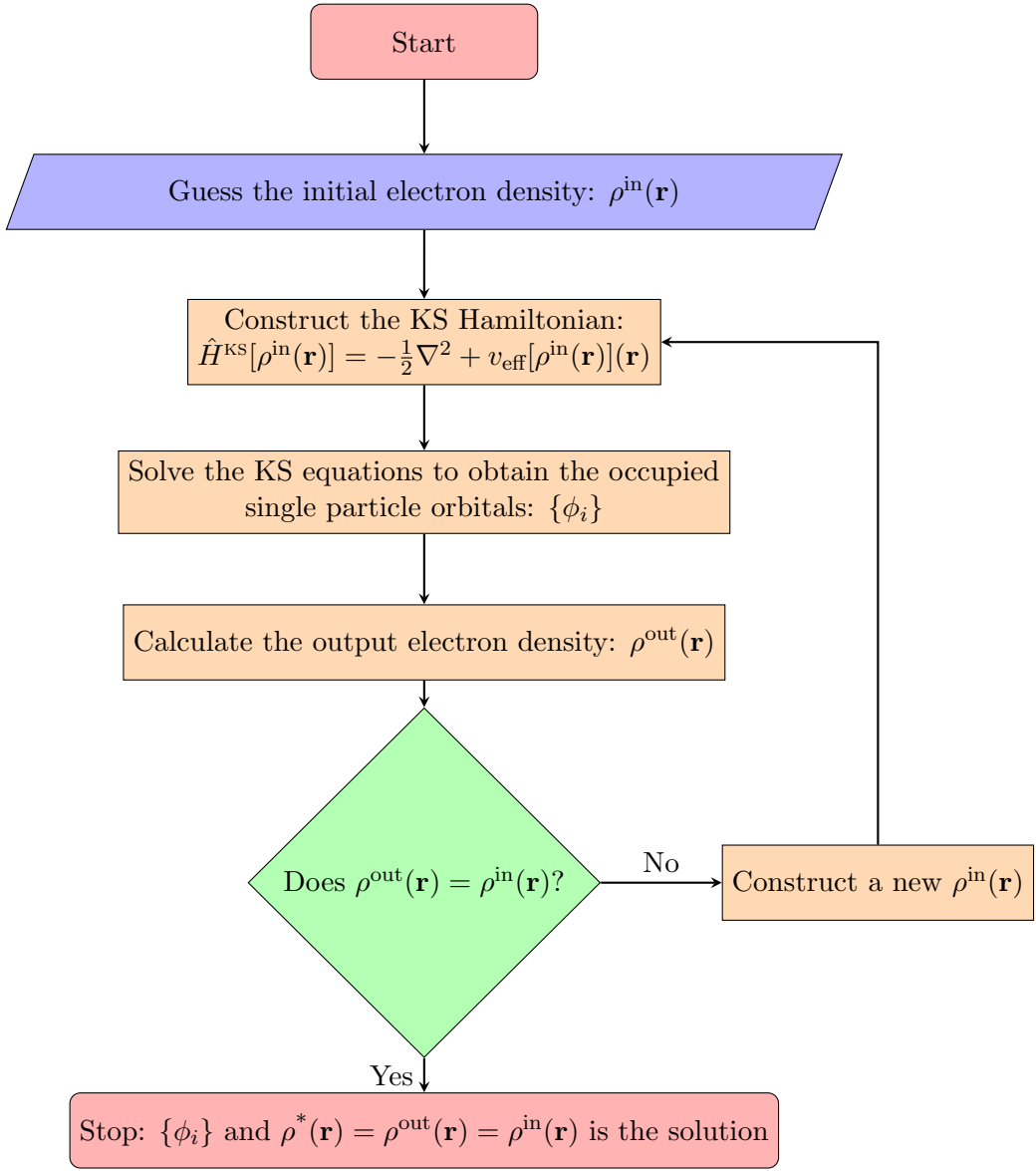
\begin{figure*}[htbp]
\centering

\begin{tikzpicture}[node distance=2cm]
\tikzstyle{startstop} = [rectangle, rounded corners, minimum width=3cm, minimum height=1cm,text centered, draw=black, fill=red!30]
\tikzstyle{io} = [trapezium, trapezium left angle=70, trapezium right angle=110, minimum width=3cm, minimum height=1cm, text centered, draw=black, fill=blue!30]
\tikzstyle{process} = [rectangle, minimum width=3cm, minimum height=1cm, text centered, draw=black, fill=orange!30]
\tikzstyle{decision} = [diamond, minimum width=3cm, minimum height=1cm, text centered, draw=black, fill=green!30]
\tikzstyle{arrow} = [thick,->,>=stealth]

\node (start) [startstop] {Start};

\node (in1) [io, below of=start] {Guess the initial electron density: $\rho^{\text{in}}(\textbf{r}$)};

\node (pro1) [process, below of=in1, align=center] {Construct the KS Hamiltonian:\\$\hat{H}^{\textsc{ks}}[\rho^{\text{in}}(\textbf{r})] = -\frac{1}{2} \nabla^2 + v_{\text{eff}}[\rho^{\text{in}}(\textbf{r})](\textbf{r})$};

\node (pro2) [process, below of=pro1, align=center] {Solve the KS equations to obtain the occupied\\single particle orbitals: $\{ \phi_i \}$};

\node (pro3) [process, below of=pro2] {Calculate the output electron density: $\rho^{\text{out}}(\textbf{r})$};

\node (dec1) [decision, below of=pro3,yshift=-1.2cm] {Does $\rho^{\text{out}}(\textbf{r})=\rho^{\text{in}}(\textbf{r})$?};

\node (pro4) [process, right of=dec1, xshift=4cm] {Construct a new $\rho^{\text{in}}(\textbf{r})$};

\node (stop) [startstop, below of=dec1, yshift=-1.3cm] {Stop: $\{ \phi_i \}$ and $\rho^{\text{*}}(\textbf{r}) = \rho^{\text{out}}(\textbf{r}) = \rho^{\text{in}}(\textbf{r})$ is the solution };

\draw [arrow] (start) -- (in1);
\draw [arrow] (in1) -- (pro1);
\draw [arrow] (pro1) -- (pro2);
\draw [arrow] (pro2) -- (pro3);
\draw [arrow] (pro3) -- (dec1);
\draw [arrow] (dec1) -- node[anchor=east] {Yes} (stop);
\draw [arrow] (dec1) -- node[anchor=south] {No} (pro4);
\draw [arrow] (pro4) |- (pro1);

\end{tikzpicture}

\caption{Flow chart depicting a self-consistent solution to the KS equations.}
\label{fig:selfcons}
\end{figure*}

The KS equations are restated here as,
\begin{gather}
\Big( -\frac{1}{2} \nabla^2 + v_{\text{ext}}(\textbf{r}) + v_{\text{h}}[\rho^{\text{in}}(\textbf{r})](\textbf{r}) + v_{\text{xc}}[\rho^{\text{in}}(\textbf{r})](\textbf{r}) \Big) \phi_i(\textbf{r}) = \epsilon_i \phi_i(\textbf{r}), \label{kohnsham2}\\
\rho^{\text{out}}(\textbf{r}) = \sum_{i \in \text{occupied}} | \phi_i(\textbf{r}) |^2,\label{elecdens2}
\end{gather}
assuming a local exchange-correlation potential. In constructing the non-interacting KS system of electrons, the fundamental structure of the quantum mechanical equations has changed due to the introduction of the Hartree and exchange-correlation potentials. That is, the input to the KS equations now depends explicitly on the output non-linearly via the electron density. Therefore, in order to even begin solving the above system of equations, one must estimate an initial electron density to use as input, $\rho^{\text{in}}$. Once an initial input density has been specified, the (linear) KS equations can be solved Eq$.$ (\ref{kohnsham2}) -- this involves diagonalisation of the KS Hamiltonian in some basis to find the occupied single particle orbitals. From these single particle orbitals, the \textit{output} electron density, $\rho^{\text{out}}$, is constructed using Eq$.$ (\ref{elecdens2}). The input and output densities are (in general) only equal if one has found the ground state electron density that solves the KS system. The computational journey one takes starting from an initial guess electron density, arriving at an electron density that solves the KS system is precisely what it means to achieve \textit{self-consistency}. This process is summarised in Fig$.$ (\ref{fig:selfcons}). 

The above set of equations defines a \textit{non-linear system}. That is, a map that takes an input $\rho^{\text{in}}$ and generates an output $\rho^{\text{out}}$ that is non-linearly related to the input,
\begin{align}
F[\rho^{\text{in}}] = \rho^{\text{out}},
\end{align}
where $F$ hereafter defines the \textit{KS map}. Solving this system amounts to finding a \textit{fixed point} of the non-linear KS map, $\rho^* =  \rho^{\text{in}} =  \rho^{\text{out}}$, which is typically done using an iterative procedure. This iterative procedure acts to define a sequence, $\{ \rho^{\text{in}}_1, \dots , \rho^{\text{in}}_n \}$, such that $\rho^* =  \rho^{\text{in}}_n$ within some defined tolerance. Ideally, this sequence is generated as robustly and efficiently as possible, which is to say, the sequence will eventually converge, and it does so such that $n$ is minimised. In literature, this is often referred to as the \textit{self-consistent field} (SCF) process, where here the self-consistent field is the density -- a 3-dimensional real scalar field. It should be noted that self-consistency can be equivalently treated by defining a converging sequence of potentials, or wavefunctions; density has simply been chosen here. The reason being that, within \textsc{castep}, the default method for achieving self-consistency explicitly operates by updating the electron density. This is not required, and indeed a reliable fallback method within \textsc{castep} updates the single particle orbitals directly in such a way that minimisation in error is guaranteed between iterations. This method is called ensemble density functional theory (EDFT) \citep{edft}, and is extremely robust due to its variational nature, but not nearly as efficient as rivalling techniques. The default method for achieving self-consistency within \textsc{castep} is called \textit{density mixing}, and combines the input and output densities at each iteration to estimate a new input density. This will be detailed in $\S$\ref{DMsec}. However, before defining and exploring density mixing, a brief discussion of how KS DFT is implemented in software will be given. This will serve to motivate the need for and scope of improved density mixing methods, and also introduce some key concepts of \textit{in silico} DFT that will be utilised in the following work, such as the planewave basis set.


\subsection{Implementation}
\label{impsec}

In order to implement a solution to the KS equations, one must first characterise the type of system one wishes to solve -- here, that is systems in the \textit{solid state}. The concept of a crystal is therefore introduced as an ordered lattice of atoms, invariant under translation by \textit{primitive lattice vectors}. The set of all integer multiples of the primitive lattice vectors form the infinite, periodic Bravais lattice,
\begin{align}
\Lambda = \{ n_1 \textbf{a}_1 +  n_2 \textbf{a}_2 + n_3 \textbf{a}_3 \text{ } | \text{ } n_i \in \mathbb{Z} \}. \label{brav}
\end{align}
One can therefore construct the full Bravais lattice out of two constituents: a \textit{primitive unit cell}, and a translation of this unit cell by the set of all integer multiples of the primitive lattice vectors (the basis). Here, the elements of the Bravais lattice are atomic positions, and the lattice itself is an approximation to the bulk crystal structure -- a method of reducing an $\sim \mathcal{O}(10^{23})$ size system to that of just a primitive cell and three translation vectors. In order to fully exploit this symmetry, one can invoke \textit{Bloch's theorem}, which vastly simplifies calculations of the wavefunction in periodic systems.

\textbf{Bloch's Theorem}. \textit{If a Hamiltonian is invariant under a translation by \textbf{R}, then the resulting wavefunction will take that of a Bloch form,}
\begin{align}
\psi_{\textbf{k}}(\textbf{r}) = e^{i \textbf{k}.\textbf{r}} u_{\textbf{k}}(\textbf{r}),
\end{align}
\textit{where}  $u_{\textbf{k}}(\textbf{r})$ \textit{is a function with the periodicity of the Hamiltonian,}  $u_{\textbf{k}}(\textbf{r}+\textbf{R}) =  u_{\textbf{k}}(\textbf{r})$.

The quantity $\textbf{k}$ labels the \textit{crystal momentum} -- a conserved quantity under translation by a `reciprocal lattice vector'. Like the Bravais lattice, the reciprocal lattice is defined by the set of all integer translations by three reciprocal lattice vectors on a reciprocal primitive cell, and these two lattices are related via a Fourier transform. Therefore, there exists a range of $\textbf{k}$ on the reciprocal lattice in which all information is contained -- a \textit{Brillouin zone}. The first Brillouin zone is a particular type of primitive unit cell on the reciprocal lattice, and is conventionally chosen to be the range of \textbf{k} for which the full wavefunction is calculated. 

The wavefunction of the infinity of electrons in the Bravais lattice can now be calculated as the wavefunction within the unit cell, multiplied by a phase labelled by $\textbf{k}$ spanning the first Brillouin zone. This infinity has been transferred from an infinity of electrons, to an infinite set of continuous values of \textbf{k} that need to be sampled within the first Brillouin zone. Fortunately, the energy (and therefore the wavefunctions) are smoothly varying with \textbf{k} \citep{smooth}, allowing for a rather course sampling of \textbf{k} within the first Brillouin zone to accurately represent the wavefunction of an infinite crystal. In the language of KS theory, the external potential is now specified by the Bravais lattice; the KS Hamiltonian is therefore invariant under translation (as the kinetic energy operator is), allowing Bloch's theorem to be utilised in practical, solid state calculations.

Given Bloch's theorem, the final ingredient required in order to computationally solve the KS equations is a method of representing the wavefunctions. This is done by choosing a complete set of basis functions, and implementations of KS DFT are typically distinguished based on this choice. As discussed, the wavefunction to be calculated is the \textit{Bloch function}, which possesses an inherent periodicity defined by the primitive lattice vectors. Hence, a natural basis to consider would be a complete set of functions with the same inherent periodicity as the unit cell. This can be done using \textit{planewaves}, which is the most popular basis set for electronic structure calculations, defined as
\begin{align}
u_{\textbf{k}}(\textbf{r}) = \sum_{\textbf{G}} C_{\textbf{k},\textbf{G}} e^{i \textbf{G}.{\textbf{r}}}.
\end{align}
The set of allowed planewaves is defined by the constraint $\textbf{G}.\textbf{R} = 2 \pi m$ for $m \in \mathbb{N}$, i.e$.$ the planewaves must have the same periodicity as the unit cell. The full wavefunction expanded in the planewave basis now takes the form
\begin{align}
\psi_{\textbf{k}}(\textbf{r}) = \sum_{\textbf{G}} C_{\textbf{k}+\textbf{G}} e^{i (\textbf{k}+\textbf{G}).{\textbf{r}}}
\end{align}
for $\textbf{k}$ in the first Brillouin zone. Other than inherent periodiciy and orthogonality, there are a few distinct advantages of a planewave basis. First, the space of vectors $\{ \textbf{G} \}$ defining the planewaves lies on the reciprocal lattice, thus unlocking the fast Fourier transform (FFT) as a powerful computational tool to switch between spaces when convenient. Moreover, there is no bias in the basis set toward a particular input system. Therefore, the accuracy of the representation converges monotonically with increasing number of planewaves ordered by increasing magnitude of the allowed $G$-vectors\footnote{This fact also makes including time-dependence particularly easy compared to system-tailored basis sets.}. But, in being unbiased, the size of the Hamiltonian can become prohibitively large in order to achieve a certain degree of accuracy\footnote{Moreover, it becomes difficult to distinguish how much charge is associated with particular nuclei in a bonded system.}. In an attempt to remedy this, one can make the observation that \textit{core electrons} -- electrons in orbitals that are extremely localised to the nuclei -- contribute little to the observable properties and behaviour of a system \citep{pseudop}. The contribution of core electrons is mostly contained in how they screen the nuclear potential from the \textit{valence electrons}. Therefore, one might think of constructing an effective potential that provides the exact behaviour for valence electrons, without explicitly considering core electrons as degrees of freedom in the system. This is called the \textit{pseudopotential approximation}. It is particularly useful when utilising a planewave basis set, as orbitals corresponding to core electrons become increasingly oscillatory, owing to the fact that the core states are constrained to be orthogonal to the valence states. Oscillatory behaviour in the wavefunction requires many planewaves to represent accurately, so the smoothing as a result of the pseudopotential approximation allows one to use a much reduced number of planewaves. There are many popular computer programs implementing the above theory: \textsc{castep} \citep{CASTEP}, \textsc{vasp} \citep{vasp}, \textsc{abinit} \citep{abinit}, \textsc{quantum espresso} \citep{qesp},  and many more.

The ensuing discussion will now restrict specifically to the methodology employed by \textsc{castep}. The KS equations expanded in terms of the planewave basis set take the form,
\begin{align}
\sum_{\textbf{G}'} \Bigg[ \frac{1}{2} |\textbf{k+G}|^2& \delta_{\textbf{GG}'} + v_{\text{ext}}(\textbf{G}-\textbf{G}') \nonumber \\
 & + v_{\text{h}}(\textbf{G}-\textbf{G}') + v_{\text{xc}}(\textbf{G}-\textbf{G}') \Bigg] C_{\textbf{k+G}',b} = \epsilon_{\textbf{k+G},b}  C_{\textbf{k+G},b},
\end{align}
where $b$ here labels the distinct single particle orbitals -- the \textit{band index}. The naive way to solve these equations would be to construct the full matrix, perform an exact diagonalisation, and use the $N$ lowest energy eigenvectors and eigenvalues to calculate the electron density and total energy. This is impractical, as exact diagonalsation here is an $\mathcal{O}(N^3_G)$ operation, where $N_G$ is the number of planewaves included in the truncated basis set. $N_G$ can become as large as $10^6$, therefore the matrix cannot be diagonalised, or even stored in RAM. Instead, an \textit{iterative diagonalisation} is implemented in order to compute only the lowest $N_b << N_G$ energy eigenvectors and eigenvalues, where $N_b$ is the number of bands calculated (typically of the order of the number of electrons). Without getting into details, this involves an algorithm (e.g$.$ the Davidson method \citep{davidson}) which acts to isolate and compute the $N_b$ lowest energy eigenvectors. However, unconstrained, this algorithm would find $N_b$ copies of the lowest energy eigenvector, so a constrained version is needed. That is, the algorithm is performed subject to the constraint that the wavefunctions remain orthogonal. Therefore, each iterative diagonalisation step requires an orthogonalisation step in the subspace of the $N_b$ lowest energy eigenvectors. This is the source of the oft-quoted `cubic-scaling' nature in KS DFT implementations. In reality, the scaling of KS DFT implementations is not quite that simple\footnote{And, in fact, not cubic at all until the system size becomes large. There is indeed an $\mathcal{O}(N_k N_b^3)$ scaling component to the orthogonalisation procedure (in inverting the band overlap matrix), but this does not become dominant over the scaling quoted until $N_b \sim 10^3$ \citep{sclark}.}. From Ref$.$ \citep{mattreport}, the dominant scaling is indeed due to the orthogonalisation, but instead scales as $\mathcal{O}(N_k N_G N_b^2)$; where $N_k$ is the number of \textbf{k} points one uses to sample the Brillouin zone\footnote{This scaling is dominant over the scaling of the application of the KS Hamiltonian on the trial vectors in the iterative diagonaliser. With a series of clever operations and FFTs, the scaling becomes $\mathcal{O}(N_k N_b N_G \mathrm{log} N_G)$. For example, one would apply the kinetic energy operator in Fourier space, and any local potential operators in real space, as they are both diagonal in their respective spaces and FFTs can move between them.}.

Iterative diagonalisation of the KS Hamiltonian is one of the two major time sinks in a typical KS DFT calculation. Referring back to the flowchart in Fig$.$ (\ref{fig:selfcons}), every time the input density fails to match the output density, the iterative diagonalisation is executed again for the updated Hamiltonian toward self-consistency. Hence, the wall clock time becomes multiplied by the number of SCF cycles\footnote{Assuming a negligible time spent doing density mixing, also excluding initialisation, finalisation, and other sundries.}. Accelerating KS DFT therefore amounts to either improving the strategies one uses to solve the KS equations, or reducing the amount that one needs to solve the KS equations in order to find a self-consistent solution. This work focuses on the latter.

\section{Density Mixing}
\label{DMsec}

The need for a self-consistent density (or equivalently, potential) was outlined in $\S$\ref{selfconssec1}. The following section, and the remainder of this thesis, restricts focus now to the methodology employed in order to find this self-consistent solution. This is done using density mixing, which is the default method of choice for most of (if not all) planewave electronic structure implementations. Density mixing can be separated into two distinct considerations: the numerical analysis one uses to choose/define an algorithm to reach convergence, and how one preconditions this algorithm. In the former, KS DFT is simply a black box. The latter consideration then acts to inject as much prior knowledge about KS DFT as possible into the black box to assist convergence. First, the methodology from a purely mathematical perspective will be outlined. 

\subsection{What is Density Mixing?}

Density mixing is defined as follows: one seeks a fixed point of the KS map,
\begin{align}
F[\rho^*] = \rho^*,
\end{align}
or, equivalently, a root of the \textit{residual},
\begin{align}
R[\rho^*] = F[\rho^*] - \rho^* = 0.
\end{align}
One could therefore devise an iterative method of reaching $\rho^*$ by combining the input and output density at the current iteration in such a way that the subsequent iterative density is closer to convergence. In fact, one can use the entire history of densities cycled through in the iterative procedure to construct the subsequent density. This leads to a \textit{density mixing scheme} being the name given to the function,
\begin{align}
\rho^{\text{in}}_{n+1} = f( \{ \rho^{\text{in}}_i, \rho^{\text{out}}_i \})  \text{ } \forall \text{ }  i \in [1,n].
\end{align}
The following work seeks to find the optimal form of $f$ specifically for the KS map $F$. A vast mathematical literature of numerical methods to solve non-linear systems exists \citep{anita1}, but, there is no such thing as a free lunch, and there is no one particular method in the literature which universally performs better than others across all applications. Therefore, $f$ will be tailored by certain facts about the KS system. For example, in KS DFT one is often privileged to reasonably accurate initial guess of the electron density (using a sum of appropriately placed atomic orbitals), leading naturally to algorithms that perform well with an accurate initial guess. Moreover, as discussed, the number of $G$-vectors, i.e$.$ the number of elements required to represent $\rho$, is typically quite large, meaning the algorithm must take particular care to avoid storing prohibitively large objects in RAM. The function $f$ should therefore define a robust and efficient scheme, executable with limited memory requirements. With this in mind, the difficulties in applying some of the `naive' approaches to the SCF process can be outlined.

\subsection{The Fixed-Point Update}

The simplest update to the density one could imagine is to treat the SCF process as a fixed point iteration, feeding the output density back in as the input density by computing $\rho^{\text{in}}_{n+1}=F[\rho^{\text{in}}_{n}]= \rho^{\text{out}}_{n}$. This turns out to be unsuitable for a variety of reasons. To elaborate, first note that the initial guess of density is not far from convergence. For analysis' sake, this allows the operation $F$ to be linearised about $\rho^*$. If one writes $\rho^{\text{in/out}}_n = \rho^* + \delta \rho^{\text{in/out}}_n$, the operation $F$ can be Taylor expanded about $\rho^*$ as such,
\begin{align}
\rho^* + \delta \rho^{\text{out}}_n =& F[\rho^* + \delta \rho^{\text{in}}_{n}] \nonumber \\
\approx& F[\rho^*] + \frac{\delta F[\rho_n^{\text{in}}]}{\delta \rho_n^{\text{in}}}\bigg\rvert_{\rho^*} \delta \rho^{\text{in}}_{n} \nonumber \\
=& \rho^* + \frac{\delta F[\rho_n^{\text{in}}]}{\delta \rho_n^{\text{in}}}\bigg\rvert_{\rho^*}  \delta \rho^{\text{in}}_{n}.
\end{align}
This implies that the change in current output density, and hence the change in the subsequent input density, is linearly related to the current input density via,
\begin{align} \label{linearresponse}
\delta \rho^{\text{in}}_{n+1} =  \delta \rho^{\text{out}}_n =\frac{\delta F[\rho_n^{\text{in}}]}{\delta \rho_n^{\text{in}}}\bigg\rvert_{\rho^*}  \delta \rho^{\text{in}}_{n}.
\end{align}
For compactness, define,
\begin{align}
M = \frac{\delta F[\rho_n^{\text{in}}]}{\delta \rho_n^{\text{in}}}\bigg\rvert_{\rho^*} = \frac{\delta \rho^{\text{out}}_n}{\delta \rho^{\text{in}}_n} \in \mathbb{C}^{N_G \times N_G}.
\end{align}
This matrix holds special meaning outside of fixed point iterations, as it is closely related to the KS \textit{dielectric}, and can be used to \textit{precondition} density mixing schemes (see $\S$\ref{precond}). For now, one can simply ask what the mathematical conditions on $M$ are, without reference to its physical meaning, to guarantee convergence within this framework. Convergence is defined as $\delta \rho^{\text{in}}_{n+1} \rightarrow 0$ as $n \rightarrow \infty$, or
\begin{gather}
\delta \rho^{\text{in}}_{n+1} = M   \delta \rho^{\text{in}}_{n} = (M)^n \delta \rho^{\text{in}}_{1} \rightarrow 0.
\end{gather}
It will be assumed here (although it can be shown \citep{MM}) that $M$ is of full rank, and can be diagonalised using $\{  \textbf{e}_i \}$ as an orthonormal eigenbasis -- $M_{ij} =  \delta_{ij} \lambda_i | \textbf{e}_i \rangle \langle \textbf{e}_j |$ with $\{ \lambda_i \}$ the spectrum of $M$\footnote{If $M^d$ is defined as the transformed diagonal version of $M$, then it shares some crucial properties with $M$. Namely, $M^n \rightarrow 0$ iff $(M^d)^n \rightarrow 0$, by nature of diagonalisation being a similarity transformation.}. Since $\langle \textbf{e}_i | \textbf{e}_j \rangle = \delta_{ij}$,
\begin{gather}
(M^n)_{ii} = (\lambda_i)^n   | \textbf{e}_i \rangle \langle \textbf{e}_i |, \\ \label{slosh10}
\implies M \delta \rho^{\text{in}}_n \rightarrow 0 \text{ iff } | \lambda_i | < 1 \hspace{0.5cm} \text{ } \forall \text{ } i.
\end{gather} 
The question now is whether, for typical DFT input systems, the condition $ | \lambda_i | < 1 $ is always satisfied. This is not the case, and for some illustrative models studied in Ref$.$ \citep{MM} $| \lambda_i | \sim 100$, causing a strong divergence from $\rho^*$ (see also $\S$\ref{precond}). This divergence is dubbed `charge sloshing'. A natural solution to this, from a numerical perspective, would be to attempt an unsophisticated form of preconditioning. That is, scale the eigenvalues $|\lambda_i|$ from Eq$.$ (\ref{slosh10}) with a parameter $\alpha$ to guarantee the convergence property of $\alpha |\lambda_i| < 1$. This leads to the most simple density mixing scheme able to converge a (not-so-wide) variety of input systems -- \textit{linear mixing}.

\subsection{Linear Mixing}
\label{linearmixing}

Instead of implementing $\rho^{\text{in}}_{n+1} = \rho^{\text{out}}_{n}$, one now incorporates a damping parameter $\alpha$ as such,
\begin{gather}
\rho^{\text{in}}_{n+1} =  \rho^{\text{in}}_{n} + \alpha ( \rho^{\text{out}}_{n} -  \rho^{\text{in}}_{n} ) =  \rho^{\text{in}}_{n} + \alpha R[\rho^{\text{in}}_{n}].
\end{gather}
The residual here defines not only the iterative error, but also the steepest descent direction. Therefore, the linear mixing scheme is effectively a weighted step in the steepest descent direction, and is constructed such that, in the case of $\alpha = 1$, no mixing is done (and a fixed point iteration update is computed). The corresponding matrix to $M$ in Eq$.$ (\ref{linearresponse}) is now defined as
\begin{align}
A = (1 - \alpha)\textbf{I} + M,
\end{align}
which has eigenvalues $\{ 1+\alpha (\lambda_i - 1) \}$. The convergence criteria is now modified to be $|1+\alpha (\lambda_i - 1)|<1$. As $\alpha$ can be arbitrarily tuned such that this is true, the problem has been (superficially) solved. Supposing $A$ is positive definite, $\alpha$ must be chosen such that the \textit{spectral radius} of $A$ is less than unity, $|1+\alpha(\lambda_{\text{max}} - 1)| < 1$, meaning $\rho_n$ is guaranteed to converge for all eigenvalues (assuming linearity). However, the speed of convergence is related to how close $|1+\alpha(\lambda_{\text{max}} - 1)|$ is to unity. If this quantity is only slightly less than unity, $|1+\alpha(\lambda_{\text{max}} - 1)|^n$ will take large $n$ to reach zero within tolerance. Moreover, if $\alpha$ is too low, the change in the $\rho^{\text{in}}_{n+1}$ becomes minuscule per iteration for eigenvalues at or close to $\lambda_{\text{min}}$, leading to slow convergence\footnote{Slow convergence for the eigenvalue $\lambda_{\text{min}}$ means the convergence will be slow for the density \textit{in the direction of} the eigenvector corresponding to $\lambda_{\text{min}}$.}. Therefore, an optimal value of $\alpha$ must be deduced, but how efficacious this choice is clearly depends on the ratio $\frac{\lambda_{\text{max}}}{\lambda_{\text{min}}}$ -- an important quantity in numerical analysis, the \textit{condition number} of $A$. The aim of sophisticated preconditioning is to \textit{compress} the eigenspectrum of $A$ as much as possible, leading to faster and more robust convergence.

In general $A$ is not well conditioned, e.g. for metallic materials the condition number of $A$ turns out to be divergent proportional to the size of the unit cell \citep{MM}, hence certain systems can take large amounts of time to converge (to the point where a user would say the calculation doesn't converge in a practical sense). The condition of the KS system is therefore intrinsically linked to the condition of $M$. This matrix encodes the density-density \textit{dielectric response} of the KS system. That is, a linear measure of how a perturbation in the density at $\textbf{r}$ affects the density at $\textbf{r}'$. An optimal method to remove ill-conditioning in the KS system would be to know this dielectric response exactly for a general input system. An analysis of the difficulty in calculating this response, and the physical meaning behind it, will be reserved for $\S$\ref{precond}, as this defines preconditioning. Until then, KS DFT can remain a black box. The two most successful numerical methods for solving KS DFT (in the absence of preconditioning) are those of Broyden \citep{broyden} and Pulay \citep{DIIS}. A large part of the following work, the Marks $\&$ Luke scheme, will depend on an extension to Broyden's method. Therefore, care will be taken in defining the mathematical and conceptual framework of Broyden's method. 

\subsection{Broyden Mixing}
\label{broydensec}

Broyden's class of methods are often referred to as being quasi-Newton, meaning they are based on the Newton-Raphson update equation, hereafter referred to as taking the Newton step. Translated into the language of KS DFT, this update becomes,
\begin{gather}
\rho^{\text{in}}_{n+1}(\textbf{r}) = \rho^{\text{in}}_{n}(\textbf{r}) - J^{-1}_R[\rho^{\text{in}}_n(\textbf{r})] R[\rho^{\text{in}}_n (\textbf{r})], \\
J_R[\rho^{\text{in}}_n] = \frac{\partial R[\rho^{\text{in}}_n]}{\partial \rho^{\text{in}}_n} \in \mathbb{C}^{N_G \times N_G},
\end{gather}
where $J$ is the \textit{Jacobian}. The mathematics in deriving the Newton step is detailed in Appendix. \ref{appA} as the update possesses many important and desirable properties, namely, its order of convergence. It can be shown that if there exists an interval $I$ where $\rho^{\text{in}}_1$ is `sufficiently close' to $\rho^*$, $R'$ is bounded away from zero and $R''$ exists (and is continuous), the rate of convergence of the Newton method is \textit{quadratic}. Hence, quadratic convergence is held as a `golden standard' for any algorithms to follow. Unfortunately, Newton's method as it appears here is unsuitable for KS DFT as the Jacobian is costly to compute and store -- it involves a derivative of the KS map (calculated, for example, with a numerical derivative), and is an $N_G \times N_G$ size object. The prohibitive computational cost in computing $J$ can be greatly alleviated by considering approximate updates toward the exact $J$ at each iteration, instead of an explicit construction of $J$ (ideally without much loss of convergence properties) -- Broyden's methods.

Broyden's methods for non-linear root finding were first published in 1965, yet the concepts and machinery he proposed remain foundational to many methods in practical use today. Broyden supposes that instead of computing and storing the exact Jacobian matrix, one can construct (or guess) an approximate Jacobian at only the initial iteration, and update it at every subsequent iteration in such a way that it approaches the exact Jacobian, and maintains convergence properties similar to that of Newton's method. Since density mixing only induces relatively small changes in $\rho^{\text{in}}_n$, the changes in the Jacobian will be (in some sense) small per iteration. This fact lends itself naturally to asking the question: what is this `small' update such that one can keep the near quadratic convergence of Newton by utilising data from previous iterations, but also avoid the computational expense of constructing the Jacobian? Broyden proposed two methods for achieving this: method one (producing the so-called `good Broyden formula') updates the Jacobian according to some assumptions and \textit{then} inverts it as is required in Newton's formula; whereas the second (`bad') method updates the inverse of the Jacobian explicitly under similar assumptions. The first method will be detailed below, whereas the second will simply be stated as it follows a similar methodology to the first. 

 Supposing one knows the Jacobian $J_{R, n-1}[\rho^{\text{in}}_{n-1}]$, an update is required to find $J_{R, n}[\rho^{\text{in}}_{n}]$ such that
\begin{gather}
J_{R, n}[\rho^{\text{in}}_{n}] = J_{R, n-1}[\rho^{\text{in}}_{n-1}] + C_n
\end{gather}
can be computed. The Jacobian $J_n$ should approach, in some sense, the exact Jacobian $J^*$. This update $C_n$ can be cleverly obtained by first enforcing a `secant' condition, i.e. force $J_{R,n}[\rho^{\text{in}}_{n}]$ to obey a finite-difference equation that is approximately true, which manifestly uses known data. This secant condition is defined as follows: for a simple one dimensional system, the definition of the derivative of $f$ at $x_{n}$ is obtained by taking the following limit,
\begin{gather}
f'(x_{n}) = \lim_{x_{n-1} \rightarrow x_{n}} \frac{f(x_{n}) - f(x_{n-1})}{x_{n} - x_{n-1}}.
\end{gather}
Meaning if $x_{n}$ and $x_{n-1}$ are sufficiently close, the secant condition is the finite-difference approximation, 
\begin{gather}
\label{secant}
f'(x_{n}) \approx \frac{f(x_{n}) - f(x_{n-1})}{x_{n} - x_{n-1}}.
\end{gather}
Importantly, in this one dimensional case, $f'(x_{n})$ is \textit{uniquely} defined by Eq$.$ (\ref{secant}). This is not the case in $>1$ dimensions, which will now be detailed. Generalising now to the $N_G$-dimensional system of KS DFT, the secant condition becomes,
\begin{gather}
\label{secant2}
J_{R,n}[\rho^{\text{in}}_{n}] (\rho^{\text{in}}_{n} - \rho^{\text{in}}_{n-1})= R[\rho^{\text{in}}_{n}] - R[\rho^{\text{in}}_{n-1}].
\end{gather}
If $J_{n}$ is an $N_G \times N_G$ matrix of full rank, it has $N_G$ linearly independent basis vectors forming a $N_G$-dimensional vector space, $V$. Enforcing the secant condition in Eq$.$ (\ref{secant2}) specifies how $J_{n}$ should act on all vectors parallel to $\rho^{\text{in}}_{n} - \rho^{\text{in}}_{n-1} \in \mathbb{C}^{N_G}$, corresponding to one basis vector in an appropriately rotated basis. Therefore, there exists a $(N_G-1)$-dimensional subspace $V' \subset V$ consisting of all elements of the vector space $V$ orthogonal to $\rho^{\text{in}}_{n} - \rho^{\text{in}}_{n-1}$:
\begin{align}
V' = \{ z  \text{ } | \text{ } z \in \mathbb{C}^{N_G}, \text{ } (\rho^{\text{in}}_{n} - \rho^{\text{in}}_{n-1}) \cdot z = 0 \}.
\end{align}
Importantly, Eq$.$ (\ref{secant2}) does \textit{not} specify how $J_{n}$ acts on elements of $V'$, of which there are $N_G-1$ linearly independent members. This means there exist $N_G$ simultaneous equations to fix $J_{n}$, but  $N_G^2$ degrees of freedom in $J_{n}$. The Broyden update in its current state is therefore underdetermined with information provided only by secant. By enforcing the secant condition, \textit{one rank} of information has been gained, leaving $N_G-1$ ranks of information unknown. What has been detailed up to this point defines an infinite \textit{class} of Broyden methods, i.e$.$ all rank one updates to the Jacobian that satisfy the most recent iterative secant condition (of which there are infinitely many). A particular method within this class will now be outlined, known as `Broyden 1 (B1)' or `Broyden's good method'. 

To recover the missing information, Broyden supposes $J_{n}$ acts on elements of $V'$ the same way as $J_{n-1}$ does
\begin{align}
J_{R,n}[\rho^{\text{in}}_{n}] z = J_{R,n-1}[\rho^{\text{in}}_{n-1}] z \text{ } \hspace{0.4cm} \forall \text{ } z \in V'.
\end{align}
This is a `minimal change' condition on $J_{n}$, motivated by the fact that changes in $\rho^{\text{in}}_{n}$ are small per iteration, and so changes in $J_{n}$ will follow suit. These two conditions, minimal change and secant, provide a unique solution to the update matrix $C_n$ which is crucially of rank one (from the secant information). To condense the above, B1 seeks the update matrix $C_n$ under two conditions,
\begin{align}
1.& \text{ } J_{R,n} \Delta \rho^{\text{in}}_{n} = \Delta R_{n}, \\ \nonumber
2.& \text{ } J_{R,n} z = J_{R,n-1} z.
\end{align}
introducing the notation,
\begin{gather}
\Delta \rho^{\text{in}}_n \coloneqq \rho^{\text{in}}_{n} - \rho^{\text{in}}_{n-1},   \nonumber \\
\Delta R_n \coloneqq R[\rho^{\text{in}}_{n}] - R[\rho^{\text{in}}_{n-1}].  \nonumber
\end{gather}
One can show the following ansatz satisfies these conditions,
\begin{align}
J_{R,n} =  J_{R,n-1} + \frac{\Delta R_{n} - J_{R,n-1} \Delta \rho^{\text{in}}_{n}}{ |\Delta \rho^{\text{in}}_{n} |^2 } (\Delta \rho^{\text{in}}_{n})^{\dagger},\label{r1u}
\end{align}
and thus provides a unique solution for $C_n$. The notation $uv^{\dagger}$ defines the \textit{outer product} of $u, v \in \mathbb{C}^{N_G}$, meaning Eq$.$ (\ref{r1u}) defines a manifestly rank one update. This update is also the solution to the constrained optimisation problem,
\begin{gather}
\underset{J_{R,n} \in S}{\text{minimise}} \text{ } ||J_{R,n} - J_{R,n-1}||_f^2 \label{opop}
\end{gather}
where $||.||_f$ is the \textit{Frobenius norm} of a matrix,
\begin{gather}
||A||_f = \sqrt{\sum_{i=1}^m \sum_{j=1}^n |a_{ij}|^2},
\end{gather}
and the set $S = \{ A \text{ } | \text{ }  A \in \mathbb{C}^{N_G \times N_G}, \text{ } A \Delta \rho^{\text{in}}_n = \Delta R_n \}$ imposes the secant condition.

 In order to finally compute the Newton step, the last operation needed is a method of inverting the expression in Eq$.$ (\ref{r1u}). An explicit formula for the update of the inverse of a rank one matrix is given by the \textit{Sherman-Morrison-Woodbury formula}, resulting in
\begin{align}
\label{sherman}
J_{R,n}^{-1} =  J_{R,n-1}^{-1} + \frac{\Delta \rho^{\text{in}}_{n} - J_{R,n-1}^{-1} \Delta R_{n} }{(\Delta \rho^{\text{in}}_{n})^T  J_{R,n-1}^{-1}  \Delta R_{n} } (\Delta \rho^{\text{in}}_{n})^{\dagger} .
\end{align}
The simplicity of this expression owes to the fact that the update of $ J_{R,n-1}$ is of rank one -- higher rank updates add algebraic, and hence computational, complexity to Eq$.$ (\ref{sherman}). Bryoden's second method follows this exact same methodology, but updates the inverse of the Jacobian directly (and therefore does not use the Sherman-Morrison-Woodbury formula), i.e$.$ solve
\begin{gather}
\underset{J^{-1}_{R,n} \in S}{\text{minimise}} \text{ } ||J^{-1}_{R,n} - J^{-1}_{R,n-1}||_f^2
\end{gather}
for $S$ defining the inverse secant condition. These methods have been shown to display superlinear convergence over a large sample of KS systems \citep{kresse2}, thus becoming established as core density mixing methods and motivating the subsequent work by Marks $\&$ Luke.

The problem with B1 and B2 as presented here is that, while the difficulty of having to construct the full Jacobian at every iteration has been solved, the method still requires storing a prohibitively large matrix. However, since Broyden first proposed these methods, much work has gone into improving them to the point where they can be implemented efficiently. A good review of these contributions can be found in Ref$.$ \citep{review111}. To highlight a few in particular, Srivastava \citep{limmem} in $1984$ was one of the first to propose a formulation of B1 and B2 such that the Jacobian in its full matrix form need never be stored, instead using a series of vector-vector products to compute the update. Moreover, Vanderbilt $\&$ Louie \citep{VL} and Eyert \citep{Eyert} formulated these methods in such a way that the entire history of iterates could be utilised to assist convergence rather than just the most recent one. Johnson in 1988 \citep{johnson} then combined the work of Vanderbilt $\&$ Louie, Eyert and Srivastava to demonstrate a memory efficient variant of Broyden's methods that uses the entire history of iterates to construct updates to the approximate Jacobian. This is the method currently employed by many state-of-the-art electronic structure codes, and is the default density mixing scheme in \textsc{castep} currently. The relevant theory has now been outlined such that the development of the Marks $\&$ Luke scheme can be understood. 

A rivalling method to the above is a mixing scheme proposed by Pulay in 1980 specifically to accelerate convergence in Hartree-Fock calculations, which will now be detailed. Importantly, the following discussion will introduce the idea of a subspace spanned by the history of iterates, and how one can utilise such a subspace to accelerate convergence.

\subsection{Pulay Mixing}

Pulay mixing is also known as a \textit{direct inversion in the iterative subspace} (DIIS) method. The essence of DIIS is to store a history of residual norms which is then used to determine (via a least squares fit) what the optimal (minimal) subsequent residual will be. The argument of this residual can then be extrapolated to give an optimal $\rho^{\text{in}}_{\text{opt}}$ which is used to construct $\rho^{\text{in}}_{n+1}$. The full details of the DIIS method are not important here, as they provide little further mathematical insight to the above analysis. Nonetheless, it is worth briefly assessing the core concepts and mathematics of Pulay's method due to its success and widespread use (and it will be used as a method of comparison in $\S$\ref{results})

To restate the problem, in KS DFT one seeks the argument of the residual such that $R[\rho^{\text{in}}_{n+1}] = 0$. To solve this, Pulay considers predicting a new residual as a linear combination of all the residuals that have been iterated through,
\begin{align}\label{linres}
R[\rho^{\text{in}}_{\text{opt}}] = \sum^n_{i=1} \alpha_i R[\rho^{\text{in}}_i]. 
\end{align}
The coefficients $\alpha_i$ are now determined by enforcing that the subsequent norm of the residual is a minimum (i.e$.$ as close to zero as the history of information allows it) -- solve
\begin{align}
\text{min}( \langle R[\rho^{\text{in}}_{\text{opt}}] | R[\rho^{\text{in}}_{\text{opt}}] \rangle) \text{ s.t. } \sum_{i=1}^n \alpha_i = 1.
\end{align}
This is a constrained optimisation problem with a single constraint, and hence can be solved by the method of Lagrange multipliers. That is, one constructs a Lagrangian as such,
\begin{align}
\mathcal{L} = \langle R[\rho^{\text{in}}_{\text{opt}}] | R[\rho^{\text{in}}_{\text{opt}}] \rangle - \lambda (\sum_i \alpha_i - 1),
\end{align} 
where $\lambda$ is the Lagrange multiplier. Minimising the residual norm is now a matter of substituting in Eq$.$ (\ref{linres}), and solving,
\begin{align}
\frac{\partial \mathcal{L}}{\partial \alpha_i}, \text{ } \frac{\partial \mathcal{L}}{\partial \lambda} = 0. 
\end{align}
Upon making this substitution, the Lagrangian takes the form,
\begin{align}\label{lagexp}
\mathcal{L} = \sum_i^n \sum_j^n \alpha_i \alpha_j \langle R[\rho^{\text{in}}_{i}] | R[\rho^{\text{in}}_{j}] \rangle -  \lambda (\sum_i \alpha_i - 1).
\end{align}
Performing the derivatives in Eq$.$ (\ref{lagexp}) requires some amount of algebraic manipulation (index gymnastics using the Kronecker delta and such), and the resultant linear system of equations one solves in order to find these coefficients (which is the least squares fitting step) is,
\begin{gather} 
\sum_i^n  \langle R[\rho^{\text{in}}_{i}] | R[\rho^{\text{in}}_{j}]  \rangle \alpha_j - \lambda = 0, \nonumber \\
\sum_i^n \alpha_i = 1.
\end{gather}
This corresponds to the following ($n$+1)-dimensional linear system in matrix notation,

\[
\begin{pmatrix}
    R_{1,1}        &  R_{1,2}  & \dots &  R_{1,n} & 1  \\
    R_{2,1}       & \ddots &  &  & 1 \\
    \vdots & & & & \vdots\\
    R_{n,1}       &  &  &  &  \\
    1 & 1 & \dots & \dots & 0
\end{pmatrix}
\begin{pmatrix}
\alpha_1 \\
\alpha_2 \\
\vdots \\
\alpha_n \\
\lambda
\end{pmatrix} =
\begin{pmatrix}
0 \\
0 \\
\vdots \\
0 \\
1
\end{pmatrix} \]

\noindent where,
\begin{align}
R_{i,j} \coloneqq \langle R[\rho^{\text{in}}_{i}] | R[\rho^{\text{in}}_{j}] \rangle. \label{DIISmatrix}
\end{align}
Solving this system defines the \underline{d}irect \underline{i}nversion in the DIIS; the \underline{i}terative \underline{s}ubspace is the space spanned by the history of densities,
\begin{align}
\mathcal{S}_n = \text{span}\{ \rho^{\text{in}}_1, \dots, \rho^{\text{in}}_n \}, \label{subspace1}
\end{align}
that one extrapolates over to find the optimal subsequent density. The dimension of this subspace increases as one iterates through the system such that $\mathcal{S}_{n-1} \subset \mathcal{S}_n$. One is therefore implicitly solving a limited memory problem if one restricts to a maximum history length $m \in [1,n]$ such that the dimension of the iterative subspace is capped at $m < n$. In \textsc{castep}, $m$ is an input parameter which is set to $m=20$ by default. Solving the above system now produces the $\{ \alpha_i \}$ that minimises the residual. A final step is then required to obtain $\rho^{\text{in}}_{\text{opt}}$. This is an extrapolation step, and assuming $R$ is linear, the input argument producing this optimised residual is,
\begin{align}
\rho^{\text{in}}_{\text{opt}} = \sum_{n-m+1}^n \alpha_i \rho_i^{\text{in}}. \label{disr}
\end{align}
This is easy to check by substituting Eq$.$ (\ref{disr}) into the definition of the residual then comparing to Eq$.$ (\ref{linres}). Initially, one might think of setting the subsequent iterative density equal to the optimal density as such,
\begin{align}
\rho^{\text{in}}_{n+1} = \rho^{\text{in}}_{\text{opt}}. 
\end{align}
However, with an update of this form, the dimensionality of the iterative subspace, Eq$.$ (\ref{subspace1}), remains the same as one has simply added a linear combination of vectors in the subspace to the set. In order to resolve this issue, the subsequent iterative density could, for example, take the form
\begin{align}
\rho^{\text{in}}_{n+1} =   \rho^{\text{in}}_{\text{opt}} - J_0 R[\rho^{\text{in}}_{\text{opt}}]
\end{align}
for some initial guess Jacobian $J_0$, as is done in Ref$.$ \citep{vasp}.

This method looks, aesthetically, quite different to Broyden's methods from the previous section. However, Kresse \textit{et al}. \citep{kresse2} were able to show that Pulay's DIIS is equivalent to a quasi-Newton step with an appropriately constructed Jacobian. An advantage of this mapping is that preconditioning becomes more transparent, as preconditioning seeks to compress the eigenspectrum of the Jacobian. Interestingly however, this mapping also reveals a noteworthy potential drawback of Pulay's method, which is that the DIIS updated Jacobian does not obey the underlying physical symmetries of the KS system. That is, the iterative Jacobian is an approximation to the KS charge dielectric, which is a symmetric object by construction. The DIIS updated Jacobian does not preserve this property, and neither does many variants of Broyden's Jacobian updates. The extent to which this observation affects the mixing schemes will be revealed in \S\ref{results}.  

Lastly, it is worth mentioning some additional potential problems with Pulay's method, which is perhaps why Broyden's methods are the default in \textsc{castep}. That is, the DIIS matrix in Eq$.$ (\ref{DIISmatrix}) can become singular in one of two scenarios: either the iterates are not strongly linearly independent, or the residual becomes extremely small (close to convergence). The latter problem seems to surface more frequently in practice, a solution to which could be to include some regularisation. A performance analysis of both Broyden's and Pulay's methods as implemented in \textsc{castep} will be given in $\S$\ref{results}, and compared to the proposed improvements of $\S$\ref{metho}.

\subsection{Honourable Mentions}
\label{honmen}

All relevant algorithms implemented in \textsc{castep} have now been presented. Before moving on to preconditioning, it is worth giving an honourable mention to a few other numerical methods that one might expect to be of utility in the present context. First is a method quite similar to Pulay's in that it searches over a subspace generated by a history of iterates -- the \textit{Krylov-Newton method} \citep{krynewton}. This methods functions by noticing that, in order to take a Newton step, one simply requires a solution to the following linear system of equations,
\begin{align}
J_{n} \delta \rho^{\text{in}}_n = R_n, \label{linsys1}
\end{align}
where,
\begin{align}
\rho^{\text{in}}_{n+1} = \rho^{\text{in}}_n + \delta \rho^{\text{in}}_n.
\end{align}
Once $J_n$ and $R_n$ have been specified, the Krylov-Newton method uses the \textit{generalised minimal residual method} (GMRES) \citep{gmres} to iteratively compute $\delta \rho^{\text{in}}_n$ such that Eq$.$ (\ref{linsys1}) is satisfied\footnote{Note that the Jacobian here does not need to be explicitly stored in RAM, as only its application on the iterative solution is required.}. This involves a search over a subspace generated by past iterates, namely the \textit{Krylov subspace} of $J_n$ and $R_n$,
\begin{align}
\mathcal{K}_m(J_n, R_n) = \text{span}(\{ R_n, J_n R_n, \dots, J_n^{m-1}R_n \} ).
\end{align}
An extrapolation over this iterative subspace is performed such that an optimised subsequent step $(\rho^{\text{in}}_n)_m \in \mathcal{K}_m$ minimises the residual, $r_m = ||J_{n} \delta (\rho^{\text{in}}_n)_m - R_n||$. Theoretically, then, this method has access to the quadratic convergence properties of Newton's method. In practice however, the approximation of $J_n$ involves a simple, first order numerical derivative of the KS map. This leads to an extra evaluation of the KS map per SCF cycle, meaning the advantage of `exact Newton' is negated by doubling the computational effort per SCF cycle. 

Another popular, but perhaps dated, density mixing scheme worth mentioning is the Anderson method \citep{anderson}. This is essentially the same as Broyden's methods, except instead of using the previous iterations' Jacobian to obtain the subsequent iterative Jacobian, one updates an initial guess of the Jacobian at every iteration. This leads to
\begin{align}
J_{R,n} =  J_{R,0} + \frac{\Delta R_{n} - J_{R,0} \Delta \rho^{\text{in}}_{n}}{ |\Delta \rho^{\text{in}}_{n} |^2 } (\Delta \rho^{\text{in}}_{n})^{\dagger},
\end{align}
which is the Broyden update under the substitution $J_{n-1} \rightarrow J_0$. As expected, this algorithm performs well if one is in possession of a good initial guess Jacobian, i.e$.$ if one knew the dielectric of the input system quite accurately.

This now covers the mathematical methodology employed by the vast majority of KS DFT implementations to reach self-consistency. The mathematics presented here is the product of over half a century of effort on the part of mathematicians and physicists, and thus there appears little scope for significant improvement short of a paradigm shift. However, the focus can now be switched to how one preconditions these methods by utilising prior knowledge of the KS system. The scope for improvement using preconditioning is far greater. It is a less researched, more specialised topic, and it will be shown that a perfect preconditioner can obtain the ground state solution in just one iteration.

\section{Preconditioning}

From a mathematical perspective, preconditioning refers to making a transformation to a system in order to reduce its condition number without changing the underlying method. If one restricts analysis to the linear response regime, it was shown in $\S$\ref{linearmixing} that the condition of the KS system is given by the eigenspectrum of the Jacobian. In fact, Ref$.$ \citep{kresse2} notes that the convergence of both Broyden's and Pulay's methods are, to a good approximation, proportional to $\sqrt{\frac{\lambda_{\text{max}}(J)}{\lambda_{\text{min}}(J)}}$. Compressing the eigenspectrum of the Jacobian therefore improves the speed of convergence through this proportionality. In order to find a preconditioning matrix $P$ such that the eigenspectrum of $J$ is minimised upon the operation $P^{-1}J$, one derives the optimal first order change in the iterative electron density. First, one seeks the matrix $P$ such that the Newton update,
\begin{align}
\rho^{\text{in}}_{n+1} = \rho^{\text{in}}_{n} + P (\rho^{\text{in}}_{n} - F[\rho^{\text{in}}_{n}]),
\end{align}
leads the system directly to convergence. The substitution $\rho^{\text{in}}_{n,n+1} = \rho^{*} + \delta \rho^{\text{in}}_{n,n+1}$ can be made, and the KS map can be linearised about $\rho^{*}$ to give
\begin{align}
\delta \rho^{\text{in}}_{n+1} \approx \delta \rho^{\text{in}}_{n} \Big( I - P \Big( 1 - \frac{\delta \rho^{\text{out}}_n}{\delta \rho^{\text{in}}_n} \Big) \Big).
\end{align}
As convergence is defined as $\delta \rho^{\text{in}}_{n+1} \rightarrow 0$, or
\begin{align}
\delta \rho^{\text{in}}_{1} \Big( I - P \Big( 1 - \frac{\delta \rho^{\text{out}}_n}{\delta \rho^{\text{in}}_n} \Big) \Big)^n \rightarrow 0,
\end{align}
the optimal definition of $P$ is given by,
\begin{align}
&I - P \Big( 1 - \frac{\delta \rho^{\text{out}}_n}{\delta \rho^{\text{in}}_n} \Big) = 0 \\
\implies &P = -  \Big( 1 - \frac{\delta \rho^{\text{out}}_n}{\delta \rho^{\text{in}}_n} \Big)^{-1} = -\epsilon^{-1}_0.
\end{align}
This is, in fact, the inverse of the \textit{density-density linear response} or the \textit{static, independent electron, charge dielectric} of the KS system of electrons. If this object were known exactly, convergence would be achieved in just one iteration, and thus accelerated density mixing schemes would be rendered redundant. That is to say, if one knew exactly how the KS electrons were going to respond to a perturbation about the ground state density, then, from any starting point within the linear regime, one could invert this behaviour to obtain self-consistency immediately. In general, the KS dielectric is not known, and is difficult to compute (as $\S$\ref{precond} will show). Therefore, one seeks a matrix $P$ that interfaces with the accelerated mixing schemes such that,
\begin{align}
P J_n^{-1} \approx  \Big( 1 - \frac{\delta \rho^{\text{out}}_n}{\delta \rho^{\text{in}}_n} \Big)^{-1}.
\end{align}
Ill-conditioning of the KS system is therefore mostly determined by the KS dielectric. It is worth then exploring the nature of this object in order to identify the sources of charge sloshing such that one can suppress them and, in turn, better condition the KS system.

\subsection{Preconditioning the Kohn-Sham System}
\label{precond}

Dropping the discretised (matrix) representation and dealing now with functional forms, the density-density linear response function is defined by,
\begin{align}
\delta \rho^{\text{out}}(\textbf{r}) = \int d\textbf{r}' \text{ } (1 - \epsilon_0(\textbf{r},\textbf{r}')) \delta \rho^{\text{in}}(\textbf{r}').
\end{align}
The response function answers the following question: if one perturbs the density at $\textbf{r}$, what is the response of the KS density across all \textbf{r}$'$ such that it is linearly related to the initial perturbation? The exact quantum mechanical dielectric (or \textit{susceptibility}, defined shortly) is ubiquitous to electronic structure theory across all levels of accuracy, containing within it the excitation properties of a system. In general, it is a time-dependent quantity, meaning a perturbation in the density is applied at time $t$, and the system responds causally for all $t'>t$. Moreover, this exact quantum mechanical dielectric is built from the \textit{interacting susceptibility}. That is, the object describing how an interacting system responds to a perturbation in the \textit{external potential}. Fortunately, the optimal preconditioning matrix is built from the static (time-independent), non-interacting KS susceptibility, which parametrises how the KS system responds to a perturbation in the \textit{effective potential}:  $\chi(t-t',\textbf{r},\textbf{r}') \rightarrow \chi_0(\textbf{r},\textbf{r}')$. The former can be recovered from the latter by means of an appropriately constructed Dyson equation (see Ref$.$ \citep{jharl}). The restriction to static susceptibilities is crucial, as time is not being considered a variable of the set-up here. Instead, the act of cycling through iterative densities is a form of explicit time dependence, rather than the addition of time as an implicit variable.

\begin{figure*}[htbp]
\centering
\includegraphics[width=2.8in]{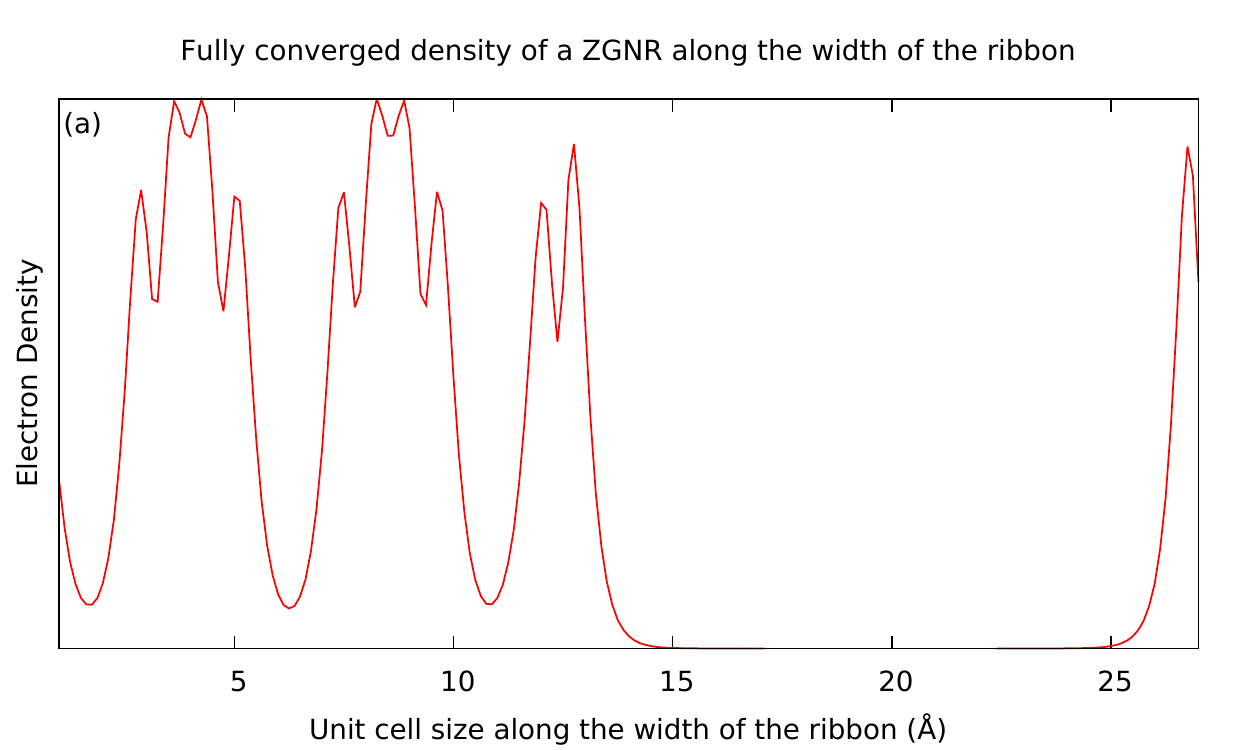}
\includegraphics[width=2.8in]{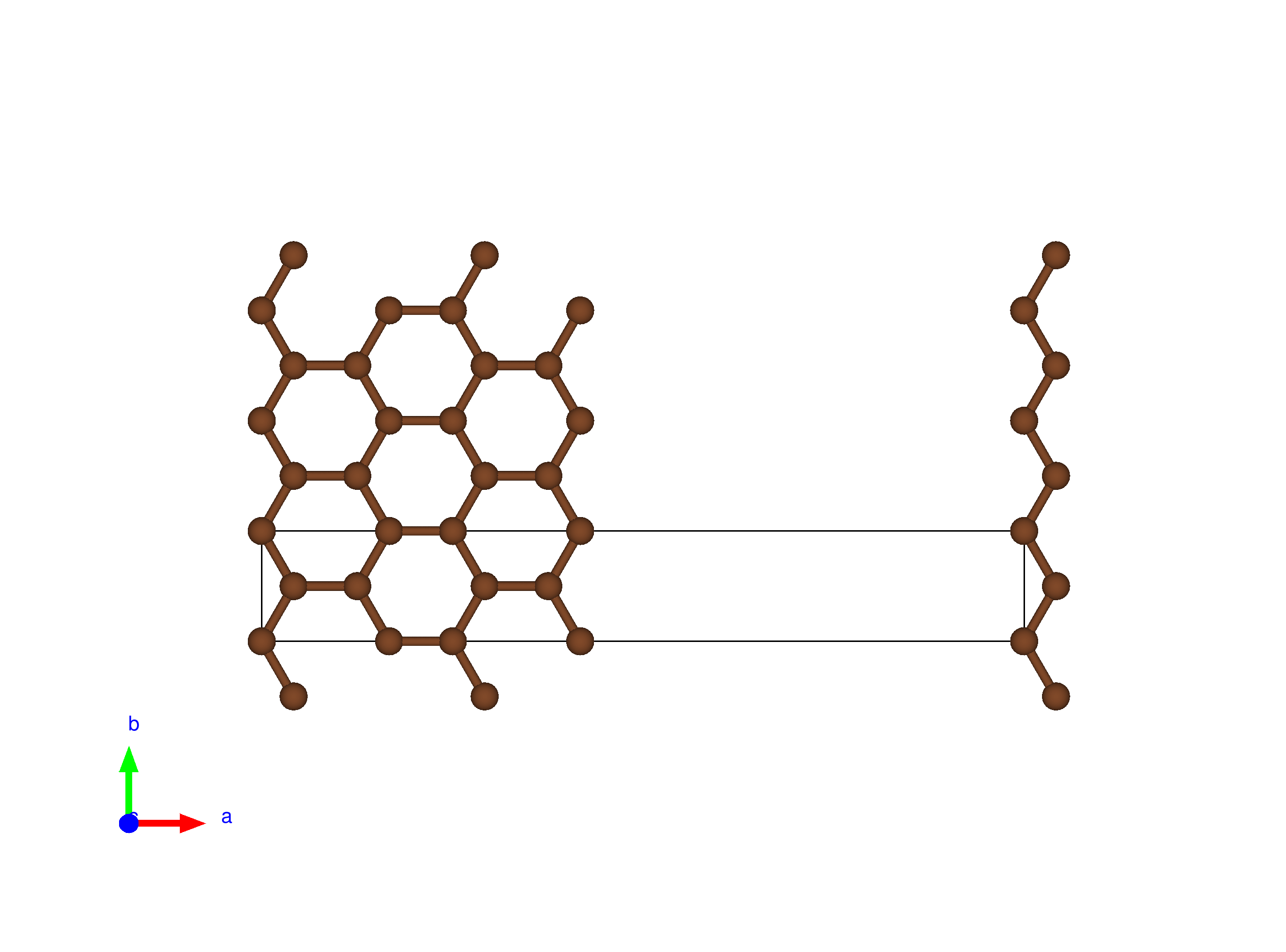}

\includegraphics[width=2.8in]{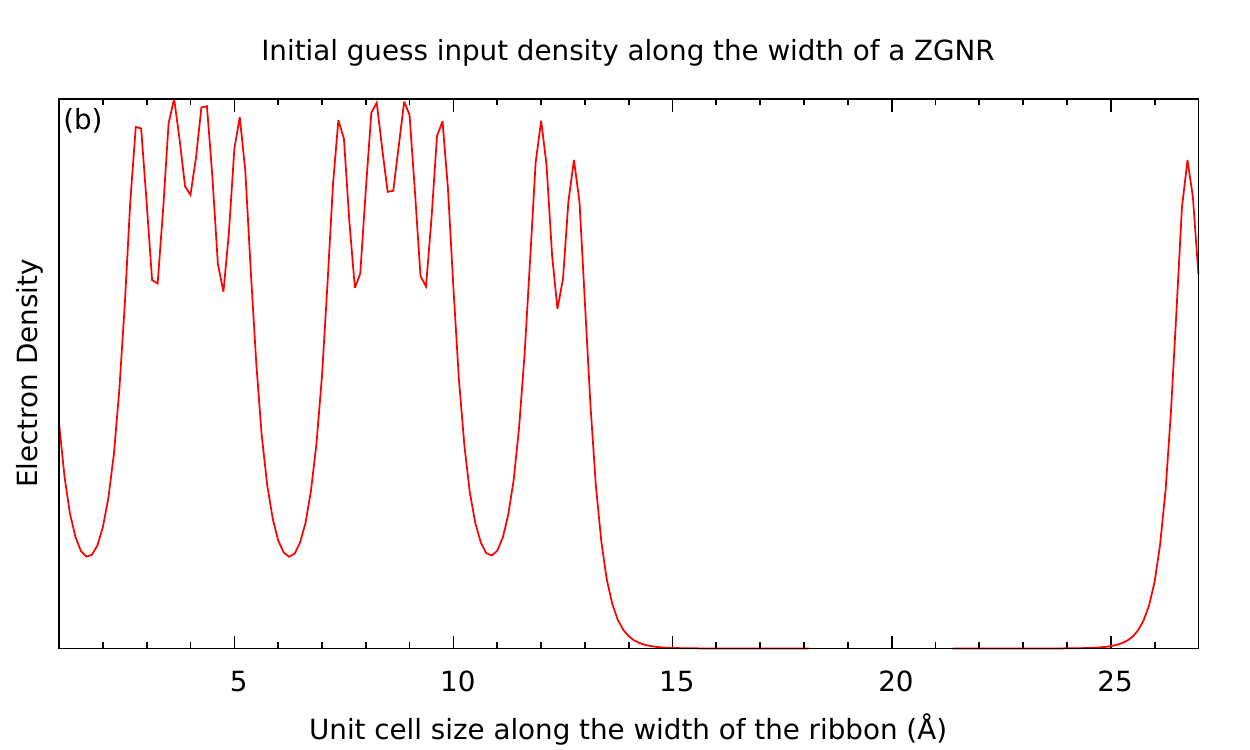}
\includegraphics[width=2.8in]{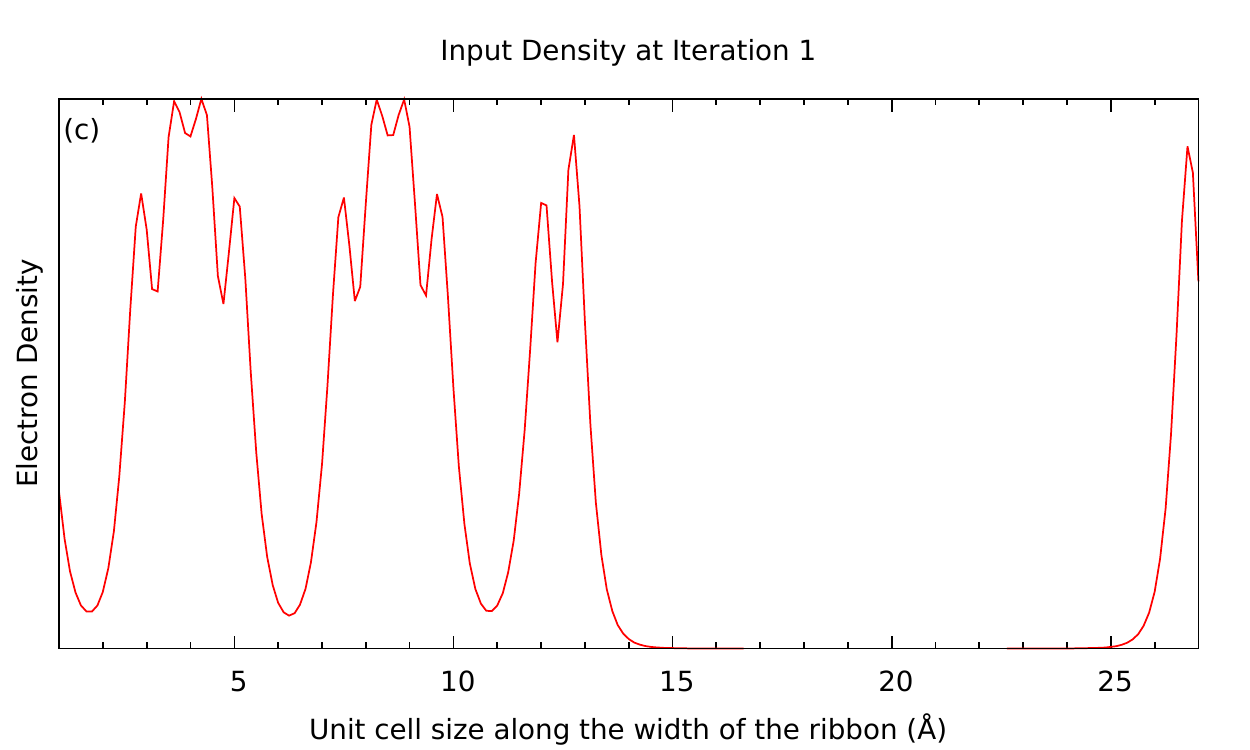}
\includegraphics[width=2.8in]{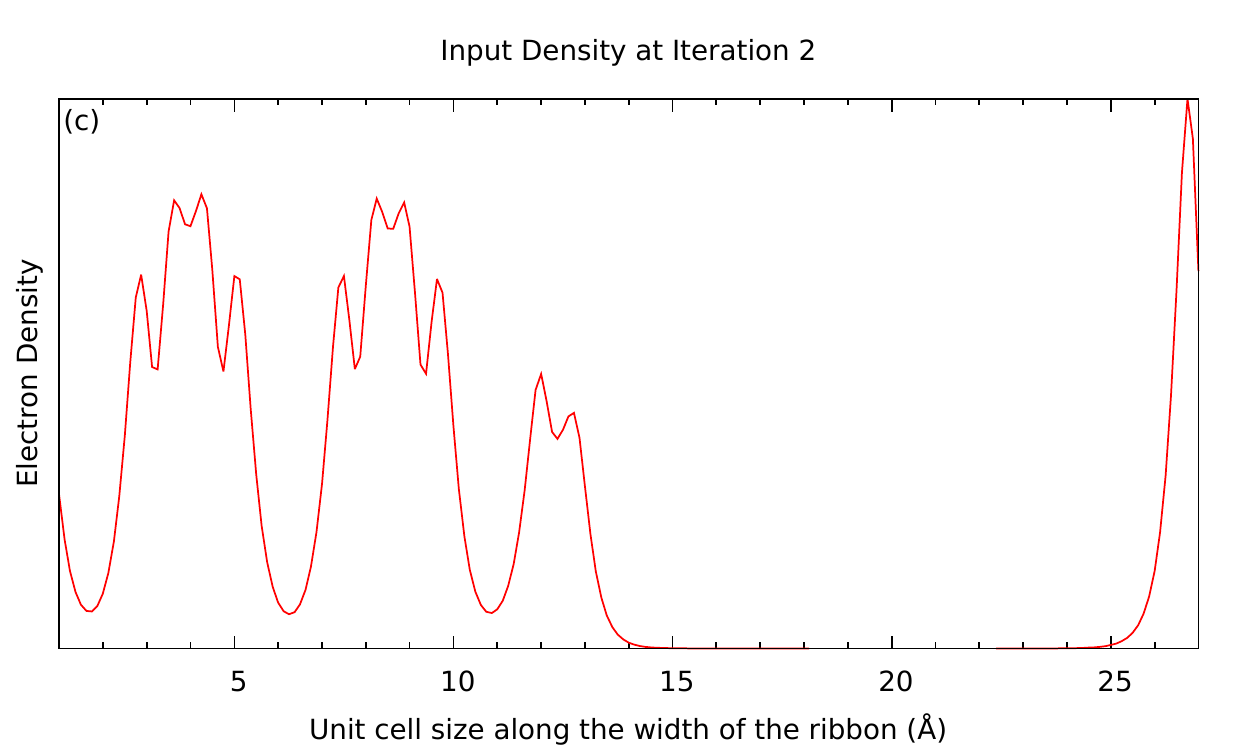}
\includegraphics[width=2.8in]{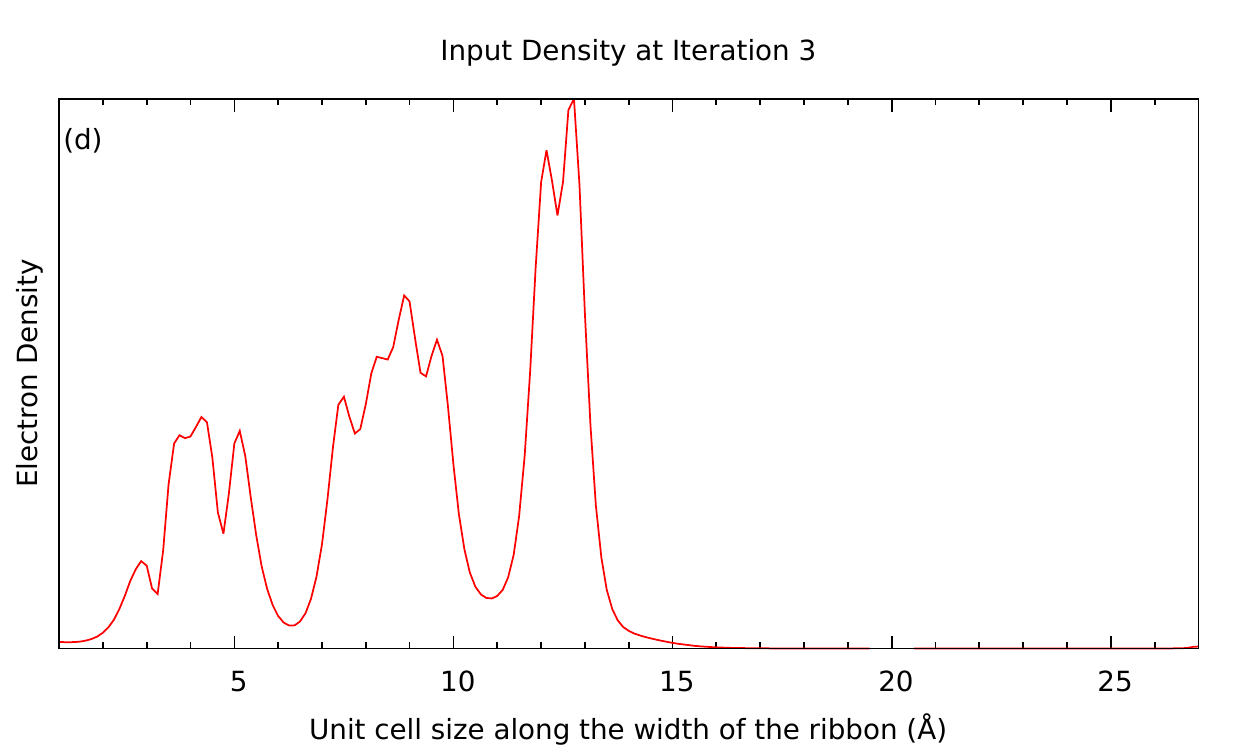}
\includegraphics[width=2.8in]{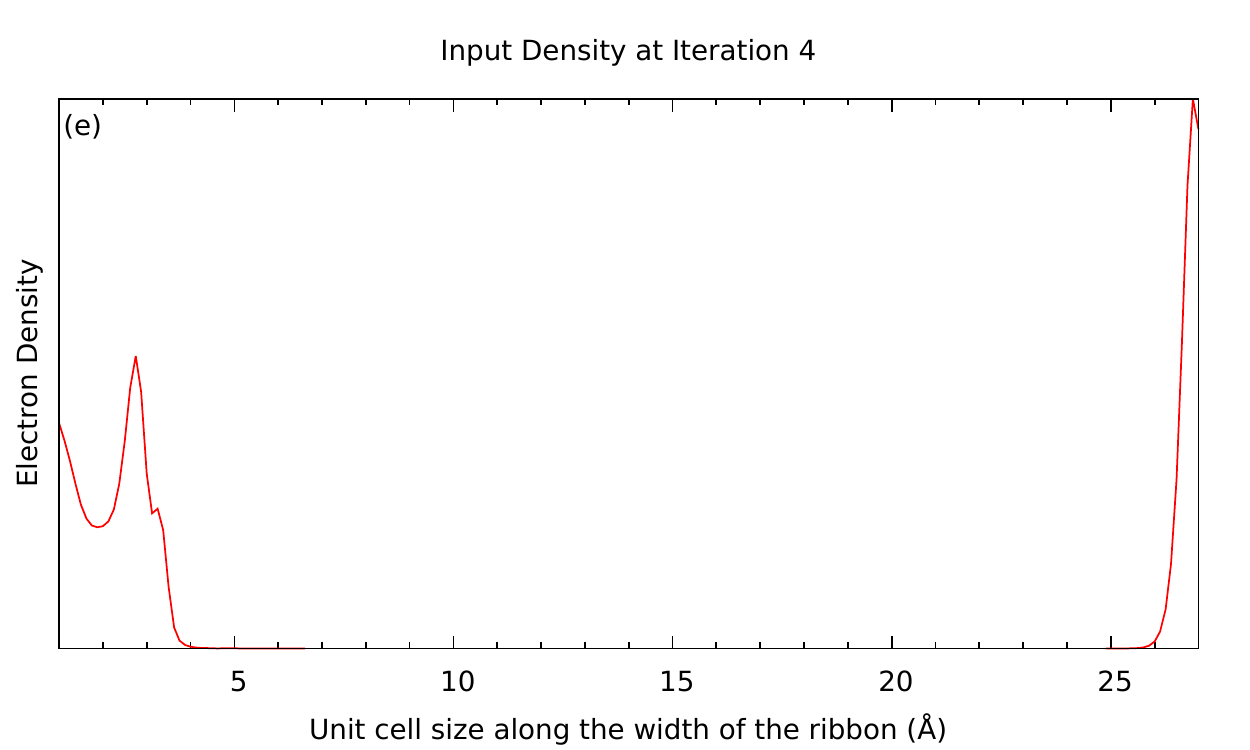}
\includegraphics[width=2.8in]{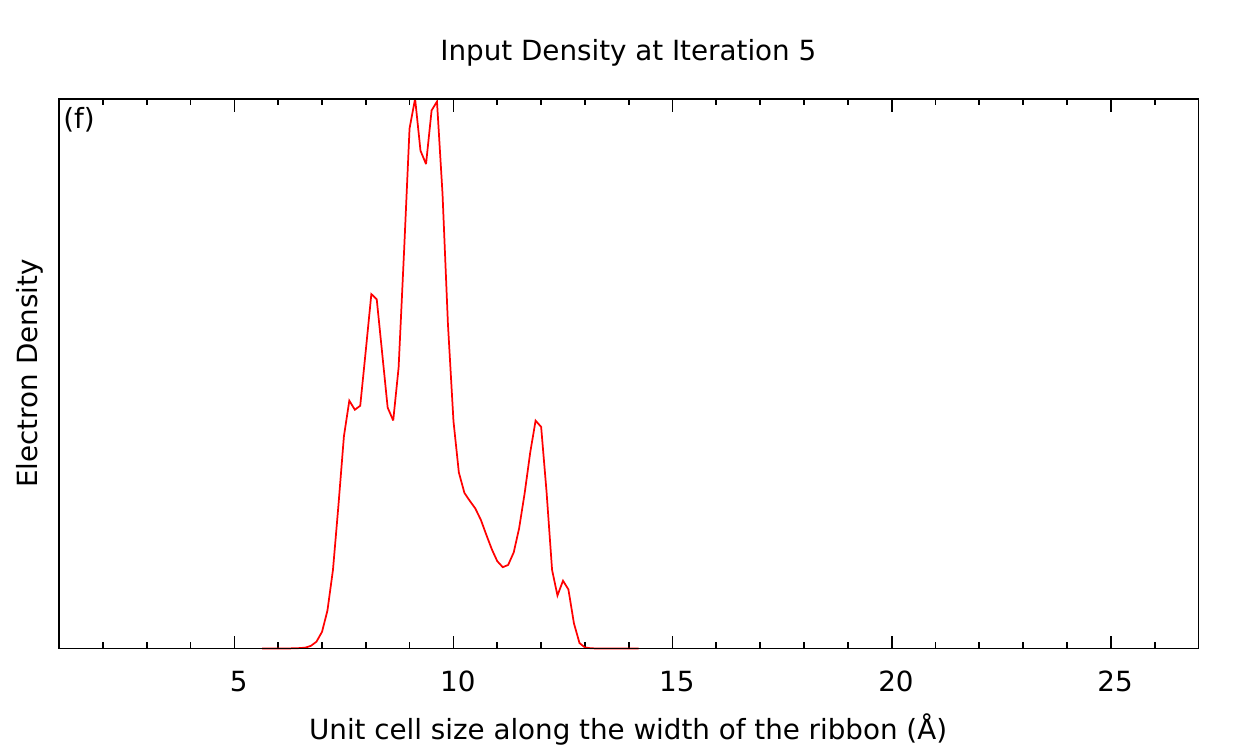}
\caption{(a) The fully converged, correct electron density for an eight atom graphene nanoribbon (top right) sliced across the width of the ribbon using preconditioned Pulay mixing. (b)-(f) Progression of the electron density through the SCF process using unpreconditioned linear mixing.}
\label{sloshfig}
\end{figure*}

The source of ill-conditioning can therefore be studied by unpacking the static KS dielectric. First, note that the entire density dependence of the KS Hamiltonian is through both the exchange-correlation and Hartree potentials. Utilising the chain rule, the following expansion can be made,
\begin{align}\label{chainrule}
 \epsilon_0(\textbf{r},\textbf{r}') = 1 - \frac{\delta \rho^{\text{out}}(\textbf{r})}{\delta \rho^{\text{in}}(\textbf{r}')} =  1 - \int d\textbf{r}'' \frac{\delta v^{\text{in}}_{\text{hxc}}(\textbf{r}'')}{\delta \rho^{\text{in}}(\textbf{r}')}\frac{\delta \rho^{\text{out}}(\textbf{r})}{\delta v^{\text{in}}_{\text{hxc}}(\textbf{r}'')}.
\end{align}
Assuming a local exchange-correlation potential, the first and second term respectively in the chain rule expansion are calculated through,
\begin{gather}
\delta \rho^{\text{out}}(\textbf{r}) = \int d\textbf{r}'' \text{ } \chi_0(\textbf{r}, \textbf{r}'') \delta v^{\text{in}}_{\text{hxc}}(\textbf{r}''),\label{RSscept}\\
 \delta v^{\text{in}}_{\text{hxc}}(\textbf{r}'') =  \int d\textbf{r}' \text{ } \big( K_{\text{c}}(\textbf{r}', \textbf{r}'')  + K_{\text{xc}}(\textbf{r}', \textbf{r}'')  \big) \delta \rho^{\text{in}}(\textbf{r}') \label{div11},
\end{gather}
where $K_{\text{c}}$ and $K_{\text{xc}}$ are the Coulomb and exchange-correlation kernels, and $ \chi_0$ now formally defines the static, independent particle susceptibility. The KS dielectric now takes the form,
\begin{align}
\epsilon_0(\textbf{r},\textbf{r}') = 1 - \int d\textbf{r}'' \big( K_{\text{c}}(\textbf{r}', \textbf{r}'')  + K_{\text{xc}}(\textbf{r}', \textbf{r}'')  \big)  \chi_0(\textbf{r}, \textbf{r}'').\label{dielec2}
\end{align}
Therefore, the condition of $\epsilon_0$ becomes divergent if either $K_{\text{hxc}}$ or $\chi_0$ become divergent (ignoring cancellations). The most prominent cause of ill-conditioning in the KS system is as a result of the Coulomb kernel. The reason for this becomes most transparent if Eq$.$ (\ref{div11}) is expressed in Fourier space as follows\footnote{As will be detailed in $\S$\ref{metho}, density mixing in \textsc{castep} is done in Fourier space.},
\begin{align}
\delta \tilde{v}^{\text{in}}_{\text{h}}(\textbf{G}) = \frac{4 \pi}{\Omega |G|^2} \delta \tilde{\rho}^{\text{in}}(\textbf{G}),
\end{align}
ignoring, for now, the exchange-correlation contribution. It is now easy to observe that finite changes in the density for long wavelength Fourier modes will be amplified by a factor of $|G|^{-2}$. Therefore, any \textit{error} in the $\delta \tilde{\rho}^{\text{in}}$ also gets amplified by this divergent factor, leading often to divergence of the scheme if one leaves this untreated. This defines \textit{Coulomb sloshing}, and an effective preconditioner to account for this behaviour would be one that suppresses low $G$-vector components of the change in the density with respect to the high $G$-vector components. In \textsc{castep}, the number of $G$-vectors increases linearly with volume of the unit cell. Thus, Coulomb sloshing is typically a problem when an input system has charge distributed across a large unit cell. For example, Figs$.$ (\ref{sloshfig})(b)-(f) depict the descent of the electron density of a graphene nanoribbon (10\AA{} gap in both aperiodic directions) into divergence as an unpreconditioned, linear mixing scheme is used to converge the electronic structure. This example illustrates some key characteristics of charge sloshing in any form. That is, in Fig$.$ (\ref{sloshfig})(b),  the charge density is initialised with only a small error with respect to the correct answer, Fig$.$ (\ref{sloshfig})(a). This error then becomes amplified, as too much charge is moved to the centre of the unit cell in Fig$.$ (\ref{sloshfig})(c). The SCF cycle then overreacts again by moving far too much charge to the edge of the unit cell, and so on. 
Note that, in the above analysis, the effect of the exchange-correlation kernel on the condition of the KS system was ignored. This is typically referred to as working in the \textit{random phase approximation} (RPA)\footnote{Although, a proper definition of the RPA within DFT is in how one recovers the interacting susceptibility from the KS susceptibility via a Dyson equation.}. In fact, the exchange-correlation kernel in, for example, the LDA, is not a source of ill-conditioning as it is a smooth, parametrised polynomial of the electron density only. An optimal preconditioner would include exchange-correlation effects, but it is not necessary. 

Interestingly, insulators (a finite KS energy gap at the Fermi energy), and metals (overlapping bands across the Fermi energy) possess crucial differences relating to self-consistency. The following argument is intended to highlight the result of this difference in relation to charge sloshing, and therefore will be heuristic in nature. First, in a simple, homogeneous metal, the susceptibility becomes \textit{constant}. That is, $\chi_0(\textbf{G},\textbf{G}') = \gamma \delta(\textbf{G}-\textbf{G}')$, meaning the electrons are mobile across the metal in response to a perturbation in the potential, parametrised by this constant. The eigenvalues of the dielectric, defined in Eq$.$ (\ref{dielec2}), for this example now become  $\sim 1 - \frac{\gamma}{|G|^2}$. Hence, a constant susceptibility leads to Coulomb sloshing being maximally pronounced in simple metals. On the other hand, insulators possess an effective `restoring force' for electrons, i.e$.$ the electrons are no longer arbitrarily mobile, which manifests in the susceptibility in the following way: $\chi_0(\textbf{G},\textbf{G}') \sim \eta |G|^2$ \citep{rex1,review111}. As a result, insulators tend to precondition themselves against Coulomb sloshing, as the eigenvalues of the dielectric become $\sim 1 - \eta$. This is constant for all Fourier modes, hence, an optimal preconditioner for simple insulators is an appropriately defined linear scaling parameter, $\alpha$.  

Ill-conditioning in the KS system now manifests in two additional cases: highly susceptible metals, and inhomogeneous (part metal part insulator) systems. The problematic nature of the former follows directly from the above arguments. The latter is problematic due to the behaviour of an optimal preconditioner needing to be fundamentally different in the metallic and insulating regions, and therefore implicitly requiring inhomogeneity. This would be the case if one were performing a surface or slab calculation, which involves a region of material and a region of vacuum.

 A final source of ill-conditioning worth mentioning is that of \textit{band sloshing}. It is fundamentally quite different to the other forms of sloshing discussed, and it occurs for systems possessing a large density of states about the Fermi level. The reason for band sloshing is as follows: the orbital occupancy is decided by identifying which bands (eigenvectors) of the KS Hamiltonian are lowest in energy. So far, occupancy has been treated as a binary variable;  orbitals are either completely occupied or not. Therefore, when many orbitals are close in energy about the Fermi level, it becomes difficult to assign such a binary occupancy. The reason for this is that the mere act of occupying an orbital results in an energy change for bands of the subsequent KS Hamiltonian. It is possible, therefore, that a previously unoccupied band becomes lower in energy than an occupied band (by nature of the bands being so close about the Fermi level). This band will then become occupied for the subsequent SCF cycle, but in doing so, the now unoccupied band could lower in energy again as a result of this change. Hence, certain systems become impossible to converge with binary occupancies, due to a constant re-shifting of band energies upon discontinuous occupancy. This is fixed in part by introducing the notion of partial occupancies of KS orbitals. The density, for example, now takes the form,
\begin{align}
\rho(\textbf{r}) = \sum_{i=1}^{N_b} f_i | \psi_i (\textbf{r}) |^2,
\end{align}
where $f_i \in \mathbb{R}$ is the occupancy of orbital $i$, now lying in the interval $[0,1]$. This has implications relating to the fundamental nature of KS DFT, that will not be discussed here\footnote{For example, the Levy-Lieb constrained optimisation now requires not only a search over all $N$-representable densities, but also over all occupancies such that $\sum_i f_i = N$ for $f_i \in [0,1]$.}. It suffices to note that one can effectively \textit{smear} this occupancy across the Fermi level according to some (not necessarily physical) smearing function in order to better condition the self-consistency. For example, using the Fermi-Dirac distribution,
\begin{align}
f_i = \frac{1}{e^{(\epsilon_i - E_f)/T}+1},
\end{align}
for some \textit{electronic temperature}, $T$. The entropic contribution to the energy as a result of this effective finite temperature can be subtracted post-convergence, leading to the zero temperature solution. This is not a concern from the point of view of preconditioning density mixing schemes as defined above, but was worth mentioning in a section devoted to ill-conditioning of the KS system.

\subsection{Dielectric Models as Preconditioners}

This section will be dedicated to reviewing attempts at preconditioning the self-consistent cycles from literature, which will ultimately serve to motivate and put into context the work done in $\S$\ref{results}. For reasons already discussed, most preconditioners typically restrict to the RPA. Hence, the remainder of this work will set $K_{\text{xc}} = 0$. It was shown in the previous section that constructing the optimal preconditioner amounts to calculating the KS susceptibility exactly, and in turn calculating the exact inverse KS dielectric. This is typically done in Fourier space, where the inverse dielectric takes the form
\begin{align}
\epsilon^{-1}_0(\textbf{G},\textbf{G}') = \Bigg[ \int d\textbf{G}'' \Big( \delta(\textbf{G}-\textbf{G}') - K_c(\textbf{G}',\textbf{G}'') \chi_0(\textbf{G},\textbf{G}'') \Big) \Bigg]^{-1}.
\end{align}
There is a subtlety introduced here: if $K_c(\textbf{G}',\textbf{G}'')$ is defined to be the Fourier transform of the Coulomb kernel, 
\begin{align}
 K_c(\textbf{G}',\textbf{G}'') = \frac{4 \pi}{|G''|^2} \delta(\textbf{G}'-\textbf{G}''),
\end{align}
then $\chi_0(\textbf{G},\textbf{G}'')$ is \textit{not} the Fourier transform of the real space susceptibility as defined above due to the convolution required in Eq$.$ (\ref{RSscept}). This is a key concern of the work to follow, and the relationship between the real and Fourier space susceptibilities will be defined more precisely in $\S$\ref{smodelmeth}. Ignoring this for now, the optimal update in the density in a discrete representation is given by
\begin{align}
\delta \rho^{\text{in}}_n = \Big( I - K_c \chi_0 \Big)^{-1} R[\rho^{\text{in}}_n], \label{exactmatrep}
\end{align}
without reference (yet) to any accelerated schemes such Pulay mixing. The most natural starting point in computing Eq$.$ (\ref{exactmatrep}) would be to consider an exact expression for the KS susceptibility. Following the mathematics of linear response theory applied to independent particle systems, Adler \citep{adler1} and Wiser \citep{wiser1} separately derived the exact expression for the KS susceptibility,
\begin{align}
\chi_0(\textbf{r},\textbf{r}') = \sum_{n=1}^{N_e} \sum_{m=N_e+1}^{N_b} \frac{\psi_n(\textbf{r}) \psi^*_m(\textbf{r}) \psi^*_n(\textbf{r}') \psi_m(\textbf{r}')}{\epsilon_n - \epsilon_m},\label{HIJ}
\end{align}
known as the Adler-Wiser equation. This equation can be expanded in a planewave basis for Bloch wavefunctions, yielding
\begin{align}
\chi_0(\textbf{G},\textbf{G}') = \sum_{\textbf{k} \in 1\text{BZ}} \sum_{n,m=1}^{N_b} ( f_{n,k} - f_{m,k} ) \frac{\langle n, \textbf{k} | e^{-i\textbf{G}.\textbf{r}} | m, \textbf{k} \rangle \langle m, \textbf{k} | e^{i\textbf{G}.\textbf{r}} | n, \textbf{k} \rangle}{\epsilon_{m,k} - \epsilon_{n,k}}.
\end{align}
Importantly, this expression involves a sum over \textit{all} unoccupied bands of the KS Hamiltonian. Therefore, preconditioning with the exact susceptibility in this form is not a  practical approach in a planewave basis, as it involves at least an $\sim \mathcal{O}(N_G^3)$ computation (see $\S$\ref{impsec}). Nonetheless, an attempt at implementing this expression exactly for small systems (with a very low cut-off energy) was made in 1982 by Ho, Ihm, and Joannopoulos -- the `HIJ' method \citep{hij}. As Ref$.$ \citep{review111} notes, constructing and applying this susceptibility is in fact an $\mathcal{O}(N_G^4)$ operation with a large prefactor. As such, it is not used in state-of-the-art electronic structure calculations. However, this method serves as a useful proof of concept for the following work. That is, if the exact linear response is known, convergence to a self-consistent solution can be achieved in $\mathcal{O}(1)$ iterations, even for the complex interfaces studied in Ref$.$ \citep{hij}\footnote{The exact linear response here implicitly includes all inhomogeneities of the system at hand, also know as the `local field effects'.}. Since 1982, multiple attempts have been made at simplifying this expression aiming to either suppress the prefactor, or the scaling \citep{gonze,gonzein1,gonzein2}. To highlight one in particular, Gonze (of \textsc{abinit}) and Anglade in 2008 were able to devise a method whereby all unoccupied bands need not be calculated -- the `extrapolar' method \citep{gonze}. Instead, the susceptibility is calculated exactly for $n,m \leq N_e$, and the unoccupied orbitals are approximated by planewaves, allowing a `closure relation' in the `infinite' sum to be utilised. The details are not important here, it being sufficient to note that the computational scaling (now $\mathcal{O}(N_e^4)$) and prefactor remain impractical, especially for large systems. Preconditioning using an explicit, exact expression for the susceptibility has thus achieved limited success so far. 

\begin{figure*}[htbp]
\centering
\includegraphics[width=5in]{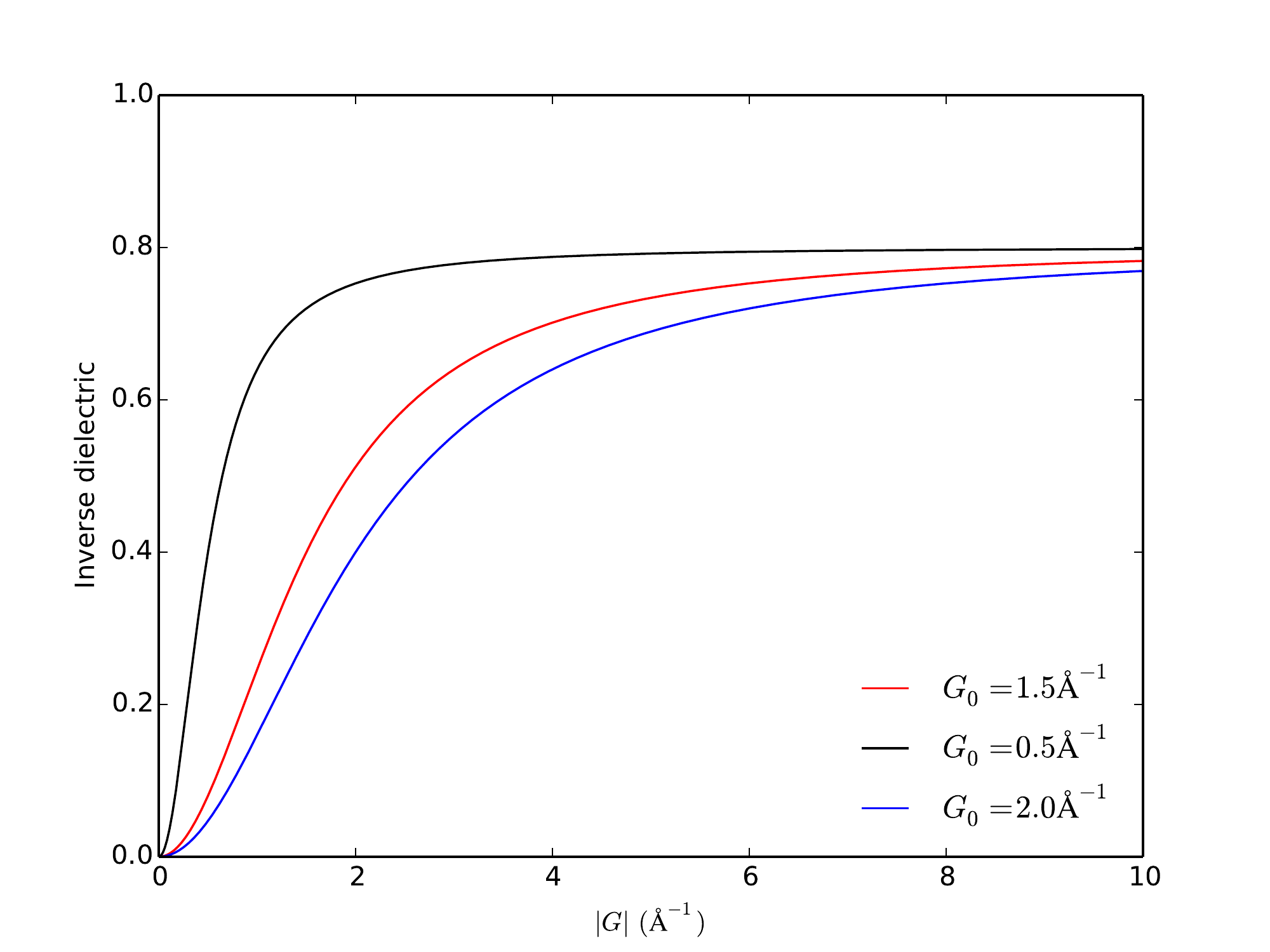}
\caption{The Kerker form across a set of typical input parameters for $\alpha=0.8$.}
\label{fig:kerker}
\end{figure*}

The Adler-Wiser equation is not the only technique to compute the susceptibility exactly. The susceptibility can also be calculated with a series of numerical derivatives of the KS map, which is perhaps a more conceptually straight forward technique. That is, the KS system is perturbed for a \textit{single} planewave \textbf{G}, then the response of the KS system is computed across \textit{all} \textbf{G}$'$. This would produce one column of the susceptibility matrix, and a full construction $\chi_0$ would require $N_G$ of these numerical derivatives. Again, this is clearly impractical, but the idea itself can be utilised by noticing that one simply needs the \textit{application} of the dielectric on a vector (the residual), \textit{not} the full matrix. This application can be performed by taking only one numerical derivative of the KS map, or by use of the \textit{Sternheimer equation} \citep{linphonon}. This concept will be elaborated further when discussing potential future work in $\S$\ref{furtherwork}. There has been, so far, no universally adopted techniques that improve the self-consistent process by treating the susceptibility exactly. This motivates the use of a dielectric \textit{model} in an attempt to capture the characteristic behaviour of charge screening in KS DFT without making explicit reference to the input system at hand. In principle, this model would be derived analytically, producing little additional computational overhead. By far the most successful dielectric model for preconditioning SCF cycles has been the so-called \textit{Kerker preconditioner} \citep{kerker}. Based on the work of Manninen \textit{et al}. \citep{mann}, an analytic expression is derived for the inverse dielectric of the homogeneous electron gas (HEG). The Kerker form can thus be obtained by first considering the linear response in real space of a homogeneous and isotropic system, $\chi_0(\textbf{r},\textbf{r}') \rightarrow \chi_0(|r-r'|)$, satisfying
\begin{align}
\delta \rho^{\text{out}} (\textbf{r}) = \int d\textbf{r}' \text{ } \chi_0(|r-r'|) \delta v^{\text{in}}_{\text{h}}(\textbf{r}').
\end{align}
This expression is a convolution over the input potential, and therefore takes the form of a product in Fourier space,
\begin{align}
\delta \tilde{\rho}^{\text{out}} (\textbf{G}) =  \tilde{\chi}_0(|G|) \delta \tilde{v}^{\text{in}}_{\text{h}}(\textbf{G}).\label{idscep}
\end{align}
This susceptibility is \textit{local} and \textit{homogeneous} in Fourier space, and the corresponding inverse KS dielectric is
\begin{align}
\epsilon^{-1}_0(|G|) = \Big( I - \frac{4 \pi}{|G|^2} \tilde{\chi}_0(|G|)    \Big)^{-1}.
\end{align}
The final step in obtaining the Kerker form is to identify the susceptibility of Eq$.$ (\ref{idscep}) with the \textit{Thomas-Fermi screening wavevector} resulting from Thomas-Fermi (TF) theory applied to the HEG \citep{tf1}\footnote{Thomas-Fermi theory applied to the HEG comes with a certain set of approximations. For example, the induced potential from an external perturbation is slowly varying in space (see Ref$.$ \citep{tf1}).}. Within this framework, the susceptibility becomes constant for all Fourier modes, leading to $\chi_0 = k_{\textsc{tf}}^2 \sim \text{constant}$. The magnitude of the TF wavevector, and hence the susceptibility, is controlled entirely by the HEG density parameter, $r_s$. The Kerker preconditioner is now given in its fully parametrised form as
\begin{align}
\epsilon^{-1}_0(|G|) =  \alpha \Big( 1 + \frac{|G_0|^2}{|G|^2}    \Big)^{-1} = \alpha \frac{|G|^2}{|G|^2 + |G_0|^2},
\end{align}
where $|G_0|^2 = 4 \pi k_{\textsc{tf}}^2$ is now a parameter of the scheme, along with the linear mixing parameter $\alpha$ from $\S$\ref{linearmixing}\footnote{This is, in fact, the static limit of the \textit{Lindhard} dielectric \citep{ashcroftmer} for the case of TF theory applied to the HEG.}. In part due to its simplicity, the Kerker preconditioner has seen astonishing success in stabilising and accelerating self-consistency in KS DFT calculations. The mixing parameter $\alpha$ serves the same purpose as in $\S$\ref{linearmixing}, which is to ensure the spectral radius of the Jacobian is below unity. The parameter $|G_0|$ now describes the degree of `metallicity' of an input system. That is, in the limit of $|G_0| = 0$, the preconditioner reduces to the linear mixing parameter for all Fourier modes, which was shown to be the optimal preconditioner for simple insulating systems. Increasing $|G_0|$ acts to further suppress low magnitude $G$-vectors components of the density update. The more susceptible an input, the more it is prone to Coulomb sloshing, and a higher $|G_0|$ is required. Kresse \textit{et al}. \citep{kresse2} suggest the parameters $\alpha=0.8$ and $|G_0|=1.5$\AA$^{-1}$ perform adequately across a range of input systems, and are therefore also the default parameters in \textsc{castep}. A plot demonstrating how the Kerker preconditioner weights components of the density to counteract Coulomb sloshing for three characteristic values of $|G_0|$ is given in Fig$.$ (\ref{fig:kerker}).

A key reason for the success of the Kerker preconditioner is that it interfaces well with accelerated mixing schemes. That is, the Kerker form often successfully compresses the eigenspectrum of the iterative Jacobian when using Broyden or Pulay mixing at each iteration. As such, the Kerker form is used almost exclusively to augment accelerated mixing schemes, rather than as a stand-alone mixing scheme. This is unlike exact calculations of the susceptibility above, which typically do not interface well with accelerated mixing schemes. Instead, direct dielectric mixing is used due to the inherent accuracy and system-dependence in the construction of the exact susceptibility. The major drawback of Kerker preconditioning is that it involves an entirely homogeneous model of the dielectric, meaning self-consistency can be hindered for highly inhomogeneous input systems. Moreover, there is no scope for systematically identifying input dependent features (e.g$.$ vacuum gaps), and adapting parameters on-the-fly. This is in part due to the complexity of the HEG dielectric in real space, involving an integral over a Yukawa-type screening kernel (see Ref$.$ \citep{mann} and \S\ref{smodelmeth}). Identifying system dependent features is most easily done in real space, thus a degree of transparency is lost with no obvious Kerker form in real space. Many simple extensions to the Kerker form for preconditioning have been proposed. These typically involve including a $|G|$-dependence in $\chi_0$ in such a way that better treats insulating and semiconducting systems \citep{LL1,penn}. For example, Zhou \textit{et al}. \citep{resta1} utilise an extension of TF theory to insulators, originally derived by Resta \citep{resta2}. This leads to a dielectric model of the form
\begin{gather}
\epsilon^{-1}_0(|G|) = \frac{|G|^2}{|G|^2 + f(|G|)}, \\
f(|G|) = \frac{\alpha \text{ sinc}({|G|\beta})}{\gamma},
\end{gather}
where $\alpha$, $\beta$, and $\gamma$ are parameters of the model. While this may provide a better treatment of screening in semiconductors and insulators from a physical perspective, the core drawbacks of the Kerker preconditioner remain. That is, the model is still homogeneous, and improvements pertaining to convergence are generally minimal and entirely parameter dependent. Therefore, the work presented in $\S$\ref{smodelmeth} and $\S$\ref{precondmodelres} will attempt to strike a middle ground between the dielectric models presented here, and an exact computation of the susceptibility. This will be done by implementing a framework that implicitly includes inhomogeneity of the input system through an on-the-fly computation of a susceptibility model in real space. This will be similar in nature to work done in Refs$.$ \citep{review111} and \citep{RSK}, which provide systematics for constructing the Kerker preconditioner in real space.


\chapter{Methodology} 

\label{metho} 

\lhead{3. \emph{Methodology}} 


This chapter will detail the methodology employed by the following work in an attempt to improve density mixing. First, a note on the numerical implementation of density mixing within \textsc{castep} will be given. Following this, the Marks $\&$ Luke and Periodic Pulay mixing schemes will be presented and discussed, including implementation specific details pertaining to \textsc{castep} in particular. Lastly, a framework for including imhomogeneity in the preconditioner will given. The scope for improvement with such a framework in place will be considered using two preliminary models.


\section{Density Mixing in CASTEP}
\label{dmincastep}

Density mixing within \textsc{castep} is done in Fourier space. This means the output density from the KS orbitals is first Fourier transformed, then the mixing scheme is applied and the subsequent iterative input density is inverse Fourier transformed in order to construct the corresponding KS Hamiltonian. Density mixing in Fourier space has a number of advantages. First, one can make the observation that the exact KS dielectric in Fourier space for both insulators and metals tends to unity in the limit of increasing $|G|$ \citep{kresse2,ashcroftmer}. Therefore, the optimal update in the density for these Fourier modes is in fact a fixed point update, i.e$.$ no mixing needs to be done for $|G| > |G|_{\text{cut-off}}$. This allows one to define a `mixing sphere' in Fourier space whereby all density components inside the sphere are fed into the mixing scheme, and the remainder are left unchanged. The computational and memory overhead of density mixing is thus drastically reduced at no loss to the mixing scheme, the only caveat being that two extra FFTs on the full grid are required.  In addition to this, the $|G|=0$ component of the density is explicitly not mixed, as fixing this component amounts to conserving net charge. A direct result of mixing in Fourier space is that all densities are now complex objects, therefore care must be taken to appropriately mix both the imaginary and real components of the density correctly. 

It was shown in $\S$\ref{precond} that for metallic systems Coulomb sloshing is pronounced, meaning the low $G$-vector components of the density update should be suppressed accordingly with a preconditioner. In this vein, Kresse \textit{et al}. \citep{kresse2} also define a \textit{mixing metric} with which to evaluate scaler products of the density,
\begin{gather}
\langle \rho_n^{\text{in}} | \rho_n^{\text{out}} \rangle = \sum_{i,j=1}^{N_b} g_{ij}  (\rho^{\text{in}}_n)_i  (\rho^{\text{out}}_n)_j, \\
g_{ij} = \frac{|G|_i^2 + |G_1|_i^2}{|G|_i^2} \delta_{ij}.
\end{gather}
The parameter $|G_1|$ describes the extent to which low $G$-vector components of the density are suppressed in the scalar product. This is typically defined such that the longest wavelength Fourier mode is weighted approximately twenty times less than the shortest wavelength Fourier mode. The introduction of the mixing metric therefore provides a systematic method of dealing with Coulomb sloshing within a mixing scheme that becomes inherent to the scheme. 

Finally, it is worth noting the effect of initial random seed on density mixing. The purpose of the random seed in \textsc{castep} is to initialise the wavefunctions for the iterative diagonaliser. One would expect therefore that the iterative diagonaliser would converge appropriately, and the output density would have no dependence on the random seed. In practice however, the wavefunctions do not need to be fully converged at each SCF cycle, as the majority of SCF cycles solve the KS equations with an unconverged (non-physical) density. Hence, convergence of the iterative diagonaliser is defined such that the general behaviour of the KS system is captured, but with a certain liberty taken in the tolerances as the interim solutions will not be used to determine physics. It is possible that this error could affect some schemes more than others, and it indeed produces some variance in the number of SCF cycles taken to converge for otherwise identical input systems. However, this difference is typically small, and it is not expected that the nature of the random seed will alter any of the conclusions drawn from the testing in $\S$\ref{results} unless stated otherwise.

\section{Marks $\&$ Luke}
\label{ML1111}

The Marks $\&$ and Luke scheme of Ref$.$ \citep{ML} is an extension to Broyden's class of methods whereby the iterative Jacobian is determined by solving a set of secant equations utilising the entire iterative subspace, rather than just the most recent iterations' data. Hence, the Marks $\&$ Luke scheme is a \textit{multisecant Broyden method}. This extension is relatively simple: one solves the same constrained optimisation problem as in Eq$.$ (\ref{opop}),
\begin{gather}
\min_{J_n \in S} || J_n - J_{n-1} ||_f^2, 
\end{gather}
but the set $S$ now defines the secant condition over the entire history of densities and residuals. That is, 
\begin{align}
S \coloneqq \{ J_n \text{ } | \text{ } J_n \in \mathbb{C}^{N_G \times N_G}, \text{ } J_n S_n = Y_n \}
\end{align}
where the objects $S_n, Y_n \in \mathbb{C}^{m \times N_G}$ now define a column of density and residual differences respectively over the $m$-dimensional subspace of iterates,
\begin{gather}
S_{n} \coloneqq [\Delta \rho^{\text{in}}_{n-m}, \dots, \Delta \rho^{\text{in}}_{n}],\label{S1} \\
Y_n \coloneqq [\Delta R_{n-m}, \dots, \Delta R_{n}].\label{Y1}
\end{gather}
The solution to this constrained optimisation follows the same procedure as in $\S$\ref{broydensec} under the substitution $\Delta \rho^{\text{in}}_n \rightarrow S_n$ and  $\Delta R_n \rightarrow Y_n$, yielding
\begin{gather}\label{MSB1}
J_n = J_{n-1} + (Y_n - J_{n-1}S_n)(S^{\dagger}_n S_n)^{-1}S^{\dagger}_n, \\
J^{-1}_n = J^{-1}_{n-1} + (S_n - J^{-1}_{n-1}Y_n)(Y^{\dagger}_nY_n)^{-1} Y^{\dagger}_n.\label{MSB2}
\end{gather}
These define the multisecant variant of Broyden's first and second methods respectively, hereafter MSB1 and MSB2. Substituting Eqs$.$ (\ref{MSB1}) and (\ref{MSB2}) into the Newton update formula gives
\begin{align}\label{MLupdate}
\rho^{\text{in}}_{n+1} = \rho^{\text{in}}_{n} + P ( I - Y_{n} A_n ) R_n - S_{n} A_n R_n,
\end{align}
for some preconditioning matrix $P$, where $A_n$ defines MSB1 and MSB2 via
\begin{align}
A^{\text{MSB1}}_n \coloneqq ( S^{\dagger}_n Y_n)^{-1} S^{\dagger}_n, \\
A^{\text{MSB2}}_n \coloneqq (Y^{\dagger}_n Y_n)^{-1} Y^{\dagger}_n.
\end{align}
There is in fact an error in the Marks $\&$ and Luke paper here, where Eq$.$ (\ref{MLupdate}) has an incorrect sign that is propagated throughout the paper. 

\begin{figure*}[htbp]
\centering
\includegraphics[width=2.8in]{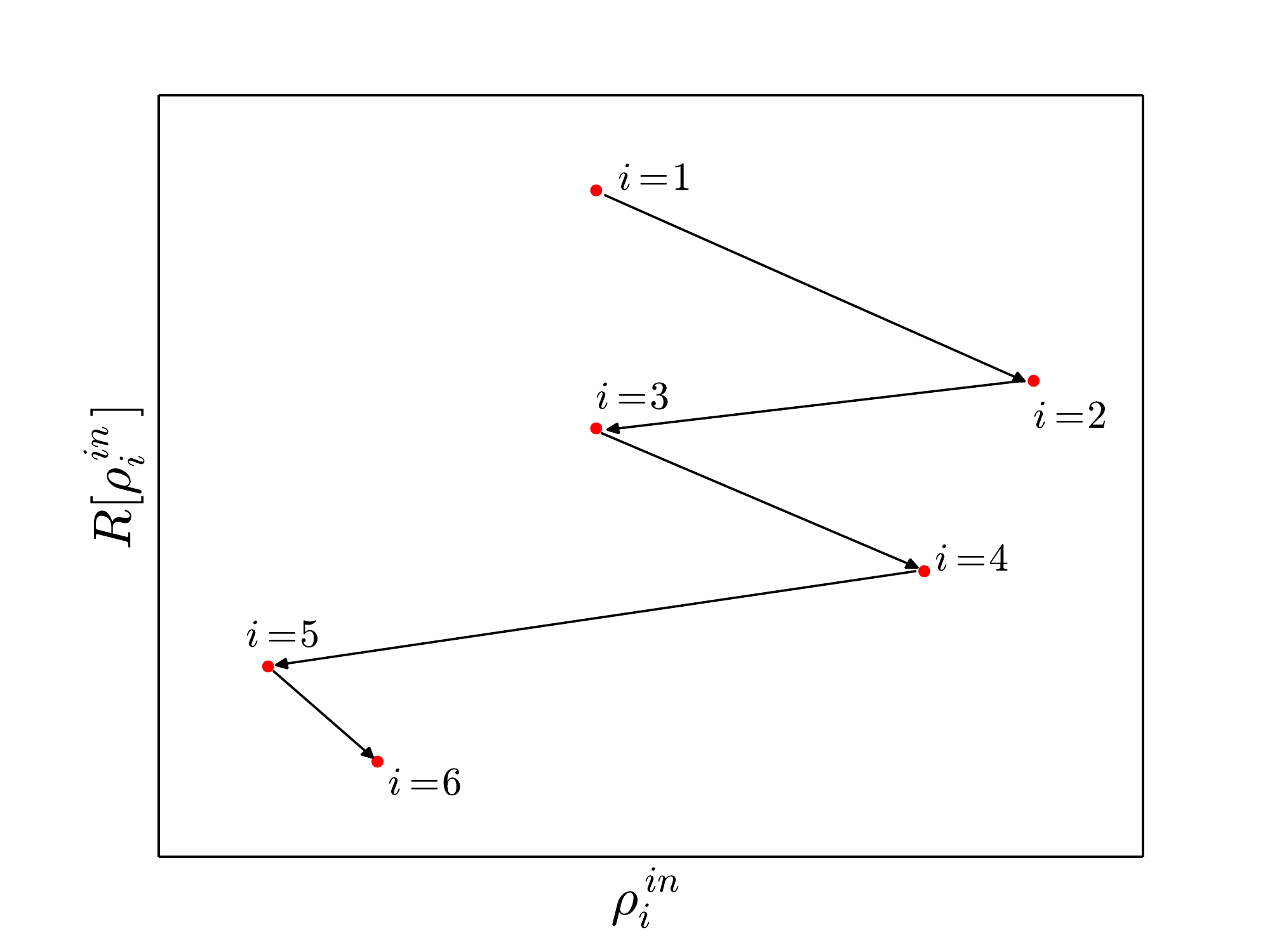}
\includegraphics[width=2.8in]{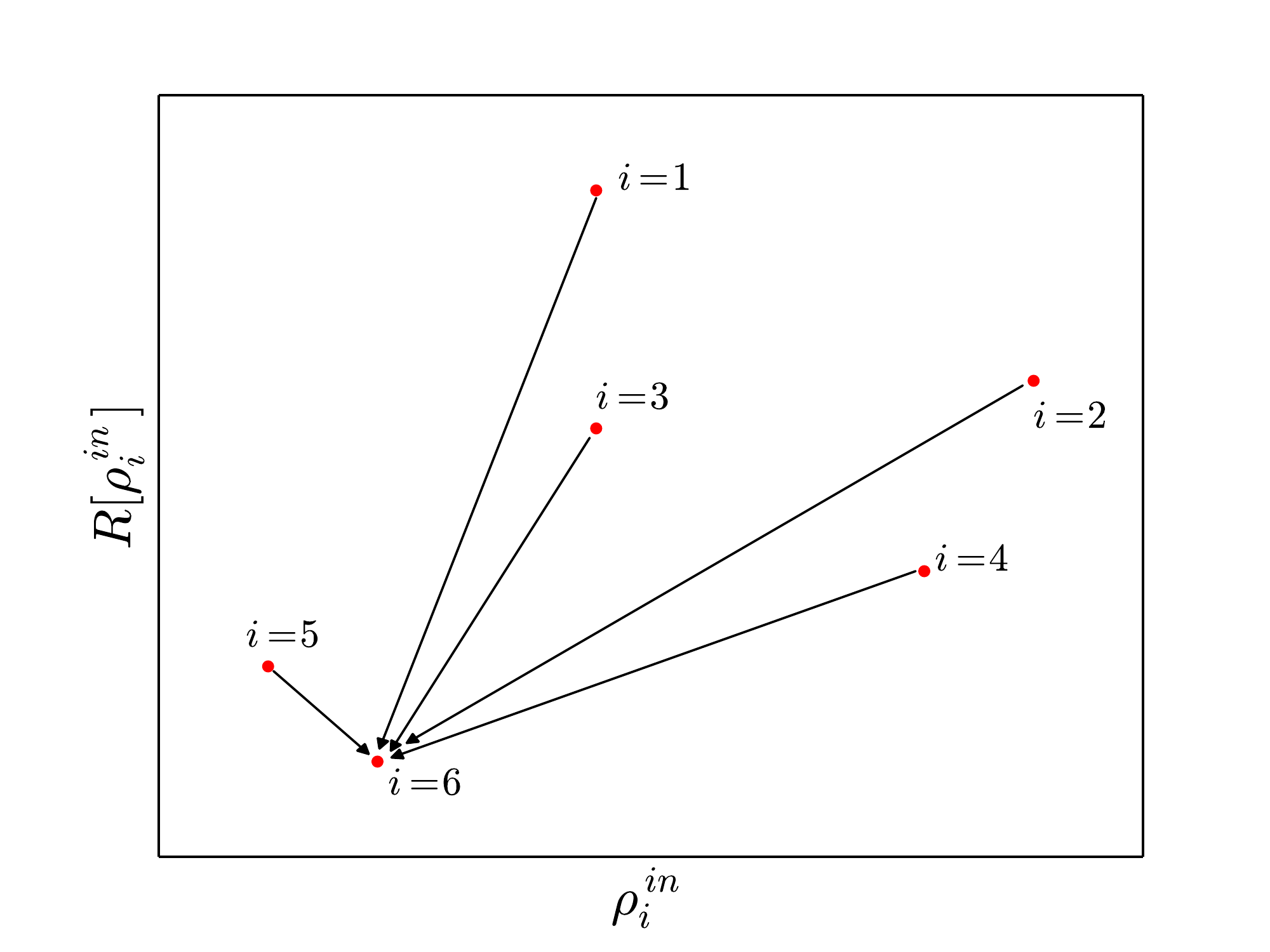}
\caption{Snapshot on the sixth iteration of a decreasing residual ($y$-axis) as $\rho^{in}_i$ ($x$-axis) moves through phase space toward convergence. Each node (\tikzcircle{2pt}) denotes the information $\{ R_i, \rho^{\text{in}}_i \}$. A secant condition is thus defined as the finite-difference equation between any two nodes in the phase space to be satisfied by the Jacobian, denoted by a black arrow. (Left) The secant conditions are \textit{fixed} and defined contiguously with increasing $i$. (Right) The re-centering of the Marks $\&$ Luke scheme now defines all secant conditions with respect to data on the most recent iteration, $i=6$. }
\label{fig:secant}
\end{figure*}

The update in Eq$.$ (\ref{MLupdate}) now needs to be adapted specifically for numerical implementation within KS DFT. First, note that there is a possibility of divergence due to the inversion in the definition of $A_n$. Similar to Pulay's DIIS, divergence occurs if the iterative densities are not strongly linearly independent, or when the residual becomes too small close to convergence. The latter can be treated by including some form of regularisation in the inversion. This is done using \textit{Tikhonov regularisation}, i.e$.$ a scalar parameter $\gamma$ is added to the diagonal of the argument of the inversion\footnote{Marks $\&$ Luke suggest a value of $\gamma=10^{-4}$ for the regularisation parameter. This value will also be used in the remainder of this work.}. Furthermore, the scheme is normalised by introducing the vector
\begin{align}
\Psi_n = \text{diag}(||\Delta R_{m-n}||^{-1}, \text{ } \dots \text{ }, ||\Delta R_{n}||^{-1} ),
\end{align}
which simply acts to scale the scheme, and will not change the underlying behaviour. With these numerical considerations in place, the update matrices take the form
\begin{gather}
A^{\text{MSB1}}_n \coloneqq \Psi_n (\Psi_n S^{\dagger}_n Y_n \Psi_n + \gamma I)^{-1} \Psi_n S^{\dagger}_n, \\
A^{\text{MSB2}}_n \coloneqq \Psi_n (\Psi_n Y^{\dagger}_n Y_n \Psi_n + \gamma I)^{-1} \Psi_n Y^{\dagger}_n.
\end{gather}
The final contribution of Marks $\&$ Luke to the scheme was to redefine the density and residual history matrices in the following way. First, notice that the current definitions of $S_n$ and $Y_n$ in Eqs$.$ (\ref{S1}) and (\ref{Y1}) populate the $i^{\text{th}}$ row with the density and residual differences defined as
\begin{gather}
 \Delta \rho^{\text{in}}_{i} =  \rho^{\text{in}}_{i} -  \Delta \rho^{\text{in}}_{i-1},\\
\Delta R_i = R_i - R_{i-1}.
\end{gather}
That is, the iterative Jacobian is required to satisfy the set of secant conditions \textit{as they were defined in the past iterations}, see Fig$.$ (\ref{fig:secant}) (left). There is no \textit{a priori} reason to suggest that this set of secant conditions provides better information than a set of secant conditions centred around the \textit{current} iteration, i.e$.$ redefine
\begin{gather}
 \Delta \rho^{\text{in}}_{n,i} \coloneqq  \rho^{\text{in}}_{n} -  \Delta \rho^{\text{in}}_{i},\\
\Delta R_{n,i} \coloneqq R_n - R_{i},
\end{gather} 
at every iteration for all $i \in [n-m-1,n-1]$, Fig$.$ (\ref{fig:secant}) (right). This re-centring of the secant condition at every iteration amounts to treating the history of data as sample points in a phase space, rather than than the contiguous path of all secant conditions cycled through in the history toward convergence.  

Lastly, a note will be given on the differences between above scheme, and the scheme originally presented by Marks $\&$ Luke in Ref$.$ \citep{ML}. Namely, the Marks $\&$ Luke scheme was implemented in \textsc{wien2k}, an all-electron, \textit{augmented} planewave approach\footnote{Augmented planewave (APW) implementations involve some mix of atomic orbital basis functions, and planewaves.} \citep{wein2k}. As this implementation involves a partially real space basis set, density mixing is also done in real space. Thus, the Marks $\&$ Luke methodology was re-derived with complex densities in mind, resulting in the conjugate-transposes defined above. Moreover, the treatment of core electrons in \textsc{wien2k} leads to an entirely different preconditioning approach that is not applicable to density mixing in \textsc{castep}. As such, the dielectric preconditioning of $\S$\ref{precond} was adapted appropriately for the Marks $\&$ Luke scheme.

\section{Periodic Pulay}

A generalisation of Pulay's DIIS has been proposed by Banerjee \textit{et al}$.$ in Ref$.$ \citep{PP}, hereafter referred to as the Periodic Pulay mixing scheme. The concept behind Periodic Pulay is as follows. Traditional Pulay mixing involves an extrapolation over the subspace of iterative densities, thus finding the best (least squares) fit to the converged density that this subspace allows. If one element of the subspace were particularly `close' to the self-consistent solution, the DIIS would weight this element in the basis  higher than other elements. Indeed, if the converged density $\rho^*$ were a member of the subspace, the DIIS would weight this element with unity, and zero the rest. Naturally, the efficacy of the DIIS depends entirely on the sample of information in the history. One can notice therefore that \textit{linear} mixing will contribute a different type of information to the history than a DIIS step. That is, a linear step tends to be efficient for \textit{either} low or high wavelength Fourier components of the density, but \textit{not both}, depending on the linear mixing parameter. Ill-conditioning of the exact dielectric prevents the linear step from being efficient in and of itself. However, linear mixing has the potential add information to the history that perhaps would not have been explored in DIIS steps, by nature of the step itself not being efficient. Thus, the idea of the Periodic Pulay method is to perform linear mixing steps in between DIIS steps, in hopes that the extrapolation over the history becomes more efficient. The algorithm to implement this is as follows.

\noindent\fbox{%
    \parbox{\textwidth}{%
\textbf{Algorithm 1}: Periodic Pulay Mixing

\vspace{0.3cm}

\indent \textbf{Input}: $\rho^{\text{in}}_1$, $\sigma$, $tol$, $k$

\vspace{0.3cm}

\indent \textbf{Output}: $\rho^{\text{in}}_n = \rho^{\text{out}}_n = \rho^*$

\vspace{0.3cm}

\begin{itemize}

\item \textbf{while} $||R_i|| < tol $

\hspace{0.5cm} $R_i = \rho^{\text{out}}_i - \rho^{\text{in}}_i$

\hspace{0.5cm} Add $\{ \rho^{\text{in}}_i, \rho^{\text{out}}_i \}$ to history

\hspace{0.5cm} \textbf{if} $(i+1)/k \in \mathbb{N}$

\hspace{1cm} $\rho^{\text{in}}_{i+1} \gets \text{ DIIS } $

\hspace{0.5cm} \textbf{else}

\hspace{1cm} $\rho^{\text{in}}_{i+1} \gets \rho^{\text{in}}_i + \sigma R_i $

\hspace{0.5cm} \textbf{end if}

\item \textbf{end while}

\end{itemize}
	}
}

In words, one takes $k-1$ linear steps between every DIIS step, where the scaling of the linear step is controlled by $\sigma$. In the limit $k \rightarrow \infty$ the scheme becomes linear mixing, whereas $k=1$ defines Pulay mixing. Ref$.$ \citep{PP} suggests the algorithm performs best with $k \sim 3$ and $\sigma \sim 0.1$; results for this scheme are given in $\S$\ref{resPP}.

\section{Susceptibility Model}
\label{smodelmeth}

Dielectric models such as Kerker form, or simple extensions to the Kerker form, were shown to be inadequate for treating inhomogeneous systems. This is largely due to an inability of the model, and implementations of the model, to include a systematic input-dependence.
Any input-dependence is required to be included explicitly by appropriately adjusting input parameters. Even with a perfect parameter set, there is no possibility for treating inhomogeneity. This aim of this section two-fold. First, the complications in constructing the Kerker preconditioner in real space will be discussed. The difficulty of including an inhomogeneous $\textbf{r}$-dependence in the susceptibility will thus become apparent. To remedy this, a framework will be detailed whereby one can propose an $\textbf{r}$-dependent model of the susceptibility, and use this to precondition the self-consistent cycles in a computationally efficient manner. This will allow one to implicitly include inhomogeneity via this \textbf{r}-dependence, while circumventing the computational expense of constructing the real space Kerker preconditioner. Finally, two preliminary, physically motivated models to be tested in $\S$\ref{precondmodelres} will be presented.

The difficulty in constructing the Kerker form in real space is highlighted well in Refs$.$ \citep{mann,RSK,review111}. To give a brief overview, one seeks the dielectric response of the HEG when subject to a perturbation in the charge density, $\delta \rho^{\text{induced}}(\textbf{r})$ (caused, for example, by placement of a test charge in the system). Under the assumptions of TF theory, this perturbation in the density results in a \textit{constant} and \textit{local} perturbation to the electrostatic field in the medium,
\begin{align}
\delta \rho^{\text{induced}}(\textbf{r}) = \frac{k_{\textsc{tf}}^2}{4 \pi}  \delta v(\textbf{r}),
\end{align}
which is to say the susceptibility is constant. Here, $\delta v(\textbf{r})$ is introduced as the \textit{change} in the potential that will propagate throughout the medium as a result of the perturbation in the density. One can thus solve the relevant Poission equation to determine how the medium will respond to this perturbation,
\begin{gather}
\nabla^2 \delta v(\textbf{r}) = -4 \pi \delta \rho^{\text{induced}}(\textbf{r}) = - k_{\textsc{tf}}^2  \delta v(\textbf{r}), \\
\implies \delta v(\textbf{r}) \sim \frac{e^{-k_{\textsc{tf}} |r|}}{|r|}.\label{yukawa}
\end{gather}
Therefore, a test body of charge $q$ placed at $\textbf{r}$ will not, as one might expect, result in the medium experiencing the Coulomb potential for all $\textbf{r}' \neq \textbf{r}$. Instead, the Coulomb potential is \textit{screened} by a factor which decays exponentially away from $\textbf{r}$ over some characteristic length scale $k_{\textsc{tf}}^{-1}$. This defines the \textit{Yukawa potential}, Eq$.$ (\ref{yukawa}). To capture the dielectric response of the HEG in real space hence requires a \textit{non-local} integral over the Yukawa kernel\footnote{Alternatively, one can solve a modified Poission equation on a real space grid, as is done in Ref$.$ \citep{RSK}.}. This is, in fact, the form in which the Kerker preconditioner was originally proposed by Manninen \textit{et al}. in Ref$.$ \citep{mann}. The non-local nature of the dielectric renders real space Kerker preconditioning inefficient. However, if the mixing were to be done in real space, one can still \textit{apply} the operators in Fourier space. A computationally efficient application of the Kerker preconditioner to a real space object is as follows. First, an accelerated mixing scheme is used to determine the unpreconditioned update in real space, $J_n^{-1}R[\rho^{\text{in}}_n(\textbf{r})] \coloneqq b(\textbf{r})$. Then, the preconditioned update, $\delta \rho_n(\textbf{r})$, solves the linear system, 
\begin{align}
\Big( I - K_c \chi_0 \Big) \delta \rho_n(\textbf{r}) = b(\textbf{r}). \label{rskk}
\end{align} 
Since $\chi_0$ is constant and $K_c$ is diagonal in Fourier space, the solution to this linear system becomes straight forward. One simply performs a Fourier transform to obtain $\tilde{b}(\textbf{G})$, inverts the dielectric, applies it to $\tilde{b}(\textbf{G})$, and then inverse Fourier transforms to obtain $ \delta \rho_n(\textbf{r})$. This procedure thus defines Kerker preconditioning when mixing is done in real space. 

The difficulty of including inhomogeneity now becomes apparent. Instead of the susceptibility being constant across the entire unit cell, leading to $\chi_0(\textbf{r},\textbf{r}') = \gamma \delta(\textbf{r}-\textbf{r}')$, the following work now seeks to include a local $\textbf{r}$-dependence to capture the inhomogeneity, $\chi_0(\textbf{r},\textbf{r}') = \psi(\textbf{r}) \delta(\textbf{r} - \textbf{r}')$. The object $\psi(\textbf{r})$ now defines some model for the susceptibility, yet to be derived. Therefore, within this framework, the susceptibility is \textit{not} diagonal in Fourier space. Hence, the dielectric is diagonal in \textit{neither} space, meaning the solution to the linear system in Eq$.$ (\ref{rskk}) has become non-trivial. 

In order to solve Eq$.$ (\ref{rskk}) with an \textbf{r}-dependent susceptibility, one could start by constructing $\tilde{\chi}_0(\textbf{G},\textbf{G}')$ directly from $\psi(\textbf{r})$. This turns out to be inefficient for various reasons, but will serve to highlight key features of the framework, and will hence be detailed. If the linear response functions are defined in the following way,
\begin{gather}
\delta \rho^{\text{out}}(\textbf{r}) = \int d\textbf{r}' \text{ } \chi_0(\textbf{r},\textbf{r}') \delta v_h(\textbf{r}'), \\
\delta \tilde{\rho}^{\text{out}}(\textbf{G}) = \int d\textbf{G}' \text{ } \tilde{\chi}_0(\textbf{G},\textbf{G}') \delta \tilde{v}_h(\textbf{G}'),
\end{gather} 
then it is clear that $\tilde{\chi}_0(\textbf{G},\textbf{G}') \neq \mathcal{F}[\chi_0(\textbf{r},\textbf{r}')]$ as a two-dimensional convolution is required. However, one can notice that if the real space response is local, then the Fourier space response becomes a \textit{Toeplitz matrix}. That is, for some local susceptibility $\psi(\textbf{r})$, the corresponding Fourier space response is
\begin{align}
\tilde{\chi}_0(\textbf{G},\textbf{G}') = \tilde{\psi}(\textbf{G}-\textbf{G}')
\end{align}
where $\tilde{\psi}(\textbf{G}) = \mathcal{F}[\psi(\textbf{r})]$, which defines the functional form of a Toeplitz matrix. Once the Toeplitz matrix of $\tilde{\psi}(\textbf{G})$ has been constructed, the linear system in Eq$.$ (\ref{rskk}) now defines a \textit{Toeplitz system}, which can be solved in $\mathcal{O}( N_G \log^2 N_G )$ operations \citep{toeplitz}. However, this method remains impractical for large systems, both in terms of memory (constructing the Toeplitz matrix) and computational (solving the resultant linear system) overhead\footnote{Initially, this was the method used to construct the preconditioner and the difficulties discussed became apparent in testing.}. 

An efficient method to solve Eq$.$ (\ref{rskk}) is thus required, suitable for $N_G \sim \mathcal{O}(10^5)$ calculations. This was done by utilising an iterative diagonaliser, and the algorithm is outlined as follows.

\noindent\fbox{%
    \parbox{\textwidth}{%
\textbf{Algorithm 2}: Solve $\Big( I - K_c \chi_0 \Big) \tilde{x}(\textbf{G}) = \tilde{b}(\textbf{G})$

\vspace{0.5cm}

\indent \textbf{Input}: The vector to be preconditioned $J_n^{-1}R[\rho^{\text{in}}_n(\textbf{G})] \coloneqq \tilde{b}(\textbf{G})$

\vspace{0.5cm}

\indent \textbf{Output}: The preconditioned vector $\tilde{x}(\textbf{G})$

\vspace{0.5cm}

\begin{itemize}

\item Initial guess: $\tilde{x}_0(\textbf{G})$

\item Store the initial guess: $\tilde{x}'_0(\textbf{G}) \leftarrow \tilde{x}_0(\textbf{G})$

\item Transform to real space:  $x_0(\textbf{r}) \leftarrow \mathcal{F}^{-1}[\tilde{x}_0(\textbf{G})]$

\item Apply the susceptibility:  $x_0(\textbf{r}) \leftarrow \int d\textbf{r}' \psi(\textbf{r}') \delta(\textbf{r} - \textbf{r}') x_0(\textbf{r}')$

\item Transform to Fourier space: $ \tilde{x}_0(\textbf{G}) \leftarrow \mathcal{F}[ x_0(\textbf{r}) ]$

\item Apply the Coulomb kernel: $\tilde{x}_0(\textbf{G}) \leftarrow K_c(\textbf{G}) \tilde{x}_0(\textbf{G})$

\item Compute and store the final result: $ \tilde{x}_0(\textbf{G}) \leftarrow \tilde{x}'_0(\textbf{G}) - \tilde{x}_0(\textbf{G})$

\item Construct the residual: $r_0 = \tilde{x}_0(\textbf{G}) - \tilde{b}(\textbf{G})$

\item \textbf{if} $||r_0|| < tol$

\hspace{0.5cm} exit: $\tilde{x}'_0(\textbf{G})$ is the solution

\item \textbf{else}

\hspace{0.5cm} update via an iterative algorithm: $\tilde{x}_1(\textbf{G}) = \tilde{x}'_0(\textbf{G}) + \dots$

\item Repeat until convergence of the iterative solver

\end{itemize}
	}
}

This algorithm requires two FFTs and $2N_G$ floating point multiplications per step in the iterative solver (ignoring solver specific constructions). If the iterative solver converges in $\mathcal{O}(10)$ iterations, there is an additional computational overhead of approximately twenty FFTs and $\sim 20N_G$ floating point multiplications. This is negligible as \textsc{castep} executes $\mathcal{O}(10^6)$ FFTs in a typical calculation \citep{phil222}. Moreover, the storage requirements are minimal as only $\mathcal{O}(1)$ $N_G$-length vectors are needed. Therefore, the methodology presented defines a computationally and memory efficient procedure for applying an inhomogeneous preconditioner in Fourier space. The preconditioner is now entirely specified by the susceptibility diagonal, $\psi(\textbf{r})$. Before determining some preliminary models for $\psi(\textbf{r})$, a note on the implementation of the above algorithm can be given. Namely, an iterative solver must be chosen  that  is appropriate for the above scheme. Initially, the (diagonally preconditioned) Jacobi and Gauss-Siedel solvers were tested \citep{jacg}. However, these were eventually found to be insufficient. The reason for this is that both the Jacobi and Gauss-Siedel solvers require the linear system to be \textit{diagonally dominant}\footnote{In fact, Gauss-Siedel requires the system to be diagonally dominant \textit{or} symmetric and positive definite. However, while the KS dielectric presented here is manifestly symmetric, it is not necessarily positive definite. Positive-definiteness depends on the model of $\psi$.}. The optimal value for the susceptibility \textit{in vacuo} is $\chi = \psi = 0$ (detailed shortly). Therefore, for input systems containing above some threshold amount of vacuum, the  Jacobi and Gauss-Siedel solvers tended to diverge. This was remedied by implementing a conjugate gradient algorithm, the method for which can be found in Ref$.$ \citep{jacg}. Note, however, that this is still a preliminary treatment designed for testing purposes only. A truly optimal solver here would require a more careful consideration of the linear system at hand. For example, to improve robustness, one could consider implementing a Krylov-subspace based method such as the GMRES detailed in $\S$\ref{honmen}. As it stands, the vanilla conjugate gradient method was found to be sufficient in converging every model and input system tested in $\S$\ref{precondmodelres}.

One now simply needs to derive an inhomogeneous model for the susceptibility, and run it through the above scheme. The first model for  $\psi$ considered in this work is derived by noticing that the exact susceptibility for the vacuum is zero. That is, the dielectric response of the vacuum is unity. Therefore, one can define $\psi$ in the following way,
\begin{align}
\psi(\textbf{r}) =& 0 \text{ for } \rho^{\text{out}}_n(\textbf{r}) \text{ in vacuum} \\ \nonumber
\psi(\textbf{r}) =& \gamma  \text{ for } \rho^{\text{out}}_n(\textbf{r}) \text{ in bulk},
\end{align}
such that the bulk is treated with the Kerker model, and the vacuum is treated exactly. It is expected that this model will perform well for surface or slab input systems, and perform approximately equivalently to Kerker preconditioning for bulk systems. This model will hereafter be  referred to as the \textit{vacuum scaled} (VS) susceptibility model.

The second model, inspired in part by the work of Scherlis \textit{et al}. \citep{woma}, parametrises the susceptibility with a smooth density dependence of the following form
\begin{align}
\psi(\textbf{r}) = \alpha \Bigg( \frac{\rho^{\text{out}}_n(\textbf{r})}{ \rho_0} \Bigg)^{\beta},\label{densparam}
\end{align}
for some $\alpha$, $\beta$, and normalisation $\rho_0$. Ideally, the parameters $\alpha$ and $\beta$ are derived from physical theory to give a well-motivated, unique model. This is done in the following work by considering an inhomogeneous variant of TF theory. That is, the TF screening wavevector is typically derived in terms of a constant HEG density $\rho^{\textsc{heg}}_0$, given by
\begin{align}
k_{\textsc{tf}}^2 = 4 \Bigg( \frac{3 \rho^{\textsc{heg}}_0}{\pi} \Bigg)^{1/3},
\end{align}
where $\rho^{\textsc{heg}}_0$ is determined by the HEG density parameter, $r_s$ (see Ref$.$ \citep{tf1}). One can now take a similar approach to $\S$\ref{precond} by identifying the susceptibility with the TF screening wavevector, and considering the following inhomogeneous extension,
\begin{align}
\psi(\textbf{r}) = 4 \Bigg( \frac{ 3 \rho^{\text{out}}_n(\textbf{r})}{ \pi \rho_0} \Bigg)^{1/3},
\end{align}
hereafter referred to as the \textit{inhomogeneous Thomas-Fermi} (ITF) model. Again, it is expected that this model will provide an improved treatment for slab and surface systems due to it possessing the correct vacuum limit, $\psi \rightarrow 0$ as $\rho^{\text{out}}_n \rightarrow 0$. Furthermore, a density dependence of this form is expected to provide a more nuanced treatment of the charge screening resulting from `rapid' (on the scale of the unit cell) variations in the density -- so-called `local field effects' \citep{ashcroftmer}. However, this preconditioner is not expected to improve ill-conditioning due material interfaces that include both metallic and insulating regions. This is because appropriate preconditioning would require determination of the electronic behaviour of each region.


\chapter{Results and Discussion} 

\label{results} 

\lhead{5. \emph{Results}} 

Density mixing schemes will be judged and compared on two criteria: efficiency and reliability. Efficiency will be measured using iteration count, rather than wall clock time, as they are equivilent in the present context. That is, when performing density mixing, the time expended is assumed to be negligible compared to that of diagonalising the Hamiltonian for all schemes presented. This was explicitly detailed in $\S$\ref{smodelmeth} for the susceptibility model, and a similar logic follows directly for the Periodic Pulay and Marks $\&$ Luke schemes. Reliability (or robustness) is defined as in the following work as the ability of a mixing scheme to converge \textit{at all}, not considering the number of iterations it takes to achieve this convergence.  Clearly, then, one needs to quantify how one weights reliability versus efficiency for density mixing. For a \textit{default} method, reliability is, in some sense, more important than efficiency. Thus, any of the schemes to follow that are able to converge a wide range of systems without failure will be compared favourably against less reliable, but perhaps more efficient, schemes. That is, of course, as long as this decrease in efficiency is not by any more than an $\mathcal{O}(1)$ factor (at risk of density mixing becoming a worse EDFT). 

With a rubric in place to judge competing methods, one can now question what improvements are realistically achievable in order to set a standard for the following schemes. The absolute best-case-scenario for denisty mixing methods would be to match the efficiency and reliability of a scheme utilising an exact susceptibility computation. It was derived in $\S$\ref{precond} that a scheme preconditioned with an exact susceptibility would converge in just one iteration from an initial guess within the linear response regime. In practice however, the initial guess is not within the linear response regime, especially for increasingly electronegative systems. For example, Magnesium Oxide consists of an ionic lattice of Mg$^{2+}$ and O$^{2-}$, but the initialisation of the charge density will not reflect this bonded structure, instead initialising the lattice with a more covalent character. Accounting for this, the best-case-scenario would be convergece across all input systems in 3-7 iterations, as concluded by Ref$.$ \citep{hij}. Increasingly sophisticated and tuned denisty mixing schemes should thus approach the limit of 3-7 iterations across all input systems.

\section{The Test Suite}

In order to properly assess the efficiency and robustness of any changes made to \textsc{castep}, a suite of representative test systems is constructed and motivated. In creating this test suite, a particular emphasis was put on sloshing prone systems (see $\S$\ref{precond}). This includes: systems with a large vacuum gap, interfaces, supercells, and metals. Furthermore, the test suite also contains certain types of difficult-to-converge systems that were not detailed in previous sections. This includes: far-from-equilibrium systems (such as those generated by AIRSS \citep{AIRSS}) and transition metal compounds. The full set of materials included in the test suite, and motivations for each, are provided in Appendix$.$ \ref{apptestsuite}.

\begin{figure*}[htbp]
\centering
\includegraphics[width=5.8in]{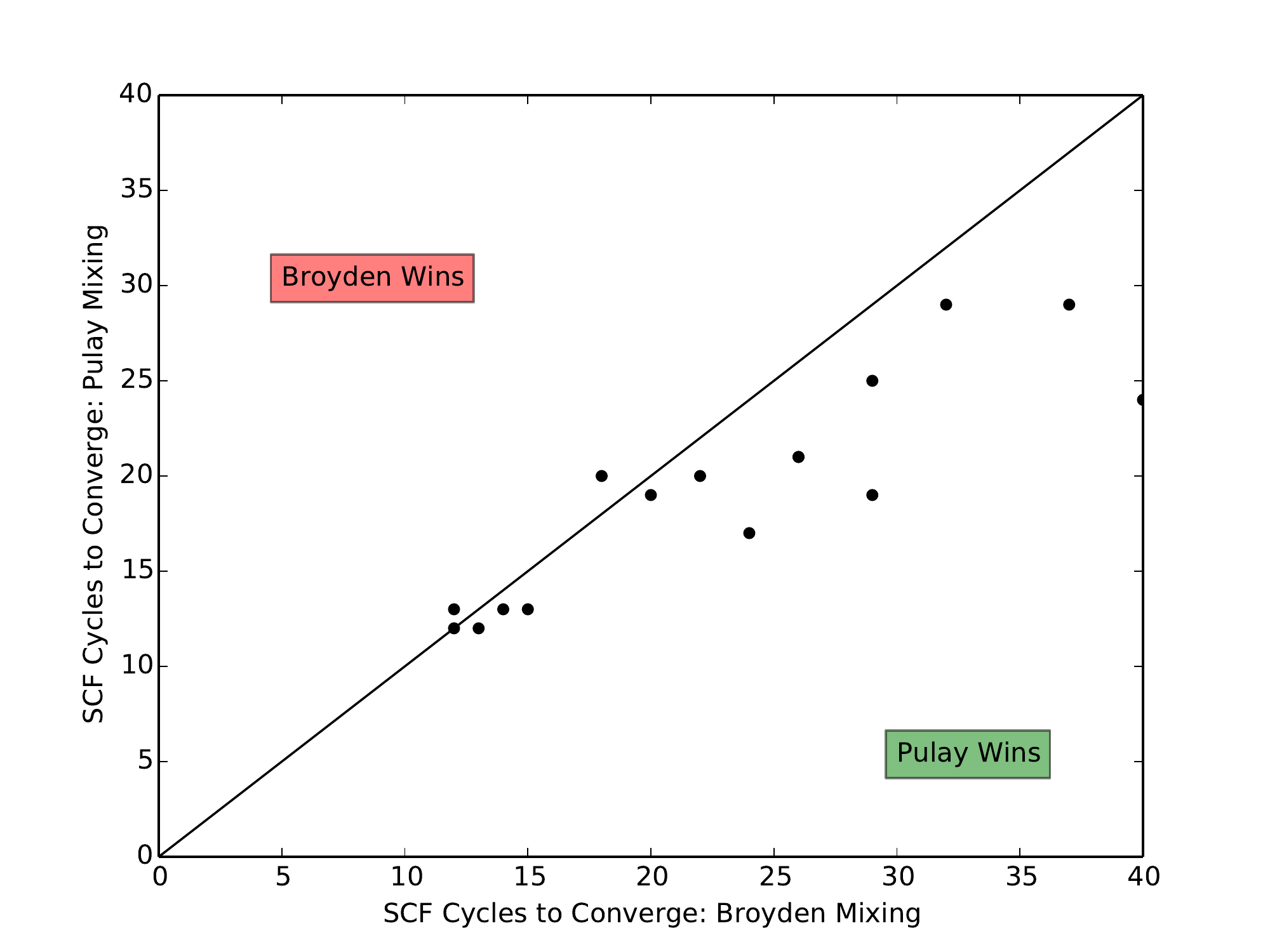}
\caption{Comparison of the SCF cycles to converge for Broyden's and Pulay's methods using the test suite. Data points in the lower triangle signify an improvement for Pulay's method, whereas data in the upper triangle signify an improvement for Broyden's method, as labelled. Data points lying on the edge of the plot have \textit{failed to converge}. The `net winner', by total number of SCF cycles over the whole test suite, is coloured green.}
\label{fig:bvp}
\end{figure*}

When a representative comparison is done utilising the full test suite, the following DFT parameters are used.
\begin{itemize}
\item[--] Exchange-correlation functional: Perdew, Burke $\&$ Ernzerhof (PBE) \citep{PBE1}
\item[--] Cut-off energy: converged to approximately $0.03 \text{ eV/atom}$
\item[--] Spin unpolarised
\item[--] Monkhorst-Pack $k$-point spacing: 0.05\AA$^{-1}$
\end{itemize}
Any analysis on individual systems will also use these parameters, unless stated otherwise. As an exemplar, the test suite can be used to compare the two standard density mixing methods in \textsc{castep}: Broyden's method (Johnson's variant \citep{johnson}), and Pulay's method, both Kerker preconditioned. From Fig$.$ (\ref{fig:bvp}) it can be seen that Pulay mixing results in a significant improvement over Broyden mixing. For one system, a far-from-equilibrium phase of TiK, Broyden mixing even diverges where Pulay mixing is able to converge\footnote{In fact, after some preliminary analysis on AIRSS output, it was often found to be the case that Pulay mixing was significantly more robust than Broyden mixing.}. The total number of SCF cycles taken to converge the test suite can be calculated. By this method, Pulay mixing was found to be approximately $14\%$ more efficient than Broyden mixing. Moreover, Pulay mixing appears to be more robust, although a methodical study of robustness here is yet to be done. The caveat, however, is that Pulay's method requires the inversion of a potentially singular matrix when close to convergence. This can be fixed with relative ease by including some form of regularisation that does not interfere with the scheme. Therefore, it is suggested that perhaps Pulay mixing be made the default density mixer in \textsc{castep}.

\section{Marks $\&$ Luke}
\label{ML1111}

The results of Fig$.$ (\ref{fig:bvp}) demonstrate that the most effective default scheme in \textsc{castep} for achieving self-consistency is Kerker preconditioned Pulay mixing. Hence, Kerker preconditioned Pulay mixing will be taken as the reference method of comparison for all subsequent testing. First, the Marks $\&$ Luke scheme of $\S$\ref{ML1111} will be tested for a fixed maximum history length of $m=30$. Furthermore, unless stated otherwise, the default \textsc{castep} Kerker preconditioner parameters will be used: $\alpha=0.8$ and $|G_0|=1.5$\AA$^{-1}$. 

\begin{figure*}[htbp]
\centering
\includegraphics[width=5.5in]{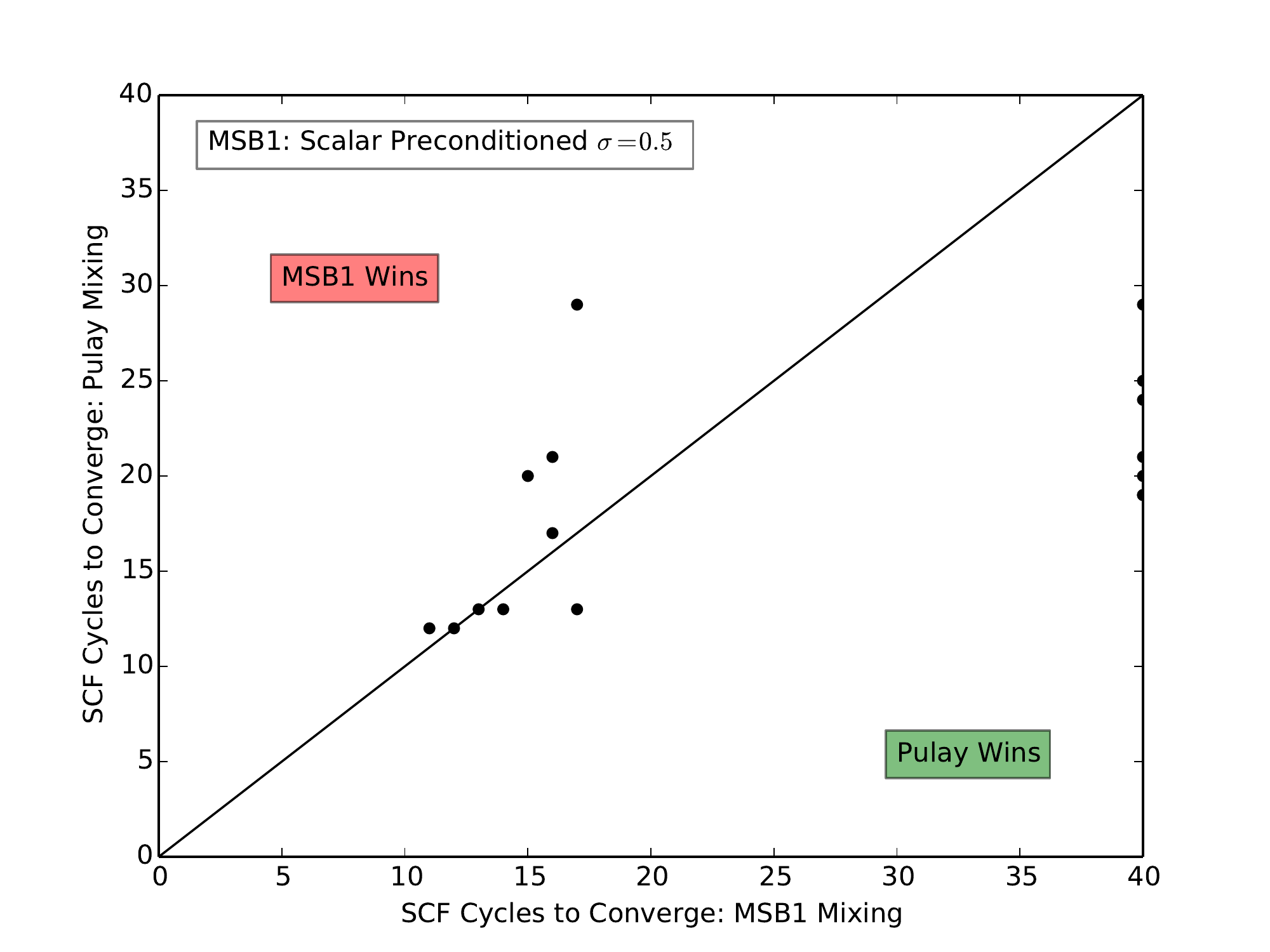}
\includegraphics[width=5.5in]{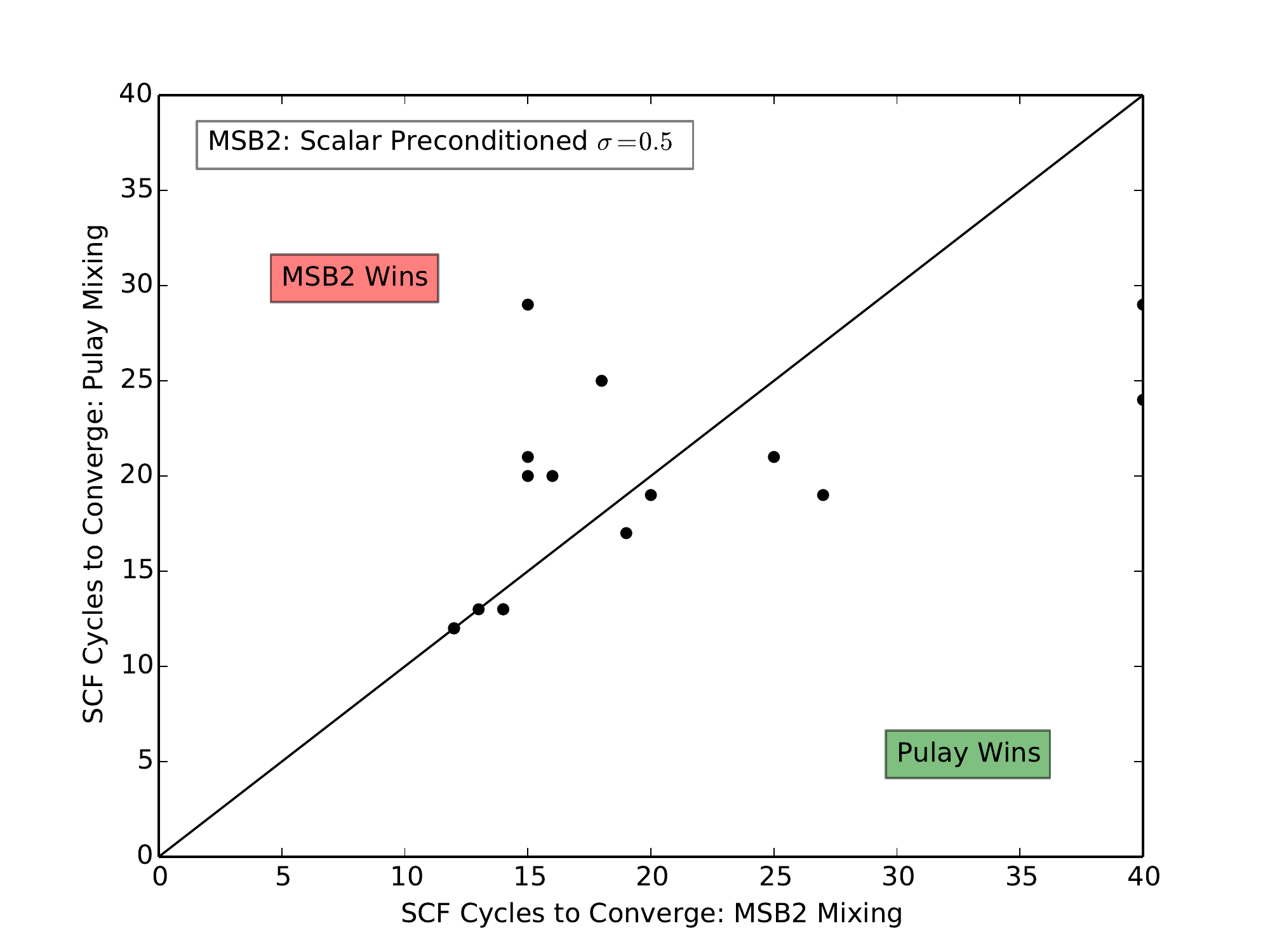}
\caption{Comparison of linear preconditioned MSB1 (top) and MSB2 (bottom) to Kerker preconditioned Pulay mixing over the full test suite.}
\label{fig:scalarpre}
\end{figure*}

In the original work of Marks $\&$ Luke, the scheme was preconditioned using a (safeguarded) linear scaling parameter. Therefore, an initial test of both MSB1 and MSB2 is provided preconditioned with a simple linear scaling parameter in an attempt to match the results of Marks $\&$ Luke, thus verifying the correctness of the implementation. This amounts to setting the preconditioning matrix $P$ in Eq$.$ (\ref{MLupdate}) equal to a scalar constant, $\sigma$. After a preliminary parameter analysis, it was found that $\sigma = 0.5$ gave the most reliable and efficient results across the test suite\footnote{Unfortunately, there is no clear map between the value of the linear scaling parameter used in \textsc{wien2k}, and the value used in the following work. This is due to inherent implementation specific differences between \textsc{wien2k} and \textsc{castep}.}. As expected, Fig$.$ (\ref{fig:scalarpre}) demonstrates that neither MSB1 nor MSB2 in this form are able to provide a systematic improvement over \textit{dielectric preconditioned} Pulay mixing. However, the nature of linear preconditioning as discussed in $\S$\ref{precond} is highlighted well here. That is, one can achieve far more efficient convergence for simple insulating systems, at the cost of robustness across the whole test suite. It appears that MSB2 is both more robust, and more efficient than MSB1, in agreement with the conclusions of Marks $\&$ Luke. The MSB2 scheme fails to converge for two systems, whereas the MSB1 scheme fails to converge six systems and is $10\%$ less efficient than MSB2. Interestingly, this behaviour possesses similarities to elementary implementations of Broyden's method. That is, B2 (`Broyden's bad method') involving updating the inverse Jacobian directly is found superior to B1 (`Broyden's good method') \citep{limmem}. The results of Fig$.$ (\ref{fig:scalarpre}) confirm the correctness of the implementation, and an attempt at dielectric preconditioning can now be made.

\begin{figure*}[htbp]
\centering
\includegraphics[width=5.5in]{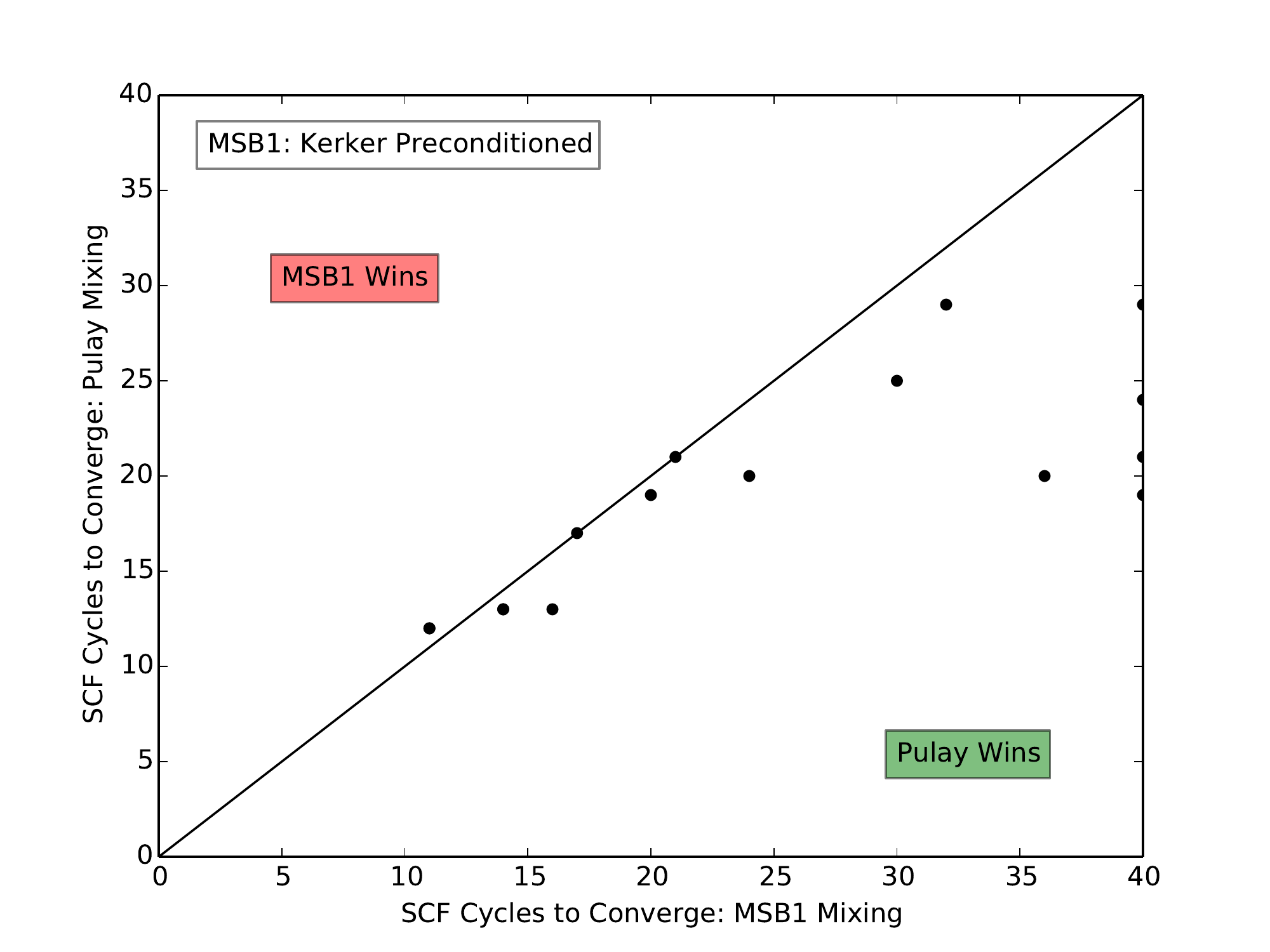}
\includegraphics[width=5.5in]{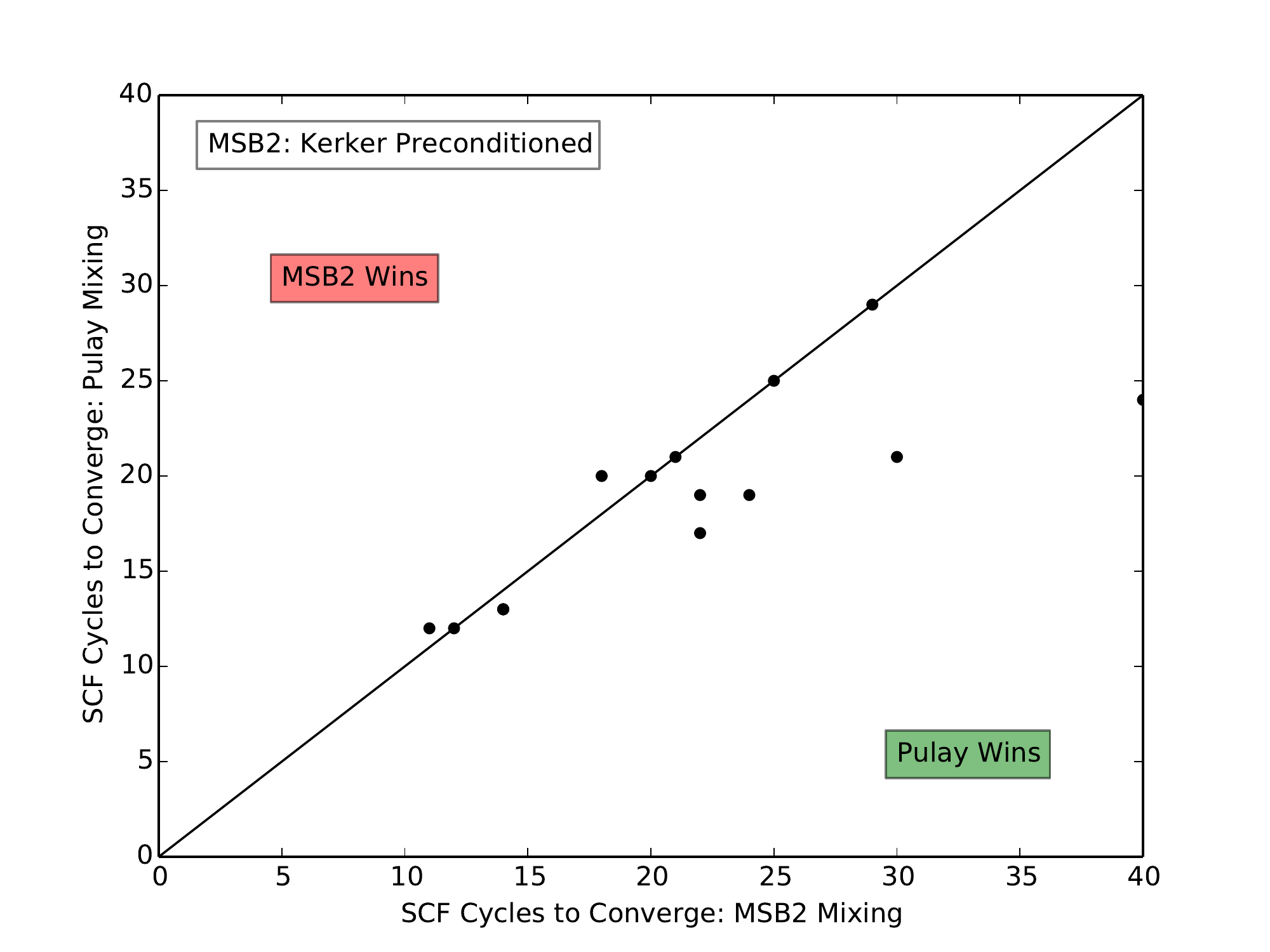}
\caption{Comparison of Kerker preconditioned MSB1 (top) and MSB2 (bottom) to Kerker preconditioned Pulay mixing over the full test suite.}
\label{fig:kerkerpre}
\end{figure*}

\begin{figure*}[htbp]
\centering
\includegraphics[width=5in]{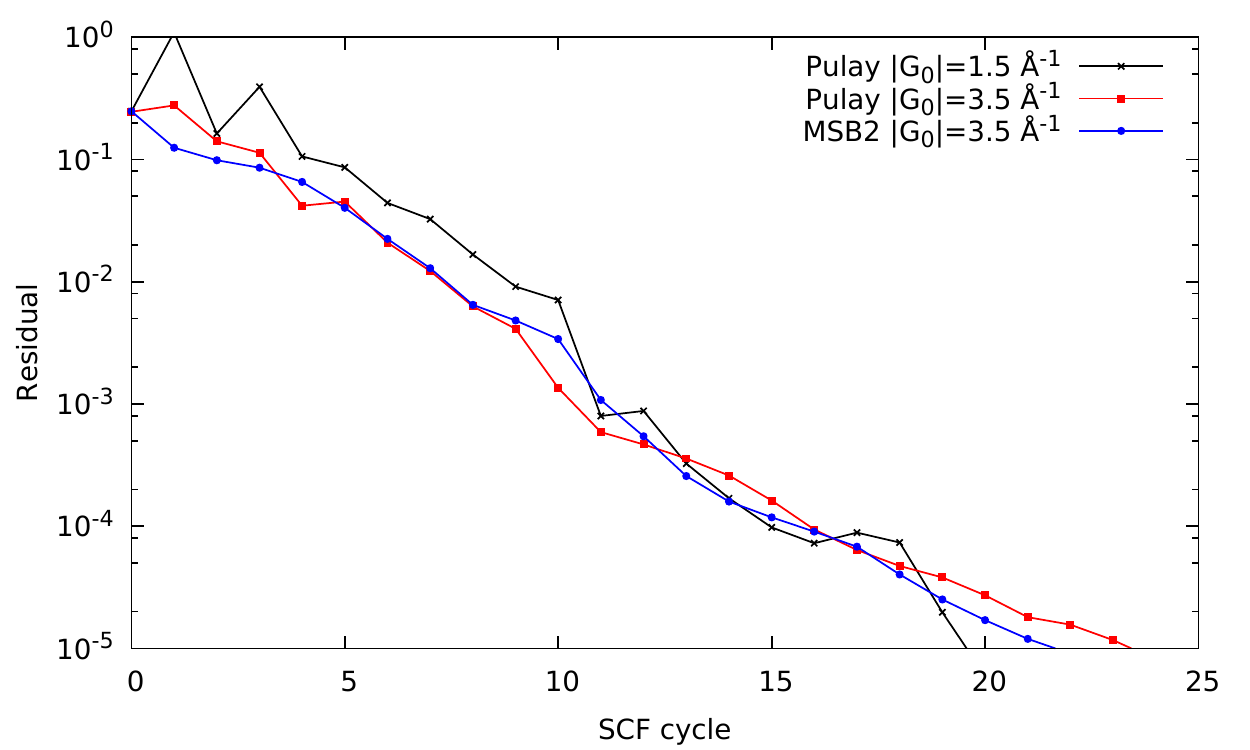}
\caption{Convergence of a far-from-equilibrium TiK phase using MSB2 and Pulay mixing. The schemes were made competitive by an appropriate adjustment of $|G_0|$.}
\label{fig:TiK1}
\end{figure*}

\begin{figure*}[htbp]
\centering
\includegraphics[width=5in]{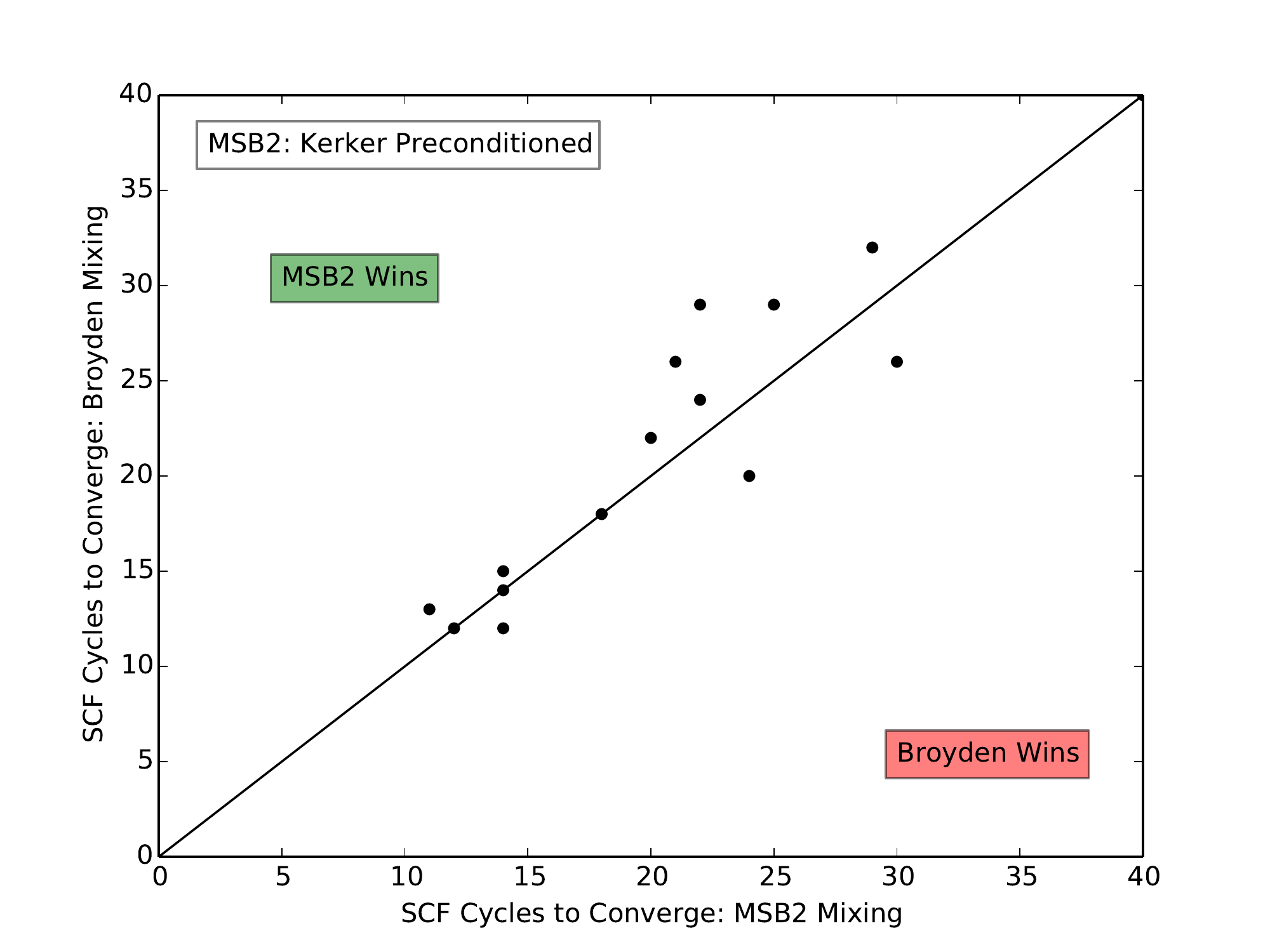}
\caption{Comparison of Kerker preconditioned MSB2 to Kerker preconditioned Broyden mixing over the full test suite.}
\label{msb2vb}
\end{figure*}

It is not obvious in the initial presentation of the Marks $\&$ Luke scheme how the update should be preconditioned for optimal performance. Hence, the scheme was reassessed taking particular care to track the position of the initial guess Jacobian. As dielectric preconditioning should iteratively pre-multiply the initial guess Jacobian, this analysis highlighted the correct placement for the preconditioning matrix $P$ in the Marks $\&$ Luke update, Eq$.$ (\ref{MLupdate}). As such, MSB1 and MSB2 are tested using the diagonal Kerker form for $P$, the results for which are displayed in Fig$.$ (\ref{fig:kerkerpre}). As expected, dielectric preconditioning leads to a performance improvement across the whole test suite, and MSB2 remains superior to MSB1. However, Kerker preconditioned MSB2 mixing is $9\%$ less efficient across the whole test suite than Pulay mixing, and still fails to converge one system -- a far-from-equilibrium phase of TiK. This is perhaps a shortcoming of the preconditioner, rather than the underlying method, as Fig$.$ (\ref{fig:TiK1}) demonstrates. That is, the TiK phase was found to be prone to band sloshing, requiring the calculation of many additional, unoccupied bands to reach converge. This suggests the material has a large density of states at the Fermi level, and is thus also prone to Coulomb sloshing. As such, MSB2 and Pulay mixing were performed on TiK with an appropriately adjusted Kerker parameter of $|G_0|=3.5$\AA$^{-1}$ to better account for Coulomb sloshing in the preconditioner. In doing so, MSB2 and Pulay mixing were found to converge at the same rate (Fig$.$ (\ref{fig:TiK1})), suggesting that, with an optimised set of Kerker parameters, MSB2 could rival the efficiency and robustness of Pulay mixing. Furthermore, it can be noted that MSB2 is in fact an improvement over Broyden mixing as implemented in \textsc{castep}, Fig$.$ (\ref{msb2vb}).

However, the results presented here demonstrate that no meaningful improvement can be made using the Marks $\&$ Luke scheme over Pulay mixing as implemented in \textsc{castep}.

\begin{figure*}[htbp]
\centering
\includegraphics[width=5in]{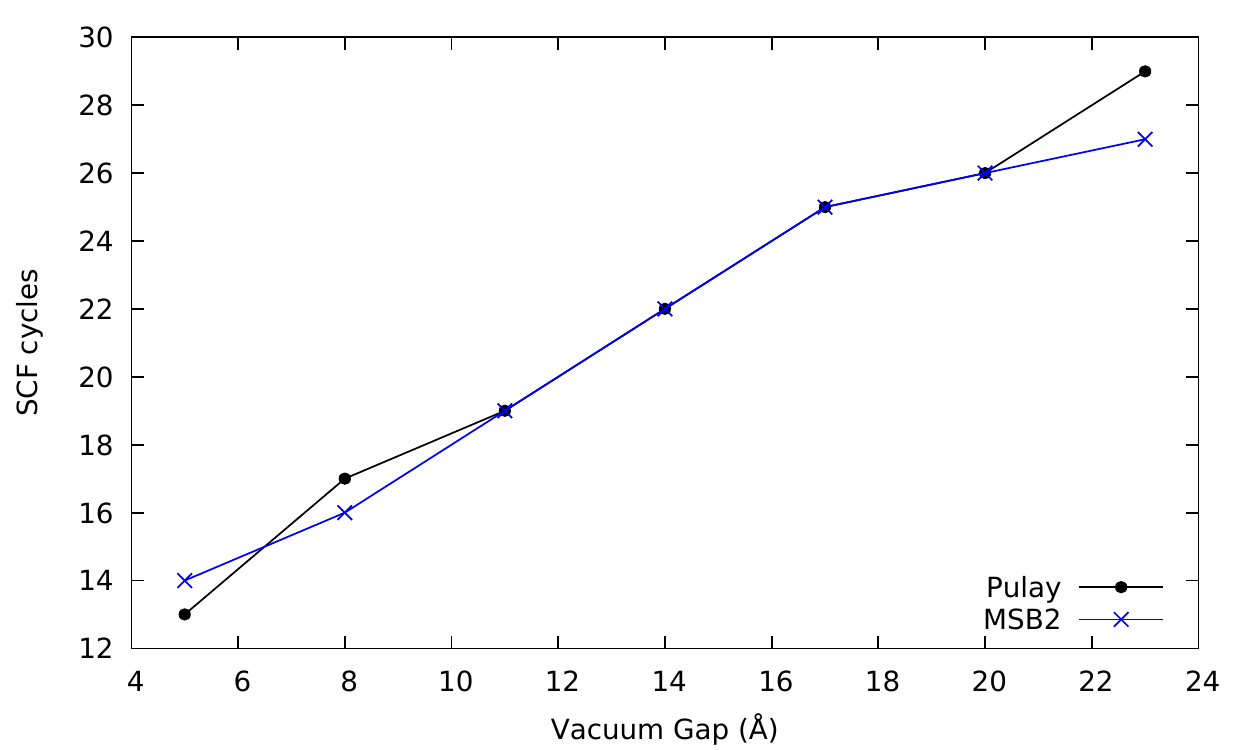}
\caption{SCF cycles to converge with increasing vacuum distance for an Aluminium surface using MSB2 and Pulay mixing.}
\label{fig:vactest1}
\end{figure*}

Lastly, it is worth also comparing the schemes for an \textit{increasingly} ill-conditioned system whereby no new physics is introduced as a result of this ill-conditioning. To do this, an Aluminium surface is converged with an increasing vacuum gap of between 5\AA{} and 23\AA, Fig$.$ (\ref{fig:vactest1}). Additional SCF cycles taken to converge \textit{after} the total energy of the structure no longer changes is purely a numerical artefact, rather than a result of the physical system at hand. In principle, then, a sophisticated scheme should be able to remove this numerical artefact completely, resulting in little-to-no additional iterations with increasing vacuum (as, for example, in Ref$.$ \citep{ADFT}). Both schemes suffer from a near linear relationship in iterations to converge versus vacuum gap, suggesting ill-conditioning of this form is best dealt with in the preconditioner. 

\newpage

\section{Periodic Pulay}
\label{resPP}

\begin{figure*}[htbp]
\centering
\includegraphics[width=5in]{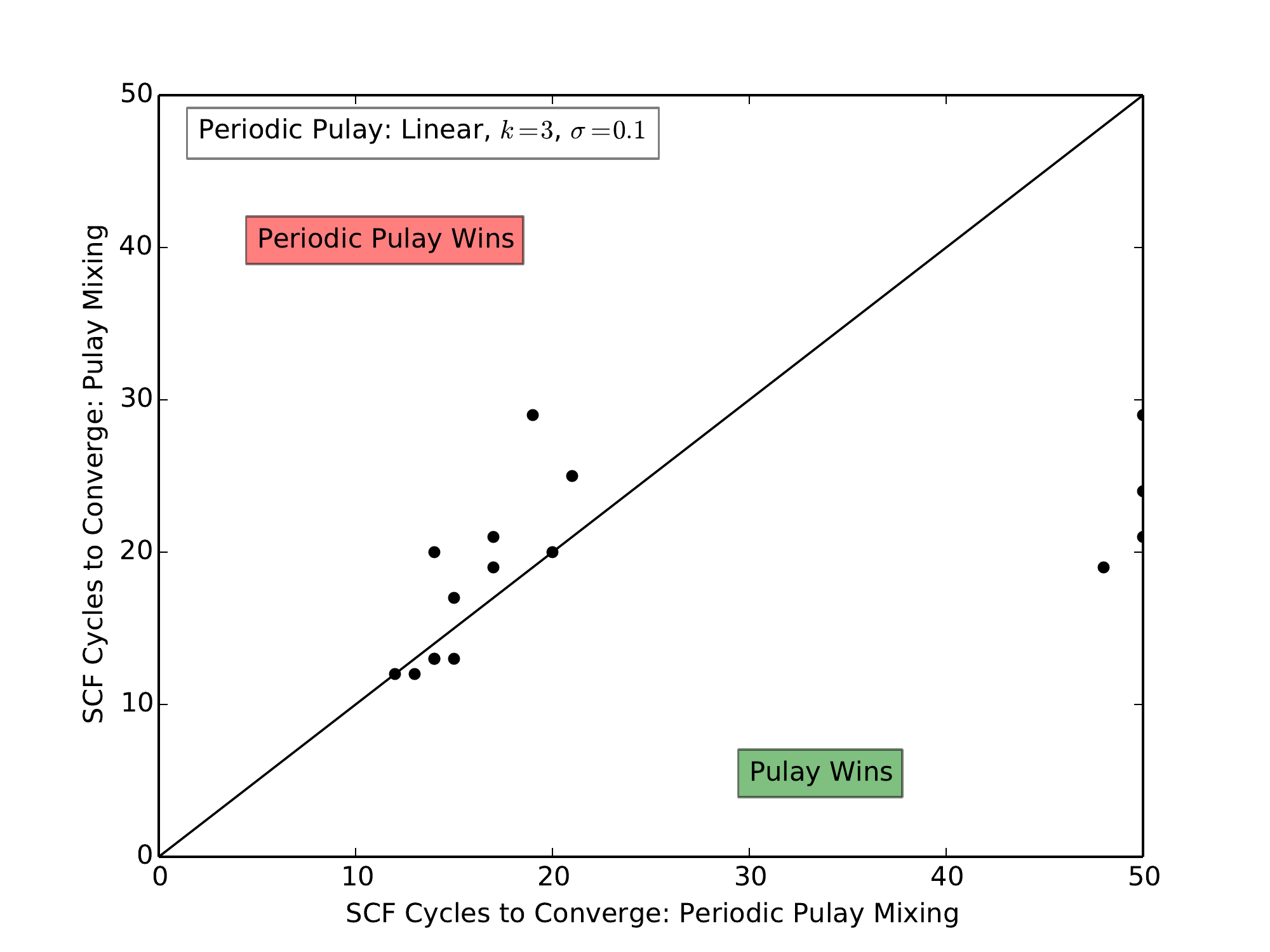}
\caption{Comparison of Periodic Pulay (two linear steps per DIIS step) and Pulay mixing.}
\label{2L1P}
\end{figure*}

Periodic Pulay mixing is tested in a similar fashion to the Marks $\&$ Luke scheme, by first comparing the method against Kerker preconditioned Pulay mixing across the test suite. The parameters suggested by Banerjee \textit{et al}. were found to give approximately the best compromise between DIIS steps and linear steps. That is, two linear steps per DIIS step, $k=3$, with a linear step length $\sigma=0.1$, Fig$.$ (\ref{2L1P}).

\begin{figure*}[htbp]
\centering
\includegraphics[width=5in]{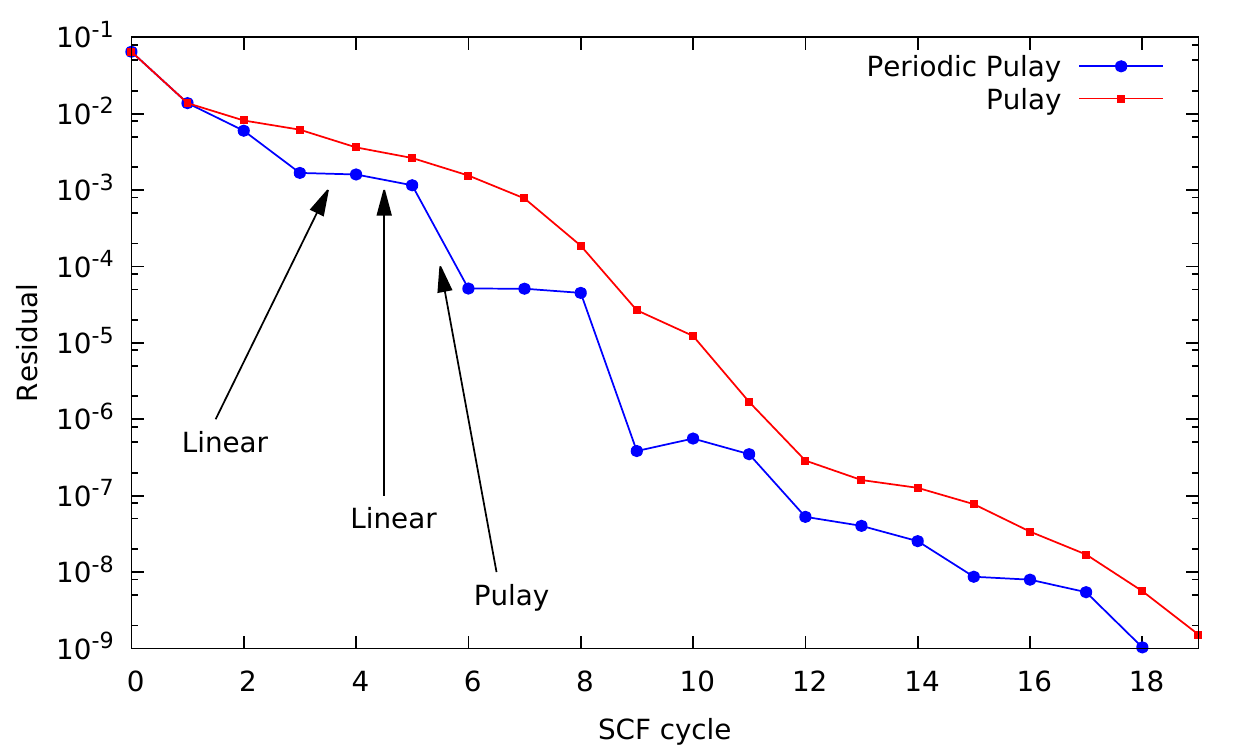}
\caption{Comparison of Periodic Pulay (two linear steps per DIIS step) and Pulay mixing for an Aluminum surface.}
\label{2L1P_Al10A}
\end{figure*}

Counter to the conclusions of Banerjee \textit{et al}., Periodic Pulay mixing is found to be far less robust than Pulay mixing, resulting in a failure to converge for three systems. However, for simple systems (particularly insulating systems), Periodic Pulay mixing is found to accelerate convergence over standard Pulay mixing. This is not a proof of concept for the Periodic Pulay scheme \textit{per se}, as simple insulating systems are best converged with linear mixing anyway. Over the whole test suite, the improvement in efficiency is minor compared to the drastic decrease in reliability. Hence, Periodic Pulay mixing is not in a position to replace Pulay mixing in \textsc{castep}, as the gain from an improved DIIS extrapolation is not enough to overcome the inefficiency of the linear step. Interestingly, however, the concept of Periodic Pulay is displayed well by studying the convergence patterns of individual systems. First, Fig$.$ (\ref{2L1P_Al10A}) displays convergence of an Aluminum surface. 

\begin{figure*}[htbp]
\centering
\includegraphics[width=5in]{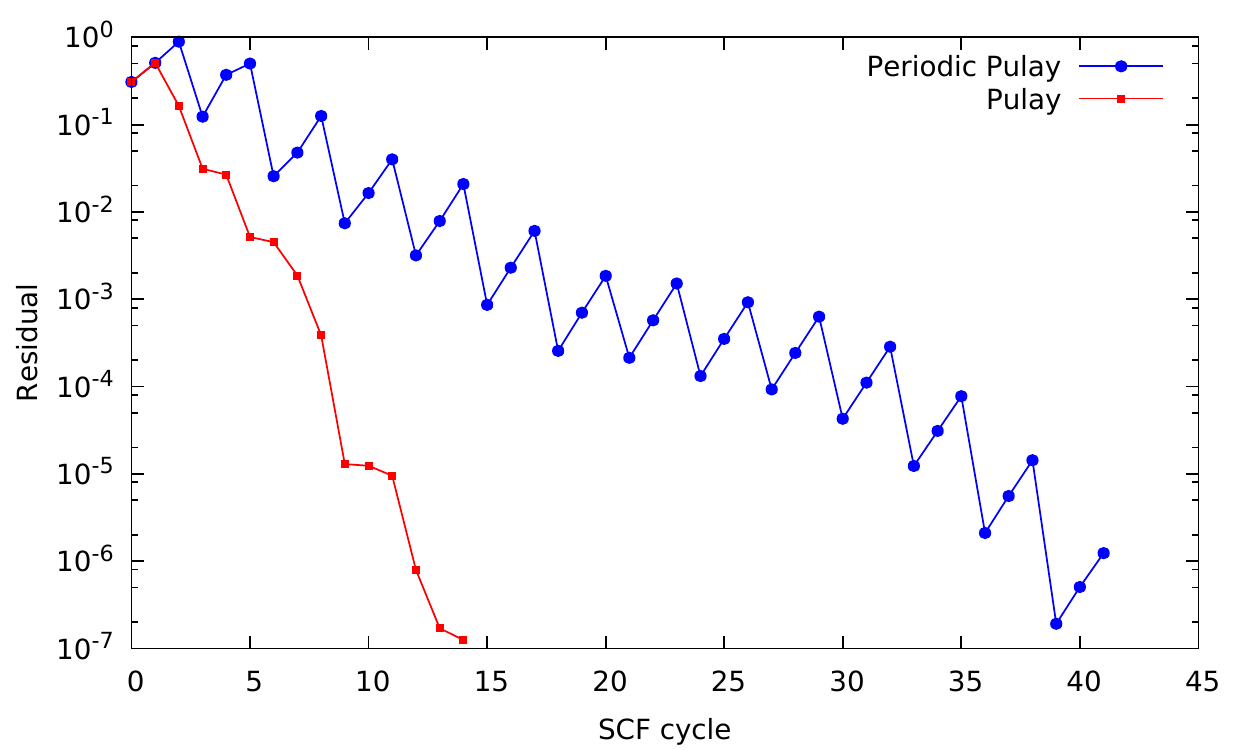}
\caption{Comparison of Periodic Pulay (two linear steps per DIIS step) and Pulay mixing for a far-from-equilibrium phase of LiCu.}
\label{2L1P_LiCu}
\end{figure*}

As predicted, the linear steps (labelled) contribute little toward reducing the residual themselves, but the DIIS extrapolation utilises this information extremely well. As such, the DIIS extrapolation is able to take a far more efficient step toward convergence than in regular Pulay mixing. This results in slightly accelerated convergence for the Aluminium surface. However, the drawback of Periodic Pulay is highlighted by studying the convergence behaviour of a more complex system -- a far-from-equilibrium phase of LiCu, Fig$.$ (\ref{2L1P_LiCu}).

\newpage

\begin{figure*}[htbp]
\centering
\includegraphics[width=5in]{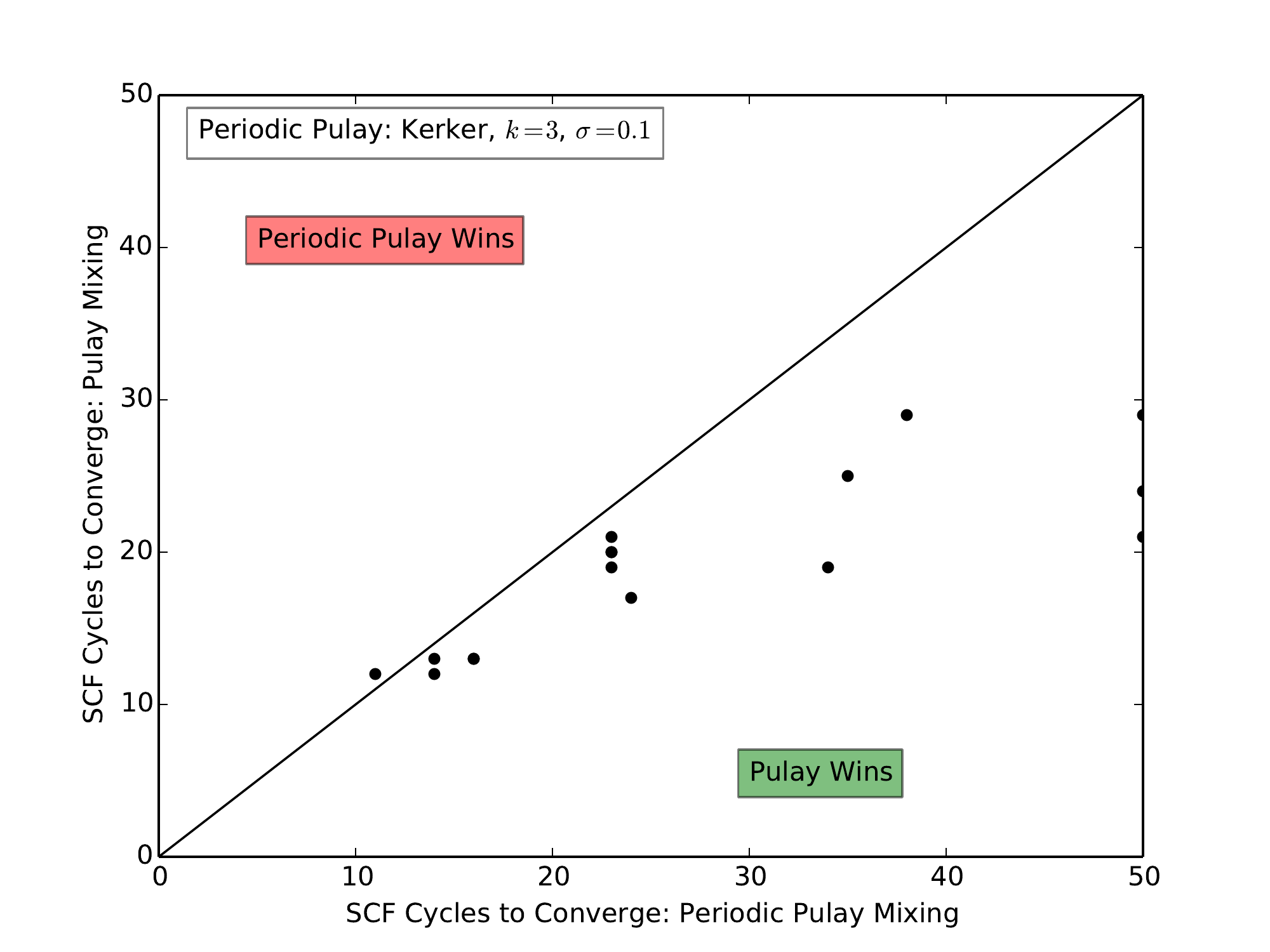}
\caption{Comparison of Periodic Pulay (two Kerker steps per DIIS step) and Pulay mixing.}
\label{2K1P}
\end{figure*}

The linear steps here act to hinder convergence so drastically that the net gain from the DIIS extrapolation is only just able to overcome the inefficiency of the linear steps. This results in a severe deceleration toward convergence compared to Pulay mixing. One might think of fixing this instability by utilising dielectric preconditioned linear steps rather than scalar preconditioned linear steps. As Fig$.$ (\ref{2K1P}) shows, this in fact results in further deceleration, and is less efficacious than using scalar preconditioned linear steps. The reason for this is that the scalar preconditioned linear steps were designed to provide the history with certain type of information that a sophisticated mixing scheme would not contribute. Conversely, dielectric preconditioned linear steps explicitly attempt to predict the behaviour of a sophisticated mixing scheme. Hence, the advantage of adding `non-sophisticated' information to the history is negated. This argument is heuristic in nature, and as Ref$.$ \citep{PP} notes, the mathematics behind Periodic Pulay mixing is poorly understood. Perhaps a more rigorous study of the mathematics would reveal how best to utilise this concept, which clearly works (to an extent) from an empirical standpoint.

\newpage 

\section{Susceptibility Model}
\label{precondmodelres}

Analysing the effectiveness of a dielectric model to precondition the SCF cycles requires some additional considerations to the previous testing. First, density mixing is now performed on the full reciprocal space grid. That is, the mixing sphere defined in $\S$\ref{dmincastep} is set equal to size of the $G$-vector grid, meaning density mixing is done for all Fourier components. Moreover, the mixing metric is also disabled as it can interfere with conclusions drawn regarding how effective a given dielectric model is at suppressing instabilities. In order to ensure the implementation of the framework in $\S$\ref{smodelmeth} is correct, an initial test was performed for the Kerker limit, $\psi(\textbf{r}) = \gamma$. The conjugate gradient solver was converged to near machine precision, and the framework indeed exactly reproduced the behaviour of the Kerker preconditioner.

\begin{figure*}[htbp]
\centering
\includegraphics[width=5in]{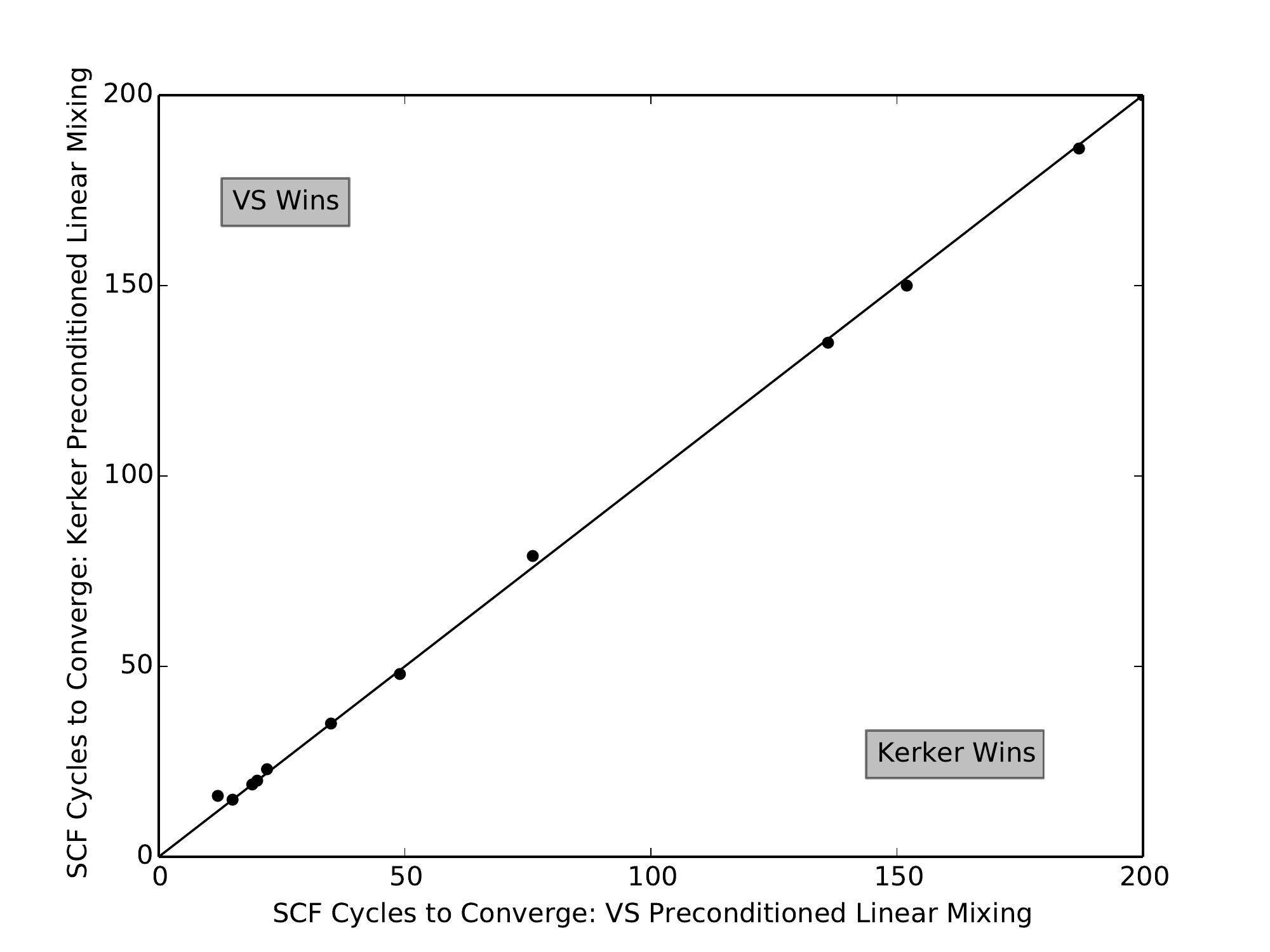}
\caption{Comparison of Kerker preconditioned linear mixing to VS preconditioned linear mixing over the test suite.}
\label{VSk}
\end{figure*}

The VS susceptibility model is first analysed. To do this, one can test the accuracy of \textit{just} the dielectric model across the test suite, without reference yet to accelerated mixing schemes. That is, compare VS preconditioned \textit{linear mixing} to Kerker preconditioned linear mixing across the test suite, Fig$.$ (\ref{VSk}).

As expected, simple dielectric mixing is far less efficient and robust compared to the accelerated mixing schemes of $\S$\ref{ML1111} and $\S$\ref{resPP}. Typical iteration counts are now $\mathcal{O}(100)$, with multiple failures from both dielectric models. Perhaps unexpectedly, the VS susceptibility model is unable to demonstrate an improved efficiency for \textit{any} system in the test suite when simple dielectric mixing is used. Instead, VS and Kerker preconditioned linear mixing perform approximately equally across the whole test suite, demonstrating the fact that the VS dielectric converges to the Kerker dielectric in the bulk limit. Notably however, there is little-to-no improvement for vacuum-containing input systems. The reason for this relatively clear; dielectric preconditioned linear mixing, or any method that can be rearranged as a quasi-Newton step, takes the general form, 
\begin{align}
\rho^{\text{in}}_{n+1} =  \rho^{\text{in}}_{n} - \epsilon^{-1}_{\text{model}} \Big( \rho^{\text{out}}_n - \rho^{\text{in}}_{n} \Big). \label{resss1}
\end{align} 
For the input vacuum region, defined as $\rho^{\text{in}}_n(\textbf{r}) = 0$, the KS map will return $\rho^{\text{out}}_n(\textbf{r}) = 0$ for the vast majority of the vacuum (this is not true close to the interface), leading to a zero residual. That is to say, the initial guess is exactly correct for most of the vacuum region, therefore treating the vacuum with an exact dielectric  will contribute little to the scheme, seen clearly in Eq$.$ (\ref{resss1}). Ill-conditioning due to the vacuum region is therefore equivalent to ill-conditioning due to an increased unit cell size. In order for the preconditioner to suppress this form of ill-conditioning, a different approach would be required to that presented in $\S$\ref{precond}. The work of $\S$\ref{precond} seeks to model the dielectric of the input material more accurately, rather than considering explicitly the numerical ill-conditioning as a result of increased unit cell size\footnote{This statement highlights the difference between numerical ill-conditioning and `physically sourced' ill-conditioning. That is, the exact dielectrics of a primitive cell and supercell of some material are formally equivalent. However, the supercell would still require more iterations to converge as a result of what is referred to here as `numerical ill-conditioning'.}.   

\begin{figure*}[htbp]
\centering
\includegraphics[width=5in]{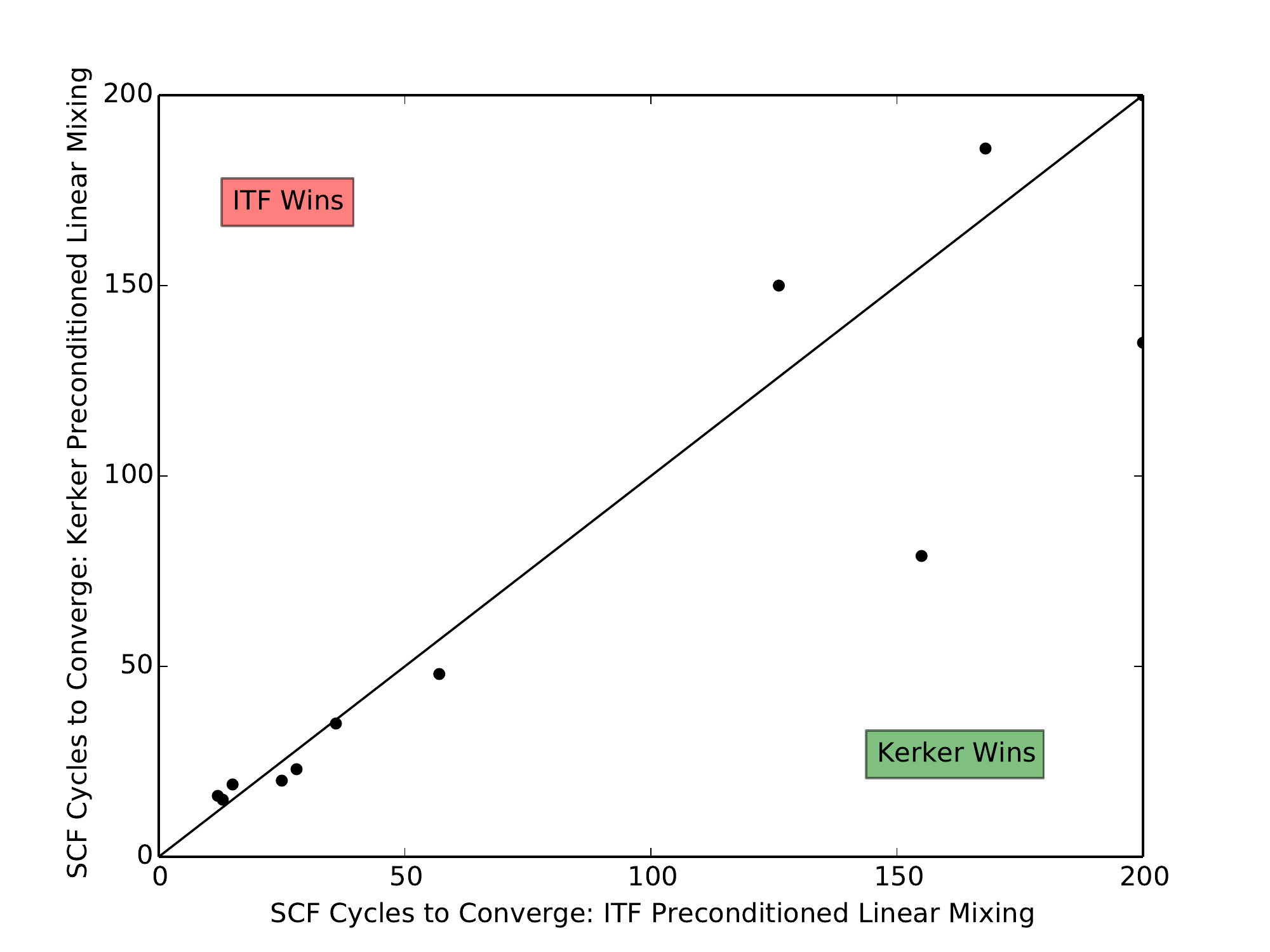}
\caption{Comparison of Kerker preconditioned linear mixing to ITF preconditioned linear mixing over the test suite.}
\label{ITFk}
\end{figure*}

\begin{figure*}[htbp]
\centering
\includegraphics[width=5in]{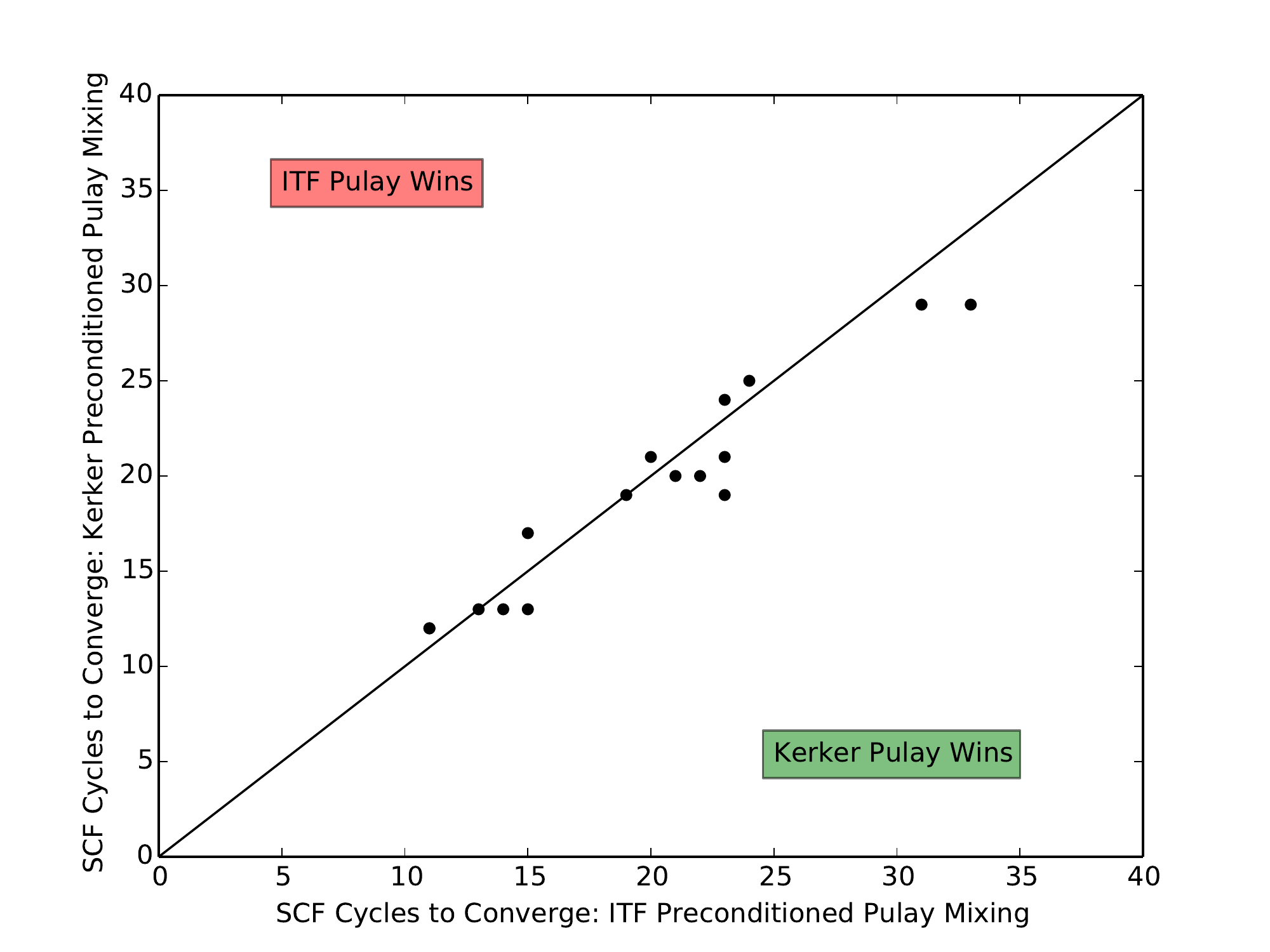}
\caption{Comparison of Kerker preconditioned Pulay mixing to ITF preconditioned linear mixing over the test suite.}
\label{ITFp}
\end{figure*}

It is clear from the above analysis that including a density dependence in the susceptibility will not remove ill-conditioning due to increased unit cell size. However, this density dependence is expected to provide a better treatment of the dielectric for regions where the initial guess is incorrect, thus requiring an accurate dielectric. The parametrisation of the density dependence studied will be that of the ITF susceptibility model, where Fig$.$ (\ref{ITFk}) displays ITF preconditioned linear mixing against Kerker preconditioned linear mixing.
As expected, there is much more variance in the iterations to converge than in the VS study. Particularly, the two largest improvements in Fig$.$ (\ref{ITFk}) for the ITF model (referring to nodes in the upper right of the ITF region) are graphene and an Aluminium surface, both containing a vacuum gap. Over the entire test suite, however, the ITF susceptibility does not provide a better model treatment of the exact dielectric than the Kerker form. This conclusion remains the same when the ITF susceptibility model is used to precondition accelerated mixing schemes such as Pulay mixing, Fig$.$ (\ref{ITFp}). When utilising an accelerated scheme, the preconditioners perform comparably, but the ITF parametrisation clearly leads to a systematic, net decrease in efficiency. This is likely due to the heuristic nature with which Thomas-Fermi theory was extended for inhomogeneous systems. However, as discussed, both models for $\psi$ presented here are preliminary in nature, and were intended to demonstrate the framework, rather than give a nuanced model treatment of the dielectirc. As such, an immediate avenue for further work will be to derive a rigorous, well-motivated model for the real space susceptibility (see $\S$\ref{furtherwork}). 

\begin{figure*}[htbp]
\centering
\includegraphics[width=5in]{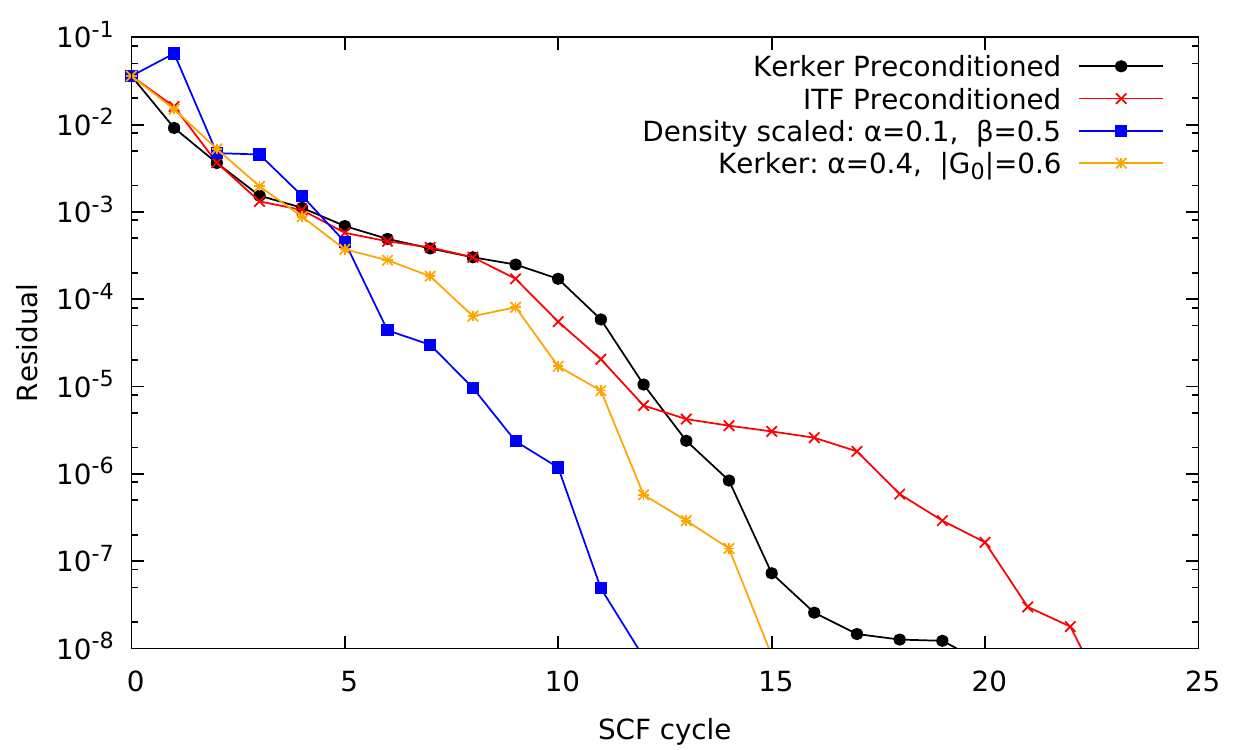}
\caption{Comparison of the convergence behaviour for an Aluminium surface using Pulay mixing with various preconditioners. First, Pulay mixing is preconditioned with default Kerker form and the ITF form. Second, both Kerker and the density scaled susceptibility model have optimised parameters for the input system.}
\label{dens_scaled_test}
\end{figure*}

The scope for improvement within the framework can be demonstrated by performing an elementary parameter analysis of the density scaling in Eq$.$ (\ref{densparam}), i.e$.$ optimise $\alpha$ and $\beta$ manually. Fig$.$ (\ref{dens_scaled_test}) illustrates the result of this parameter analysis, whereby setting $\alpha=0.1$ and $\beta=0.5$ leads to a two-fold increase in efficiency over unoptimised Kerker preconditioning for an Aluminium surface. Moreover, if the Kerker parameters are optimised in a similar fashion, the `peak performance' of the density scaled susceptibility is revealed better than that of optimised Kerker. This suggests there is potential for the efficiency of an inhomogeneous susceptibility model to be greater than that of the Kerker model, especially if a user is willing to optimise parameters\footnote{Although, a note regarding on-the-fly determined parameters will be given in $\S$\ref{furtherwork}.}. To conclude, both the VS and ITF models presented here lacked the depth required in order to produce an improvement over Kerker preconditioning. However, the framework was demonstrated to work, and the potential of the framework to provide an improved peak performance when both preconditioners are optimised was demonstrated.

\chapter{Conclusion} 

\label{conc} 

\lhead{6. \emph{Conclusion}} 

\section{Concluding Remarks}

This thesis has described and presented three separate approaches that attempt to improve efficiency and reliability in which self-consistency can be  attained in KS DFT implementations. These approaches were implemented within the planewave pseudopotential electronic structure software, \textsc{castep}. The default methods of \textsc{castep} were used as a benchmark for the proposed improvements of this thesis.

First, a multisecant variant of Broyden's methods for non-linear root finding were considered, referred to as MSB1 and MSB2. These schemes, based on the work of Marks $\&$ Luke, were made suitable for numerical implementation by the introduction of an appropriate parametrisation (including regularisation and normalisation). Most importantly, the scheme was adapted for complex electron densities and dielectric preconditioning. With these augmentations in place, the methods were revealed to be insufficient in providing a systematic improvement over the best-performing default method in \textsc{castep} -- Pulay mixing. However, MSB2 mixing was only slightly less efficient than Pulay mixing, and notably outperformed Broyden mixing as implemented in \textsc{castep}. This is a proof of concept for the work of Marks $\&$ Luke, but the methods remain unable to improve on the current standard without further adaptation. Second, the Periodic Pulay mixing scheme of Ref$.$ \citep{PP} was implemented. This scheme defines quite a general concept in numerical analysis, namely, how to construct the iterative subspace such that the DIIS is best utilised \citep{concref1}. It was revealed that Periodic Pulay mixing could provide an improvement in efficiency over Pulay mixing when two linear steps are added to the history per DIIS step. However, this resulted in a far less robust scheme than Pulay mixing, as the hindrance from taking the linear steps caused divergence for increasingly ill-conditioned input systems. Nonetheless, there is promise for Periodic Pulay mixing if a rigorous study on how best to utilise the concept can be done, as discussed in Ref$.$ \citep{PP}.

Finally, a computationally and memory efficient framework was implemented whereby a system-dependent, local and inhomogeneous model for the real space KS susceptibility could be proposed and subsequently used to precondition density mixing schemes. This was an attempt to remedy the shortcomings of popular preconditioning strategies such as Kerker preconditioning. The framework was demonstrated to function as intended, although the susceptibility models considered in this work lacked the depth required to provide an improved treatment of the KS dielectric over the Kerker form. As such, any improvements were found to be mostly parameter dependent. However, when the parameters of both the Kerker form and the density scaled susceptibility were optimised manually, the peak performance of the latter was shown to be superior to the peak performance of the former. Thus, with a more sophisticated and well-motivated susceptibility model, this framework may eventually provide the means for including inhomogeneity and system-dependence directly into the preconditioner, leading eventually to a more robust and efficient default method for achieving self-consistently.

\section{Further Work}
\label{furtherwork}

The real space susceptibility models presented in this thesis were preliminary and lacking the sophistication required to improve the efficiency of density mixing. As such, immediate future work will be to derive a rigorous, well-motivated model for the real space susceptibility that can be used to precondition the self-consistent cycles. Furthermore, Kresse \textit{et al}. in Ref$.$ \citep{kresse2} detail how an on-the-fly optimisation of the linear scaling parameter, $\alpha$, can be done utilising the iterative subspace of densities. This will be implemented within \textsc{castep}, and on-the-fly updates of other parameters will be studied in a similar fashion.

In an effort to implement the GW approximation in \textsc{castep}, exact methods for calculating the KS susceptibility have also been implemented \citep{vincent}. In this implementation, the KS susceptibility is constructed either using numerical derivatives, or alternatively by use of the Sternheimer equation. Therefore, these exact computations of the KS susceptibility will be adapted to precondition the SCF cycles of density mixing. In doing so, an up-to-date analysis of the effectiveness of exact dielectric preconditioning can performed.

Throughout this work spin was ignored. Density mixing for spin-polarised input systems is far more poorly dealt with than in charge density mixing. The reason for this is that the linear response function for \textit{spin densities} is entirely defined by the exchange-correlation contribution to the KS Hamiltonian. However, the current preconditioning strategy is to simply apply the Kerker preconditioner, constructed with no reference to exchange and correlation, to the spin-up and spin-down densities respectively. As expected, this leads to far less robust and efficient convergence in spin-polarised systems. An analysis of the spin density-density response function can thus be performed for simple exchange-correlation approximations such as the LSDA in an attempt to better condition spin-dependent systems.





\addtocontents{toc}{\vspace{2em}} 

\appendix 



\chapter{Derivation of Newton's Method} 

\label{appA} 

\lhead{Appendix A. Derivation of Newton's Method} 

Newton's method for finding $\textbf{x}^*$ such that $\textbf{f}(\textbf{x}^*) = 0$ starting from an initial guess $\textbf{x}_0$ will be derived. Firstly, the assumption that $\textbf{x}_0$ is close to the root will be made, such that the addition of a small perturbation vector $\textbf{e} \in \mathbb{R}^n$ yields the root, $\textbf{f}({\textbf{x} + \textbf{e}}) = 0$. The goal is to therefore find the $\textbf{e}$ that satisfies this equation, or in practice, an $\textbf{e}$ that will drive subsequent values of $\textbf{x}$ closer to satisfying it. To find this $\textbf{e}$, the Taylor expansion of $f_i(\textbf{x}+\textbf{e})$ about $\textbf{x}$ is computed,
\begin{align}
f_i(\textbf{x}+\textbf{e}) =& f_i(\textbf{x}) + \sum_j \frac{\partial f_i( \textbf{x})}{\partial x_j} h_j + \mathcal{O}(||\textbf{e}||^2), \\ \label{taylor}
\textbf{f}(\textbf{x}+\textbf{e}) =& \textbf{f}(\textbf{x}) + J_f(\textbf{x}) \textbf{e} + \mathcal{O}(||\textbf{e}||^2).
\end{align}
where $J_f(\textbf{x})$ is the Jacobian matrix of $f$ at $\textbf{x}$ -- $J_f(\textbf{x})_{ij} = \frac{\partial f_i}{\partial x_j}$. The larger $||\textbf{e}||$ is, the further away the Taylor expansion will deviate from the exact evaluation of the function. Ideally, the initial $\textbf{e}$ that is calculated will bring $\textbf{x}$ much closer to the root (provided it was `close' to the root to begin with), but will over or undershoot due to the first order nature of the Taylor expansion. Taylor expansions are then repeatedly done as $\textbf{e}$ (the absolute error in $\textbf{x} - \textbf{x}^*$) tends to zero. This defines the self-consistent process to find  $\textbf{x}^*$ from a suitable initial value,
\begin{align}
\textbf{x}_{n+1} = \textbf{x}_{n} + \textbf{e}.
\end{align} 
Where $\textbf{e}$ is obtained from Eq$.$ (\ref{taylor}) via
\begin{gather}
\textbf{f}(\textbf{x}+\textbf{e}) = \textbf{0} \implies \textbf{e} = -J^{-1}_f(\textbf{x})\textbf{f}(\textbf{x}),
\end{gather}  
producing
\begin{align}
\textbf{x}_{n+1} = \textbf{x}_{n} -J^{-1}_f(\textbf{x})\textbf{f}(\textbf{x}).
\end{align}
This the famous Newton-Raphson update formula.


\chapter{Test Suite} 

\label{apptestsuite} 

\lhead{Appendix B. Test Suite} 

Many of the materials presented here have been collated from literature on ill-conditioned KS systems. Hence, further motivation for the set of systems to follow can be found in \citep{review111,ADFT,RSK,gonze,TFmixing,ML,PP}. However, key features of interest and potential sources of ill-conditioning are listed below each input.

\begin{enumerate}

\item \textbf{Aluminium}. 
\begin{itemize}
\item Metallic 
\end{itemize}

\item \textbf{Aluminium Surface}. 
\begin{itemize}
\item Inhomogeneous ($10$\AA{} vacuum gap)
\item Interface of metallic and insulating regions
\end{itemize}

\item \textbf{Aluminium} $4 \times 1 \times 1$ \textbf{Supercell}. 
\begin{itemize}
\item Large, asymmetric unit cell
\end{itemize}

\item \textbf{Argon}
\begin{itemize}
\item Wide-gap insulator
\end{itemize}

\item \textbf{Gallium Arsenide}
\begin{itemize}
\item Inhomogeneous
\item Semiconductor
\end{itemize}

\item \textbf{Graphene}
\begin{itemize}
\item Semimetal
\item Inhomogeneous ($10$\AA{} vacuum gap)
\end{itemize}

\item \textbf{Isolated Oxygen}
\begin{itemize}
\item Localised
\item Inhomogeneous ($10$\AA{} vacuum gap in all dimensions)
\end{itemize}

\item \textbf{AIRSS Output: Potassium, Phosphorous, and Tin}
\begin{itemize}
\item Atomic configuration far-from-equilibrium
\end{itemize}

\item \textbf{AIRSS Output: Lithium and Copper}
\begin{itemize}
\item Atomic configuration far-from-equilibrium
\item Contains a transition metal
\end{itemize}

\item \textbf{AIRSS Output: Titanium and Potassium}
\begin{itemize}
\item Atomic configuration far-from-equilibrium
\item Contains a transition metal
\end{itemize}

\item \textbf{Magnesium Oxide}
\begin{itemize}
\item Prototypical polar oxide
\item Wide-gap insulator
\end{itemize}

\item \textbf{Palladium}
\begin{itemize}
\item Transition metal
\end{itemize}

\item \textbf{Silicon}
\begin{itemize}
\item Prototypical simple semiconductor
\end{itemize}

\item \textbf{Strontium}
\begin{itemize}
\item Metal
\end{itemize}

\item \textbf{Titanium Oxide}
\begin{itemize}
\item Transition metal oxide
\end{itemize}

\item \textbf{8-Zigzag Graphene Nanoribbon}
\begin{itemize}
\item Metallic
\item Inhomogeneous ($10$\AA{} vacuum gap in two dimensions)
\end{itemize}

\end{enumerate}



\addtocontents{toc}{\vspace{2em}} 

\backmatter



\label{Bibliography}

\lhead{\emph{References}} 

\bibliographystyle{unsrtnat} 

\bibliography{Bibliography} 

\begin{thebibliography}{80}
\providecommand{\natexlab}[1]{#1}
\providecommand{\url}[1]{\texttt{#1}}
\expandafter\ifx\csname urlstyle\endcsname\relax
  \providecommand{\doi}[1]{doi: #1}\else
  \providecommand{\doi}{doi: \begingroup \urlstyle{rm}\Url}\fi

\bibitem[Martin((2008))]{martin_electruc}
R.~M. Martin.
\newblock {Electronic Structure: Basic Theory and Practical Methods}.
\newblock \emph{Cambridge University Press; 1st edition}, (2008).

\bibitem[{R. A. Friesner}((2005))]{qchem}
{R. A. Friesner}.
\newblock {Ab initio quantum chemistry: Methodology and applications}.
\newblock \emph{Proc. Nat. Academy. Sci. 102, 19}, (2005).

\bibitem[{M. Sob \textit{et al}.}((2004))]{matsci}
{M. Sob \textit{et al}.}
\newblock {The role of ab initio electronic structure calculations in studies
  of the strength of materials}.
\newblock \emph{Mat. Sci. Eng. A, 387, 148-157}, (2004).

\bibitem[{G. Hagen \textit{et al}.}((2014))]{nucphys}
{G. Hagen \textit{et al}.}
\newblock {Coupled-cluster computations of atomic nuclei}.
\newblock \emph{Reports on Progress in Physics, 77, 9}, (2014).

\bibitem[\textit{et al}.((2014))]{phil_dft}
P.~Hasnip \textit{et al}.
\newblock {Density functional theory in the solid state}.
\newblock \emph{Phil. Trans. R. Soc. A 372: 20130270}, (2014).

\bibitem[{D. R. Bowler}((2016))]{dft111}
{D. R. Bowler}.
\newblock {Density functional theory: a tale of success in three codes}.
\newblock \emph{J. Phys. Cond. Mat., 28, 42}, (2016).

\bibitem[Burke((2012))]{111122}
K.~Burke.
\newblock {Perspective on density functional theory}.
\newblock \emph{J. Chem. Phys. 136, 150901}, (2012).

\bibitem[{K. Lejaeghere \textit{et al}.}((2016))]{repdft}
{K. Lejaeghere \textit{et al}.}
\newblock {Reproducibility in density functional theory calculations of
  solids}.
\newblock \emph{Science, 351, 6280}, (2016).

\bibitem[{C.-K. Skylaris \textit{et al}.}((2005))]{onetep}
{C.-K. Skylaris \textit{et al}.}
\newblock {Introducing ONETEP: Linear-scaling density functional simulations on
  parallel computers}.
\newblock \emph{J. Chem. Phys., 122, 084119}, (2005).

\bibitem[Anglade and Gonze((2008))]{gonze}
P.~M. Anglade and X.~Gonze.
\newblock {Preconditioning of self-consistent-field cycles in
  density-functional theory: The extrapolar method}.
\newblock \emph{Phys. Rev. B 78, 045126}, (2008).

\bibitem[\textit{et al}((2005))]{CASTEP}
S.~J.~Clark \textit{et al}.
\newblock {First principles methods using CASTEP}.
\newblock \emph{Z. Kristallogr. 220, 567–570}, (2005).

\bibitem[Marks and Luke((2008))]{ML}
L.~D. Marks and D.~R. Luke.
\newblock {Robust mixing for ab initio quantum mechanical calculations}.
\newblock \emph{Phys. Rev. B 78, 075114}, (2008).

\bibitem[\textit{et al}((2001))]{wein2k}
P.~K. S.~Blaha \textit{et al}.
\newblock {WIEN2K, An Augmented Plane Wave + Local Orbitals Program for
  Calculating Crystal Properties}.
\newblock \emph{Technische Universität Wien, Wien}, (2001).

\bibitem[\textit{et al}((2016){\natexlab{a}})]{PP}
A.~S.~Banerjee \textit{et al}.
\newblock {Periodic Pulay method for robust and efficient convergence
  acceleration of self-consistent field iterations}.
\newblock \emph{Chem. Phys. Lett 647, 31–35}, (2016){\natexlab{a}}.

\bibitem[\textit{et al}.((1996){\natexlab{a}})]{dmft}
A.~Georges \textit{et al}.
\newblock {Dynamical mean-field theory of strongly correlated fermion systems
  and the limit of infinite dimensions}.
\newblock \emph{Rev. Mod. Phys. 68, 13}, (1996){\natexlab{a}}.

\bibitem[\textit{et al}((2016){\natexlab{b}})]{martin_intelec}
R.~M.~Martin \textit{et al}.
\newblock {Interacting Electrons: Theory and Computational Approaches}.
\newblock \emph{Cambridge University Press}, (2016){\natexlab{b}}.

\bibitem[Hedin((1965))]{hedin}
L.~Hedin.
\newblock {New Method for Calculating the One-Particle Green's Function with
  Application to the Electron-Gas Problem}.
\newblock \emph{Phys. Rev. 139, A796}, (1965).

\bibitem[Salpeter and Bethe((1951))]{BSE}
E.~E. Salpeter and H.~A. Bethe.
\newblock {A Relativistic Equation for Bound-State Problems}.
\newblock \emph{Phys. Rev. 84, 1232}, (1951).

\bibitem[Cremer((2013))]{configI}
D.~Cremer.
\newblock {From configuration interaction to coupled cluster theory: The
  quadratic configuration interaction approach}.
\newblock \emph{WIREs Comput. Mol. Sci., 3: 482-503}, (2013).

\bibitem[Acioli((1998))]{qmc}
P.~H. Acioli.
\newblock {Review of quantum Monte Carlo methods and their applications}.
\newblock \emph{Journal of Molecular Structure, 394 3, 75-85}, (1998).

\bibitem[Cremer((2011))]{mppt}
D.~Cremer.
\newblock {Møller–Plesset perturbation theory: from small molecule methods
  to methods for thousands of atoms}.
\newblock \emph{WIREs Comput. Mol. Sci. 1 509–530}, (2011).

\bibitem[Hohenberg and Kohn((1964))]{HKthe}
P.~Hohenberg and W.~Kohn.
\newblock {Inhomogeneous Electron Gas}.
\newblock \emph{Phys. Rev. 136, B864}, (1964).

\bibitem[\textit{et al}((1979))]{levy}
M.~Levy \textit{et al}.
\newblock {Universal variational functionals of electron densities, first-order
  density matrices, and natural spin-orbitals and solution of the
  v-representability problem}.
\newblock \emph{PNAS, 76, 12}, (1979).

\bibitem[Gilbert((1975))]{nrep}
T.~L. Gilbert.
\newblock {Hohenberg-Kohn theorem for nonlocal external potentials}.
\newblock \emph{Phys. Rev. B 12, 2111}, (1975).

\bibitem[Capele((2006))]{birdseye}
K.~Capele.
\newblock {A bird's-eye view of density-functional theory}.
\newblock \emph{arXiv:cond-mat/0211443}, (2006).

\bibitem[Kohn and Sham((1965))]{KSeq}
W.~Kohn and L.~J. Sham.
\newblock {Self-Consistent Equations Including Exchange and Correlation
  Effects}.
\newblock \emph{Phys. Rev. 140, A1133}, (1965).

\bibitem[Schindlmayr and Godby((1995))]{vrepprob}
A.~Schindlmayr and R.~W. Godby.
\newblock {Density-functional theory and the v-representability problem for
  model strongly correlated electron systems}.
\newblock \emph{Phys. Rev. B 51, 10427}, (1995).

\bibitem[Solovej((1986))]{tf1}
J.~P. Solovej.
\newblock {A new look at Thomas–Fermi theory}.
\newblock \emph{Mol. Phys. 114, 7-8}, (1986).

\bibitem[Janak((1978))]{janak}
J.~F. Janak.
\newblock {Proof that $dE/dn_i = \epsilon_i$ in density-functional theory}.
\newblock \emph{Phys. Rev. B 18, 7165}, (1978).

\bibitem[M.~Levy and Sahni((1984))]{highoc}
J.~P.~Perdew M.~Levy and V.~Sahni.
\newblock {Exact differential equation for the density and ionization energy of
  a many-particle system}.
\newblock \emph{Phys. Rev. A 30, 2745}, (1984).

\bibitem[\textit{et al}.((2016))]{lda1}
M.~T.~Entwistle \textit{et al}.
\newblock {Local density approximations from finite systems}.
\newblock \emph{Phys. Rev. B 94, 205134}, (2016).

\bibitem[Perdew and Schmidt((2001))]{jladder}
J.~P. Perdew and K.~Schmidt.
\newblock {Jacob’s ladder of density functional approximations for the
  exchange-correlation energy}.
\newblock \emph{AIP Conf. Proc. 577, 1}, (2001).

\bibitem[N.~Marzari and Payne((1997))]{edft}
D.~Vanderbilt N.~Marzari and M.~C. Payne.
\newblock {Ensemble Density-Functional Theory for Ab Initio Molecular Dynamics
  of Metals and Finite-Temperature Insulators}.
\newblock \emph{Phys. Rev. Lett. 79 7}, (1997).

\bibitem[L.~P.~Bouckaert and Wigner((1936))]{smooth}
R.~Smoluchowski L.~P.~Bouckaert and E.~Wigner.
\newblock {Theory of Brillouin zones and symmetry properties of wave functions
  in crystals}.
\newblock \emph{Phys. Rev. 50, 58}, (1936).

\bibitem[Heine((1970))]{pseudop}
V.~Heine.
\newblock {The Pseudopotential Concept}.
\newblock \emph{Solid State Physics 24, 1-36}, (1970).

\bibitem[Kresse and Furthmuller((1996))]{vasp}
G.~Kresse and J.~Furthmuller.
\newblock {Efficiency of ab-initio total energy calculations for metals and
  semiconductors using a plane-wave basis set}.
\newblock \emph{Comp. Mat. Sci. 6, 1}, (1996).

\bibitem[\textit{et al}.((2002))]{abinit}
X.~Gonze \textit{et al}.
\newblock {First-principles computation of material properties : the ABINIT
  software project}.
\newblock \emph{Comp. Mat. Sci. 25, 478-492}, (2002).

\bibitem[\textit{et al}.((2009){\natexlab{a}})]{qesp}
P.~Giannozzi \textit{et al}.
\newblock {QUANTUM ESPRESSO: a modular and open-source software project for
  quantum simulations of materials}.
\newblock \emph{J. Phys. Cond. Mat., 21, 39}, (2009){\natexlab{a}}.

\bibitem[Davidson((1975))]{davidson}
E.~R. Davidson.
\newblock {The iterative calculation of a few of the lowest eigenvalues and
  corresponding eigenvectors of large real symmetric matrices}.
\newblock \emph{J. Comput. Phys. 17, 87-94}, (1975).

\bibitem[scl((2017))]{sclark}
{Private communication with Stewart Clark}.
\newblock (2017).

\bibitem[Smith((2017))]{mattreport}
M.~Smith.
\newblock {Accelerating first principles electronic structure calculations}.
\newblock \emph{First Year PhD Report (Unpublished)}, (2017).

\bibitem[Faul((2016))]{anita1}
A.~C. Faul.
\newblock {A Concise Introduction to Numerical Analysis}.
\newblock \emph{Chapman and Hall/CRC}, (2016).

\bibitem[Dederichs and Zeller((1983))]{MM}
P.~H. Dederichs and R.~Zeller.
\newblock {Self-consistency iterations in electronic-structure calculations}.
\newblock \emph{Phys. Rev. B 28, 5462}, (1983).

\bibitem[Broyden((1965))]{broyden}
C.~G. Broyden.
\newblock {A class of methods for solving nonlinear simultaneous equations}.
\newblock \emph{Math. Comp. 19, 577-593}, (1965).

\bibitem[Pulay((1980))]{DIIS}
P.~Pulay.
\newblock {Convergence acceleration of iterative sequences. the case of scf
  iteration}.
\newblock \emph{Chem. Phys. Lett. 73, 2}, (1980).

\bibitem[Kresse and Furthmüller((1996))]{kresse2}
G.~Kresse and J.~Furthmüller.
\newblock {Efficient iterative schemes for ab initio total-energy calculations
  using a plane-wave basis set}.
\newblock \emph{Phys. Rev. B, 54, 11169}, (1996).

\bibitem[Lin and Yang((2012))]{review111}
L.~Lin and C.~Yang.
\newblock {Elliptic preconditioner for accelerating the self consistent field
  iteration in Kohn-Sham density functional theory}.
\newblock \emph{arXiv:1206.2225}, (2012).

\bibitem[Srivastava((1984))]{limmem}
G.~P. Srivastava.
\newblock {Broyden's method for self-consistent field convergence
  acceleration}.
\newblock \emph{J. Phys. A. 17, 6}, (1984).

\bibitem[Vanderbilt and Louie((1984))]{VL}
D.~Vanderbilt and S.~G. Louie.
\newblock {Total energies of diamond (111) surface reconstructions by a linear
  combination of atomic orbitals method}.
\newblock \emph{Phys. Rev. B. 30, 6118-6130}, (1984).

\bibitem[Eyert((1996))]{Eyert}
V.~Eyert.
\newblock {A comparative study on methods for convergence acceleration of
  iterative vector sequences.}
\newblock \emph{J. Comp. Phys. 124, 271-285}, (1996).

\bibitem[Johnson((1988))]{johnson}
D.~D. Johnson.
\newblock {Modified Broyden’s method for accelerating convergence in
  self-consistent calculations}.
\newblock \emph{Phys. Rev. B, 38, 12807–12813}, (1988).

\bibitem[\textit{et al}.((2004))]{krynewton}
D.~A.~Knoll \textit{et al}.
\newblock {Jacobian-free Newton–Krylov methods: a survey of approaches and
  applications}.
\newblock \emph{J. Comp. Phys. 193, 2, 357-397}, (2004).

\bibitem[Saad and Schultz((1986))]{gmres}
Y.~Saad and M.H. Schultz.
\newblock {GMRES: A generalized minimal residual algorithm for solving
  nonsymmetric linear systems}.
\newblock \emph{J. Sci. Stat. Comput., 7, 856-869}, (1986).

\bibitem[Anderson((1965))]{anderson}
D.~G. Anderson.
\newblock {Iterative Procedures for Nonlinear Integral Equations}.
\newblock \emph{Journal of the ACM 12, 4}, (1965).

\bibitem[Harl((2008))]{jharl}
J.~Harl.
\newblock {The linear response function in density functional theory}.
\newblock \emph{PhD Thesis (unpublished)}, (2008).

\bibitem[\textit{et al}.((1997))]{rex1}
P.~Ghosez \textit{et al}.
\newblock {Long-wavelength behavior of the exchange-correlation kernel in the
  Kohn-Sham theory of periodic systems}.
\newblock \emph{Phys. Rev. B. 56, 12811-12817}, (1997).

\bibitem[Adler((1962))]{adler1}
S.~L. Adler.
\newblock {Quantum theory of the dielectric constant in real solids}.
\newblock \emph{Phys. Rev. 126, 413-420}, (1962).

\bibitem[Wiser((1963))]{wiser1}
N.~Wiser.
\newblock {Dielectric constant with local field effects included}.
\newblock \emph{Phys. Rev. 46, 1002-1011}, (1963).

\bibitem[K.~M.~Ho and Joannopoulos((1982))]{hij}
J.~Ihm K.~M.~Ho and J.~D. Joannopoulos.
\newblock {Dielectric matrix scheme for fast convergence in self-consistent
  electronic-structure calculations}.
\newblock \emph{Phys. Rev. B. 25, 4260-4262}, (1982).

\bibitem[Sawamura and Kohyama((2004))]{gonzein1}
A.~Sawamura and M.~Kohyama.
\newblock {A Second-Variational Prediction Operator for Fast Convergence in
  Self-Consistent Electronic Structure Calculations}.
\newblock \emph{Mater. Trans. 45, 1422}, (2004).

\bibitem[Auer and Krotscheck((1999))]{gonzein2}
J.~Auer and E.~Krotscheck.
\newblock {A rapidly converging algorithm for solving the Kohn-Sham and related
  equations in electronic structure theory}.
\newblock \emph{Comput. Phys. Com. 118, 139}, (1999).

\bibitem[\textit{et al.}((2017))]{linphonon}
L.~Lin \textit{et al.}
\newblock {Adaptively Compressed Polarizability Operator for Accelerating Large
  Scale Ab Initio Phonon Calculations}.
\newblock \emph{Multis. Mod. Sim. 15, 29-55}, (2017).

\bibitem[Kerker((1981))]{kerker}
G.~P. Kerker.
\newblock {Efficient iteration scheme for self-consistent pseudopotential
  calculations}.
\newblock \emph{Phys. Rev. B. 23, 6}, (1981).

\bibitem[\textit{et al}((1975))]{mann}
M.~Mannien \textit{et al}.
\newblock {Electrons and positrons in metal vacancies}.
\newblock \emph{Phys. Rev. B 12, 4012}, (1975).

\bibitem[Ashcroft and Mermin((1976))]{ashcroftmer}
N.~W. Ashcroft and N.~D. Mermin.
\newblock {Solid State Physics}.
\newblock \emph{Thomson Learning, Toronto}, (1976).

\bibitem[Levine and Louie((1982))]{LL1}
Z.~H. Levine and S.~G. Louie.
\newblock {New model dielectric function and exchange-correlation potential for
  semiconductors and insulators}.
\newblock \emph{Phys. Rev. B. 25, 10}, (1982).

\bibitem[Penn((1962))]{penn}
D.~R. Penn.
\newblock {New model dielectric function and exchange-correlation potential for
  semiconductors and insulators}.
\newblock \emph{Phys. Rev. 128, 2093}, (1962).

\bibitem[\textit{et al}.((2017))]{resta1}
Y.~Zhou \textit{et al}.
\newblock {On the applicability of Kerker precoditioning scheme to the
  self-consistent density functional theory calculations of inhomogeneous
  systems}.
\newblock \emph{arXiv:1707.00848}, (2017).

\bibitem[Resta((1977))]{resta2}
R.~Resta.
\newblock {Thomas-Fermi dielectric screening in semiconductors}.
\newblock \emph{Phys. Rev. B 16, 2717}, (1977).

\bibitem[\textit{et al}((2008))]{RSK}
Y.~Shiihara \textit{et al}.
\newblock {Real-space Kerker method for self-consistent calculation using
  non-orthogonal basis functions}.
\newblock \emph{Modelling and Simulation in Materials Science and Engineering,
  16, 3}, (2008).

\bibitem[\textit{et al}.((2013))]{toeplitz}
H.~Khalil \textit{et al}.
\newblock {Superfast solution of Toeplitz systems based on syzygy reduction}.
\newblock \emph{arXiv:1301.5798}, (2013).

\bibitem[\textit{et al}.((2009){\natexlab{b}})]{phil222}
P.~Hasnip \textit{et al}.
\newblock {Band Parallelism in CASTEP: Scaling to More Than 1000 Cores}.
\newblock \emph{Cray User Group 2009 Proceedings, 1-5}, (2009){\natexlab{b}}.

\bibitem[Saad((2003))]{jacg}
Y.~Saad.
\newblock {Iterative Methods for Sparse Linear Systems}.
\newblock \emph{Society for Industrial and Applied Mathematics; 2 edition},
  (2003).

\bibitem[\textit{et al}.((2006))]{woma}
D.~A.~Scherlis \textit{et al}.
\newblock {A unified electrostatic and cavitation model for first-principles
  molecular dynamics in solution}.
\newblock \emph{J. Chem. Phys 124, 074103}, (2006).

\bibitem[Pickard and Needs((2011))]{AIRSS}
C.~J. Pickard and R.~J. Needs.
\newblock {Ab initio random structure searching}.
\newblock \emph{Journal of Physics: Condensed Matter, 23, 5}, (2011).

\bibitem[\textit{et al}.((1996){\natexlab{b}})]{PBE1}
J.~P.~Perdew \textit{et al}.
\newblock {Generalized Gradient Approximation Made Simple}.
\newblock \emph{Phys. Rev. Lett. 77, 3865}, (1996){\natexlab{b}}.

\bibitem[{P. Hasnip and M. I. J. Probert}((2015))]{ADFT}
{P. Hasnip and M. I. J. Probert}.
\newblock {Auxiliary density functionals: a new class of methods for efficient,
  stable density functional theory calculations}.
\newblock \emph{arXiv:1503.01420}, (2015).

\bibitem[{P. P. Pratapa \textit{et al}.}((2016))]{concref1}
{P. P. Pratapa \textit{et al}.}
\newblock {Anderson acceleration of the Jacobi iterative method: An efficient
  alternative to Krylov methods for large, sparse linear systems}.
\newblock \emph{J. Comp. Phys., 306, 43-54}, (2016).

\bibitem[vin((2017))]{vincent}
{Private communication with Vincent Sacksteder}.
\newblock (2017).

\bibitem[{D. Raczkowski, A. Canning, and L. W. Wang}((2001))]{TFmixing}
{D. Raczkowski, A. Canning, and L. W. Wang}.
\newblock {Thomas-Fermi charge mixing for obtaining self-consistency in density
  functional calculations}.
\newblock \emph{Phys. Rev. B. 64, 121101}, (2001).

\end{thebibliography}

\end{document}